\documentclass[longauth]{aa}

\newcommand{\teff}{\mbox{$T_{\rm eff}$}\xspace}

\newcommand{\vsini}{\mbox{$v \sin i$}\xspace}
\newcommand{\kms}{\mbox{km\,s$^{-1}$}\xspace}
\newcommand{\ms}{\mbox{m\,s$^{-1}$}\xspace}
\newcommand{\cms}{\mbox{cm\,s$^{-1}$}\xspace}

\newcommand{\Msun}{\mbox{$\mathrm{M_\odot}$}\xspace}
\newcommand{\Rsun}{\mbox{$\mathrm{R_\odot}$}\xspace}
\newcommand{\Mearth}{\mbox{$\mathrm{M_\oplus}$}\xspace}
\newcommand{\Rearth}{\mbox{$\mathrm{R_\oplus}$}\xspace}

\usepackage[colorlinks=true,linkcolor=blue,
citecolor=blue,filecolor=blue,urlcolor=blue]{hyperref}

\def\modif{}
\usepackage{soul}
\usepackage{graphicx}
\usepackage{txfonts}

\begin{document}

   \title{NIRPS joining HARPS at ESO 3.6m}

   \subtitle{On-sky performance and science objectives}

\author{
Fran\c{c}ois Bouchy\inst{1,*},
Ren\'e Doyon\inst{2,3},
Francesco Pepe\inst{1},
Claudio Melo\inst{4},
\'Etienne Artigau\inst{2,3},
Lison Malo\inst{2,3},
Fran\c{c}ois Wildi\inst{1},
Fr\'ed\'erique Baron\inst{2,3},
Xavier Delfosse\inst{5},
Jose Renan De Medeiros\inst{6}
% ...
}

\institute{
\inst{1}Observatoire de Gen\`eve, D\'epartement d’Astronomie, Universit\'e de Gen\`eve, Chemin Pegasi 51, 1290 Versoix, Switzerland\\
\inst{2}Institut Trottier de recherche sur les exoplan\`etes, D\'epartement de Physique, Universit\'e de Montr\'eal, Montr\'eal, Qu\'ebec, Canada\\
\inst{3}Observatoire du Mont-M\'egantic, Qu\'ebec, Canada\\
\inst{4}European Southern Observatory (ESO), Karl-Schwarzschild-Str. 2, 85748 Garching bei M\"unchen, Germany\\
\inst{5}Univ. Grenoble Alpes, CNRS, IPAG, F-38000 Grenoble, France\\
\inst{6}Departamento de F\'isica Te\'orica e Experimental, Universidade Federal do Rio Grande do Norte, Campus Universit\'ario, Natal, RN, 59072-970, Brazil\\
\inst{7}Instituto de Astrof\'isica de Canarias (IAC), Calle V\'ia L\'actea s/n, 38205 La Laguna, Tenerife, Spain\\
\inst{8}Departamento de Astrof\'isica, Universidad de La Laguna (ULL), 38206 La Laguna, Tenerife, Spain\\
\inst{9}Consejo Superior de Investigaciones Cient\'ificas (CSIC), E-28006 Madrid, Spain\\
\inst{10}Instituto de Astrof\'isica e Ci\^encias do Espa\c{c}o, Universidade do Porto, CAUP, Rua das Estrelas, 4150-762 Porto, Portugal\\
\inst{11}Departamento de F\'isica e Astronomia, Faculdade de Ci\^encias, Universidade do Porto, Rua do Campo Alegre, 4169-007 Porto, Portugal\\
\inst{12}Department of Physics and Space Science, Royal Military College of Canada, PO Box 17000, Station Forces, Kingston, ON, Canada\\
\inst{13}European Southern Observatory (ESO), Av. Alonso de Cordova 3107,  Casilla 19001, Santiago de Chile, Chile\\
\inst{14}University Observatory, Faculty of Physics, Ludwig-Maximilians-Universit\"at M\"unchen, Scheinerstr. 1, 81679 Munich, Germany\\
\inst{15}Centre of Optics, Photonics and Lasers, Universit\'e Laval, Qu\'ebec, Canada\\
\inst{16}Herzberg Astronomy and Astrophysics Research Centre, National Research Council of Canada\\
\inst{17}Instituto de Astrof\'isica e Ci\^encias do Espa\c{c}o, Faculdade de Ci\^encias da Universidade de Lisboa, Campo Grande, 1749-016 Lisboa, Portugal\\
\inst{18}Departamento de F\'isica da Faculdade de Ci\^encias da Universidade de Lisboa, Edif\'icio C8, 1749-016 Lisboa, Portugal\\
\inst{19}Space Research and Planetary Sciences, Physics Institute, University of Bern, Gesellschaftsstrasse 6, 3012 Bern, Switzerland\\
\inst{20}Center for Space and Habitability, University of Bern, Gesellschaftsstrasse 6, 3012 Bern, Switzerland\\
\inst{21}Department of Physics, University of Toronto, Toronto, ON M5S 3H4, Canada\\
\inst{22}Aix Marseille Univ, CNRS, CNES, LAM, Marseille, France\\
\inst{23}Department of Physics \& Astronomy, McMaster University, 1280 Main St W, Hamilton, ON, L8S 4L8, Canada\\
\inst{24}Department of Physics, McGill University, 3600 rue University, Montr\'eal, QC, H3A 2T8, Canada\\
\inst{25}Department of Earth \& Planetary Sciences, McGill University, 3450 rue University, Montr\'eal, QC, H3A 0E8, Canada\\
\inst{26}Departamento de F\'isica, Universidade Federal do Cear\'a, Caixa Postal 6030, Campus do Pici, Fortaleza, Brazil\\
\inst{27}Centro de Astrobiolog\'ia (CAB), CSIC-INTA, ESAC campus, Camino Bajo del Castillo s/n, 28692, Villanueva de la Ca\~nada (Madrid), Spain\\
\inst{28}Centre Vie dans l’Univers, Facult\'e des sciences de l’Universit\'e de Gen\`eve, Quai Ernest-Ansermet 30, 1205 Geneva, Switzerland\\
\inst{29}Plan\'etarium de Montr\'eal, Espace pour la Vie, 4801 av. Pierre-de Coubertin, Montr\'eal, Qu\'ebec, Canada\\
\inst{30}Lund Observatory, Division of Astrophysics, Department of Physics, Lund University, Box 118, 221 00 Lund, Sweden\\
\inst{31}York University, 4700 Keele St, North York, ON M3J 1P3\\
\inst{32}University of British Columbia, 2329 West Mall, Vancouver, BC, canada, v6t 1z4\\
\inst{33}Western University, Department of Physics \& Astronomy and Institute for Earth and Space Exploration, 1151 Richmond Street, London, ON N6A 3K7, Canada\\
\inst{34}Light Bridges S.L., Observatorio del Teide, Carretera del Observatorio, s/n Guimar, 38500, Tenerife, Canarias, Spain\\
\inst{35}Hamburger Sternwarte, Gojenbergsweg 112, D-21029 Hamburg, Germany\\
\inst{36}Subaru Telescope, National Astronomical Observatory of Japan (NAOJ), 650 N Aohoku Place, Hilo, HI 96720, USA\\
\inst{37}Department of Astronomy \& Astrophysics, University of Chicago, 5640 South Ellis Avenue, Chicago, IL 60637, USA\\
\inst{38}Bishop's Univeristy, Dept of Physics and Astronomy, Johnson-104E, 2600 College Street, Sherbrooke, QC, Canada, J1M 1Z7\\
\inst{39}Laboratoire Lagrange, Observatoire de la C\^ote d’Azur, CNRS, Universit\'e C\^ote d’Azur, Nice, France\\
\inst{*}\email{Francois.Bouchy@unige.ch}
}

   \date{Received 7 December 2024 / Accepted 4 June 2025}

 
  \abstract
{The Near-InfraRed Planet Searcher (NIRPS) is a high-resolution, high-stability near-infrared (NIR) spectrograph equipped with an adaptive optics (AO) system. Installed on the ESO 3.6-m telescope at La Silla Observatory, Chile, it was developed to enable radial velocity (RV) measurements of low-mass exoplanets around M dwarfs and to characterise exoplanet atmospheres in the NIR.} 
{This paper provides a comprehensive design overview and characterisation of the NIRPS instrument, reporting on its on-sky performance, advising on how to carry out observations, and presenting its guaranteed time observation (GTO) programme.} 
{Intensive on-sky testing phases were conducted between November 2019 and March 2023. The instrument started its operations on 1 April 2023.}
{The spectral range continuously covers the $Y$, $J$, and $H$ bands from 972.4 to 1919.6\,nm. The thermal control system maintains 1\,mK stability over several months, thereby minimising drift. The NIRPS's AO-assisted fibre link improves coupling efficiency and offers a unique high-angular resolution capability with a fibre acceptance of only 0.4\arcsec. A high spectral resolving power of R$\sim$90\,000 and R$\sim$75\,000 is provided in high-accuracy (HA) and high-efficiency (HE) modes, respectively. The overall throughput from the top of the atmosphere to the detector peaks at 13\%. The RV precision, measured on the bright star Proxima with a known exoplanetary system, is {\modif 77} {\cms}. NIRPS and HARPS can be used simultaneously, offering unprecedented spectral coverage for spectroscopic characterisation and stellar activity mitigation. Modal noise can be aptly mitigated by the implementation of fibre stretchers and AO scanning mode.} 
{Initial results confirm that NIRPS opens new possibilities for RV measurements, stellar characterisation, and exoplanet atmosphere studies with high precision and high spectral fidelity. NIRPS demonstrated stable RV precision at the level of 1\,{\ms} over several weeks. The instrument’s high throughput, particularly in the $H$ band, offers a notable improvement over previous spectrographs, enhancing our ability to detect small exoplanets.}

   \titlerunning{NIRPS first light}
   \authorrunning{F. Bouchy et al.}
   \maketitle
%


\section{Introduction}

   M dwarfs, with masses in the range of 0.08 – 0.57 {\Msun}, are the most abundant type of stars in our Galaxy \citep[$\sim$75 \% of stars][]{Reyle2021}. These stars lie at the bottom of the main sequence and, consequently, they are small, cool, and intrinsically faint. The physical characteristics of M dwarfs offer many advantages when looking for smaller and cooler planets. An Earth-mass planet in the habitable zone of an M4 dwarf with a 0.23\,{\Msun} produces a Doppler wobble of 64\,{\cms} that is greater by a factor of 7 than that of the Earth's effect on our Sun. When caught in transit, such an Earth-sized planet decreases the flux of a 0.27\,{\Rsun} dwarf by 14 times as much as the Earth when it crosses the Sun. This makes planets around M dwarfs easier to detect, but even more importantly also easier to characterise. Therefore, obtaining statistics of planetary system occurrence and architecture for these stars is of great importance for constraining processes of planet formation and evolution. Occurrence rate studies based on transits \citep{Dressing2015} and radial velocity (RV) measurements \citep{Bonfils2013,Mignon2024} all point to the fact that small planets are abundant around M dwarfs, with about 15\% of them having exoplanets in their habitable zone (HZ). RV surveys of M dwarfs are gaining momentum as an important complement to surveys of solar-type stars and as a method to discover and possibly characterise warm and temperate rocky exoplanets. Radial-velocity searches for planets around M dwarfs benefit from a larger signal and a shorter orbital period for planets in the HZ. These advantages, along with the larger transit depths, have been exploited to find some of the lowest mass exoplanets known so far, both with the RV method (e.g. Proxima system - \cite{Anglada2016}, \cite{Faria2022}; YZ\,Cet - \cite{Astudillo-Defru2017}; Teegarden system - \cite{Zechmeister2019}, \cite{Dreizler2024}; Barnard system - \cite{Gonzalez2024}; GJ1002 system - \cite{Suarez2023}) and with photometric transits (e.g. L98-59=TOI-175 system, \cite{Kostov2019}, \cite{Demangeon2021}). The growing number of well-characterised exoplanets around M dwarfs allows for more detailed population studies. \cite{Parc2024} showed that smaller Sub-Neptunes (1.8 {\Rearth} < Rp < 2.8 {\Rearth}) seem to have a slightly lower bulk density than those in orbit around FGK dwarfs. This can be attributed to the fact that these planets are ice-rich, having accreted most of their solids beyond the ice line before their migration \citep{Alibert2017,Venturini2020,Burn2021}. 
   
   A subset of transiting planets orbiting bright M dwarfs has received extensive attention as ideal targets for atmospheric studies. A recent highlight is the detection of water vapour and carbon-bearing molecules in the atmosphere of temperate sub-Neptunes (also called Hycean worlds) K2-18\,b \citep{Benneke2019,Madhusudhan2023} and TOI-270d \citep{Holmberg2024,Benneke2024}. Low-mass exoplanets (< 10 {\Mearth}) orbiting M dwarfs are also the most promising candidates for future atmospheric characterisation with direct imaging due to their lower contrast in luminosity and radius compared to their host stars. For example, it can be possible to characterise their reflected light by combining high-dispersion spectroscopy with high-contrast imaging with ELT \citep[e.g.][]{Snellen2015,Lovis2024}. However, this requires such targets to be detected or identified beforehand for characterisation. The present RV surveys cover M-type stars of spectral types mainly earlier than M3 or M4, corresponding to stellar masses larger than 0.25 to 0.4 {\Msun}. The faintness of the targets in the visible wavelength range and the intrinsic stellar jitter have up to now limited the investigation on late-M dwarfs. This limitation motivated a shift towards near-infrared (NIR) facilities. 

In the past decade, high-dispersion cross-correlation spectroscopy has emerged as an extremely powerful method to identify the main carbon- and oxygen-bearing molecular species in hot Jupiter atmospheres and to estimate atmospheric elemental abundance ratios such as C/H, O/H, and C/O. These quantities  can provide fundamental insight into planet’s formation history. A key aspect of the technique is the fact that short-period planets experience large line-of-sight velocity changes that induce Doppler shifts corresponding to multiple resolution elements per hour on a high-resolution spectrograph. This can allow for the atmospheric signatures of exoplanets to be disentangled from contributions coming from both their host star and Earth’s transmittance, which (in contrast) are essentially stationary in
wavelength \citep{Birkby_2018}. This technique has successfully been used to produce a wealth of unambiguous molecular detections in both transiting and non-transiting exoplanet atmospheres using NIR spectrographs such as CRIRES \citep[e.g.][]{Snellen2010}, Keck NIRSPEC \citep[e.g.][]{Rodler2013}, {\modif GIANO-B \citep[e.g.][]{Brogi2018,Giacobbe2021,Basilicata2025} }, CARMENES \citep[e.g.][]{Alonso-Floriano2019}, IGRINS \citep[e.g.][]{Flagg2019}, Subaru IRD \citep[e.g.][]{Nugroho2021}, and SPIRou \citep[e.g.][]{Pelletier2021}. Moreover, this technique led to in-depth studies of the exoplanets' upper atmospheres through the NIR helium triplet \citep{Oklopcic2018} with CARMENES \citep[e.g.][]{Allart2018,Nortmann2018}, {\modif GIANO-B \citep[e.g.][]{Guilluy2020,Guilluy2024} }, SPIRou \citep[e.g.][]{Allart2023,Masson2024}, and Keck NIRSPEC \citep[e.g.][]{Kirk2020, Spake2021}. Characterising the upper atmosphere mass-loss rate, temperature, and dynamics informs us about the planets' evolution.
   
In response to an ESO call for new instruments for the New Technology Telescope (NTT) in February 2015, the Near-InfraRed Planet Searcher (NIRPS) {\modif consortium\footnote{{\modif The NIRPS consortium is composed of: Université de Montréal (co-project lead, Canada), Observatoire Astronomique de l'Université de Genève (co-project lead, Switzerland), Instituto de Astrofisica e Ciencias do Espaco hosted by the Centro de Astrofísica da Universidade do Porto and the Faculdade de Ciências da Universidade de Lisboa (Portugal), Instituto de Astrof\'isica de Canarias (Spain), Université de Grenoble-Alpes (France) and Universidade Federal do Rio Grande do Norte (Brazil). ESO participated in the NIRPS project as an associated partner. In total, more than 140 people contributed to the project and the science team includes 80 members.}}} proposed a dedicated NIR spectrograph to undertake an ambitious survey of planetary systems around M dwarfs. This would complement the surveys that have been running for two decades on HARPS \citep[e.g.][]{Bonfils2013,Mignon2024} by enlarging the sample of M dwarfs that can be observed while providing an improved filtering for the stellar activity \citep[e.g.][]{Carmona2023}. After the selection of the SOXS \citep[Son Of X-Shooter - ][]{Schipani2018} spectrograph for the NTT, it was clear that the exoplanet community of the ESO member states needed a new facility to maintain the leadership built over the last decades. Therefore, in May 2015, ESO invited the NIRPS team to adapt the original NIRPS design to the Cassegrain focus of the ESO 3.6-m Telescope in La Silla for simultaneous observation with the HARPS spectrograph. NIRPS builds upon the lessons learned from the first generation of nIR velocimeters GIANO \citep{Oliva2012}, CARMENES \citep{Quirrenbach2014}, and SPIRou \citep{Donati2020}, as well as the success of exoplanet optical hunters HARPS \citep{Mayor2003} and ESPRESSO \citep{Pepe2021}. 
   
   The present paper aims at providing a general description of the NIRPS instrument and its performance as delivered to the community on 1 April 2023. Section \ref{sec:project} summarises the project history. In Section \ref{sec:instrum}, we present a first-level description of the instrument and its subsystems to provide a general overview. Section \ref{sec:OPO} describes the observations and operation. Section \ref{sec:on-sky} presents the on-sky performance derived from the commissioning phases as well as from the daily calibrations obtained during first semester of operation. In Section \ref{sec:GTO}, we describe our guaranteed time observation (GTO) programme and give an overview of suitable NIRPS science cases. We present our conclusions in Section \ref{sec:conclu}. \\ 


\section{The NIRPS project}
\label{sec:project}

\begin{table*}
    \centering                          
    \caption{Commissioning runs and technical missions on NIRPS}             
    \label{comm}      
    \begin{tabular}{lcl}  \hline\hline 
    Run & Date & Activities \\   \hline 
    Commissioning 1 &  29 Nov - 6 Dec 2019  &  Front end
 installation + HARPS recommissioning\\
    Commissioning 2 &  8 - 17 June 2021  & AO+guiding tests \\
    Commissioning 3 &  22 Sep - 1 Oct 2021  & AO+guiding+ADC tests   \\ 
    Commissioning 4 &  26 Nov - 6 Dec 2021  & Fibre link installation and tests \\
    Technical activities & Mar - Apr 2022 & Back end installation \\
    First light  & 17 May 2022 &  First light with the whole instrument  \\
    Commissioning 5 &  7 - 19 June 2022 &  \\
    Technical activities  &  Jul - Aug 2022 &  Replacement of echelle grating \\ 
    Commissioning 6 &  19 Sept - 1 Oct 2022  &  \\
    Technical activities  &  Oct 2022 &  Blaze angle adjustment + fibre stretchers addition \\ 
    Commissioning 7 &  26 Nov - 7 Dec 2022 &  \\
    Commissioning 8 &  16 - 27 Jan 2023  &  \\
    Commissioning 9 &  1 - 11 Mar 2023 &  \\ 
    Start of science operation & 1 April 2023 & \\
    Technical activities &  12 - 21 Apr 2023 & Spectrograph thermal enclosure installation \\\hline                                   
    \end{tabular}
\end{table*}

The {\modif NIRPS} project kick-off was held in January 2016 and the design phase ended with the final design review in May 2017. The procurement of components and the manufacturing of subsystems took between two and four years. The integration of the front end started in early 2017 with the first subsystem, the adaptive optics (AO), 
in the integration hall of the Geneva Observatory, which overlapped with the procurement phase of other subsystems in the various consortium partner institutes. The memorandum of understanding (MoU) between ESO and NIRPS consortium was signed in June 2017. The Provisional Acceptance Europe (PAE) document was completed in May 2019 for the Fibre Link, in September 2019 for the front end, and in October 2021 for the back end.

The assembly, integration, and verification (AIV) of the NIRPS front end was completed November 2019 followed by the first commissioning on sky. 
In response to the COVID-19 pandemic, La Silla's technical and scientific operations were completely suspended from March 2020 to November 2020. They  gradually restarted, but with strong restrictions lasting until June 2021. Two additional commissioning phases of the front end took place in June and September 2021. The installation and commissioning of the fibre link was done in December 2021. The AIV of the cryogenic spectrograph in La Silla started in March 2022. The first light of NIRPS took place in May 2022 and was followed by a first commissioning of the entire instrument in June 2022. In July 2022, the echelle grating, suffering from too many ghosts and a low throughput, was replaced by a new one developed by IOF-Fraunhofer\footnote{\href{https://www.iof.fraunhofer.de/en.html}{www.iof.fraunhofer.de}}. This was followed by four commissioning runs in September and November 2022, then in January and March 2023. The different technical phases and commissioning runs are reported in Table~\ref{comm}.  

The official start of operations took place on April 1, 2023, the date on which NIRPS was offered by ESO both to the community for open-time observations and to the NIRPS consortium for {\modif GTOs (see Section~\ref{sec:GTO})}. A thermally-controlled insulation box was installed around the cryogenic spectrograph in April 2023, a few weeks after the start of operation to improve the thermal stability of the spectrograph, the double scrambler, and the Fabry-Pérot etalon.

\section{Technical description of the NIRPS instrument}
\label{sec:instrum}

To address its main science cases, the NIRPS design has been derived from a set of top-level requirements:
\begin{itemize}
\item a spectrograph operating in the $Y$, $J$, $H$ band;
\item a spectral resolution higher than 80\,000;
\item an overall efficiency at blaze peaks at the level of 10\%;
\item RV precision at the level of 1\,{\ms} and high spectral fidelity;
\item simultaneous operation with HARPS.
\end{itemize}

To satisfy the requirements, NIRPS has been conceived as an AO-assisted, fibre-fed, cross-dispersed, high-resolution, high-stability, NIR cryogenic echelle spectrograph. A description of the instrument and all its sub-systems is given in \cite{Wildi2022} and references therein. In the following sub-sections, we summarise the main characteristics of each of the five NIRPS sub-systems (the front end,  fibre link,  calibration unit,  spectrograph, and  detector) without entering into technical details, which have previously been given in the cited reference papers. 

\subsection{The front end with its AO system}
\label{sec:front-end}

\begin{figure}[h]
  \centering
  \includegraphics[width=0.50\textwidth]{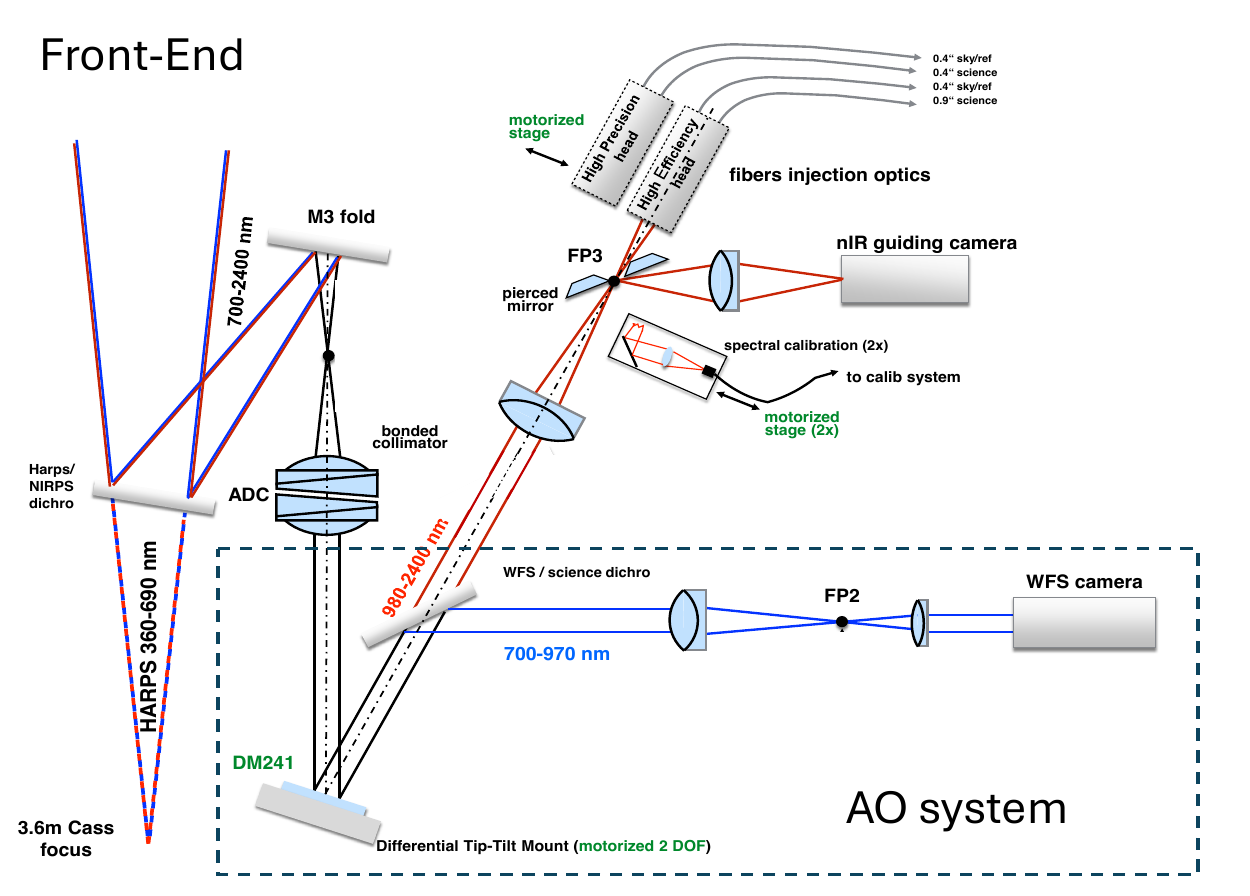}
  \caption{Schematic view of optical design and layout of the NIRPS front end
 and its components}
\label{fig:NIRPS_FE}
\end{figure}

The front end (also called the Cassegrain unit) is composed of different sub-systems located in a base plate attached directly to the Cassegrain rotator of the 3.6m telescope. The NIRPS front end design and performance are described in \cite{Blind2022}. It is composed of the following sub-systems, as given in Fig.~\ref{fig:NIRPS_FE}: 
\begin{itemize}
    \item VIS-NIR dichroic splitting the telescope beam between NIRPS and HARPS;
    \item M3 motorised fold mirror that serves to align the telescope pupil and compensate the wobble of the pupil introduced by the atmospheric dispersion corrector (ADC); 
    \item ADC \citep{Cabral2022}, which provides correction for an airmass range of 1--1.8 over the entire NIR domain;
    \item AO system \citep{Conod2016, Wildi2017} composed of an $15\times15$ actuator deformable mirror, mounted on a long stroke tip-tilt mount, and a $14\times14$ Shack Hartmann wave-front sensor; 
    \item IRTCCD guiding camera operating in the NIR; 
    \item Fibre injection head;
    \item Calibration fibre head. 
\end{itemize}

NIRPS is specifically a conjugate AO-assisted instrument, allowing for the main fibre (so-called high-accuracy fibres, HA) to be only 0.4\arcsec\ in diameter on the sky and for the spectrograph to be much more compact than seeing-limited instruments. A larger fibre (high-efficiency fibre, HE) with a field of view (FoV) of 0.9\arcsec\ is also available, delivering only 15\% lower spectral resolution thanks to a pupil slicer. The front end first splits the light between HARPS and NIRPS with a removable NIR-VIS dichroic. It allows for simultaneous observations of the two instruments, as well as to operate HARPS alone or in its polarimetric mode when removed. The dichroic's bluest part (including H \& K Ca lines) presents a very good and flat transmission (> 96\%), which is only reduced to $\sim$80\% between 435 and 455 nm. Observations that require the highest signal-to-noise ratio (S/N) in this domain must then consider using the HARPS-only mode. Furthermore, due to the non-smooth transmission of the dichroic around the Calcium H\&K lines (390nm), their indices are slightly offset when it is inserted. The VIS-NIR dichroic does not affect the HARPS RV precision and does not introduce any RV offset above a level of 19\,cm/s in high-accuracy mode (HAM) and 60\,cm/s in high-efficiency mode (EGGS). This was verified during the commissioning phase \#1 with a short sequence made on the RV standard star HD20794, {\modif known to harbor a RV scatter no larger than 1.2\,{\ms} over seven years \citep{Pepe2011}}, alternatively with and without the VIS-NIR dichroic.  

\ The second dichroic splits the light between the wave-front sensor (WFS) (700-950\,nm) and the science channel (980-1800\,nm). The adaptive optics is a classical on-axis system based on a $14\times14$ sub-apertures Shack-Hartmann WFS with a FoV of $\pm$2\arcsec, featuring a state-of-the-art sub-electron read-out, deep-depleted EMCCD, FLI OCAM2 camera, and an ALPAO DM241 $15\times15$ deformable mirror (DM). It runs with a loop frequency from 250\,Hz to 1\,kHz and corrects efficiently on stars up to $I=14.5$. The motorised tip-tilt mount on which the DM is installed allows us to compensate for the differential tip-tilt between HARPS and NIRPS, as well as regularly offloading the DM. The NIR guiding camera steers the beam to the input lens of the fibre link by introducing slope offsets in the AO feedback loop. The AO system is also used to homogeneously scan the fibre tip at a relatively low frequency (0.3\,Hz) to couple the diffraction-limited PSF into as many fibre modes as possible to minimise the modal noise (see Section~\ref{sec:modalnoise}). 

\begin{figure}[h]
  \centering
  \includegraphics[width=0.50\textwidth]{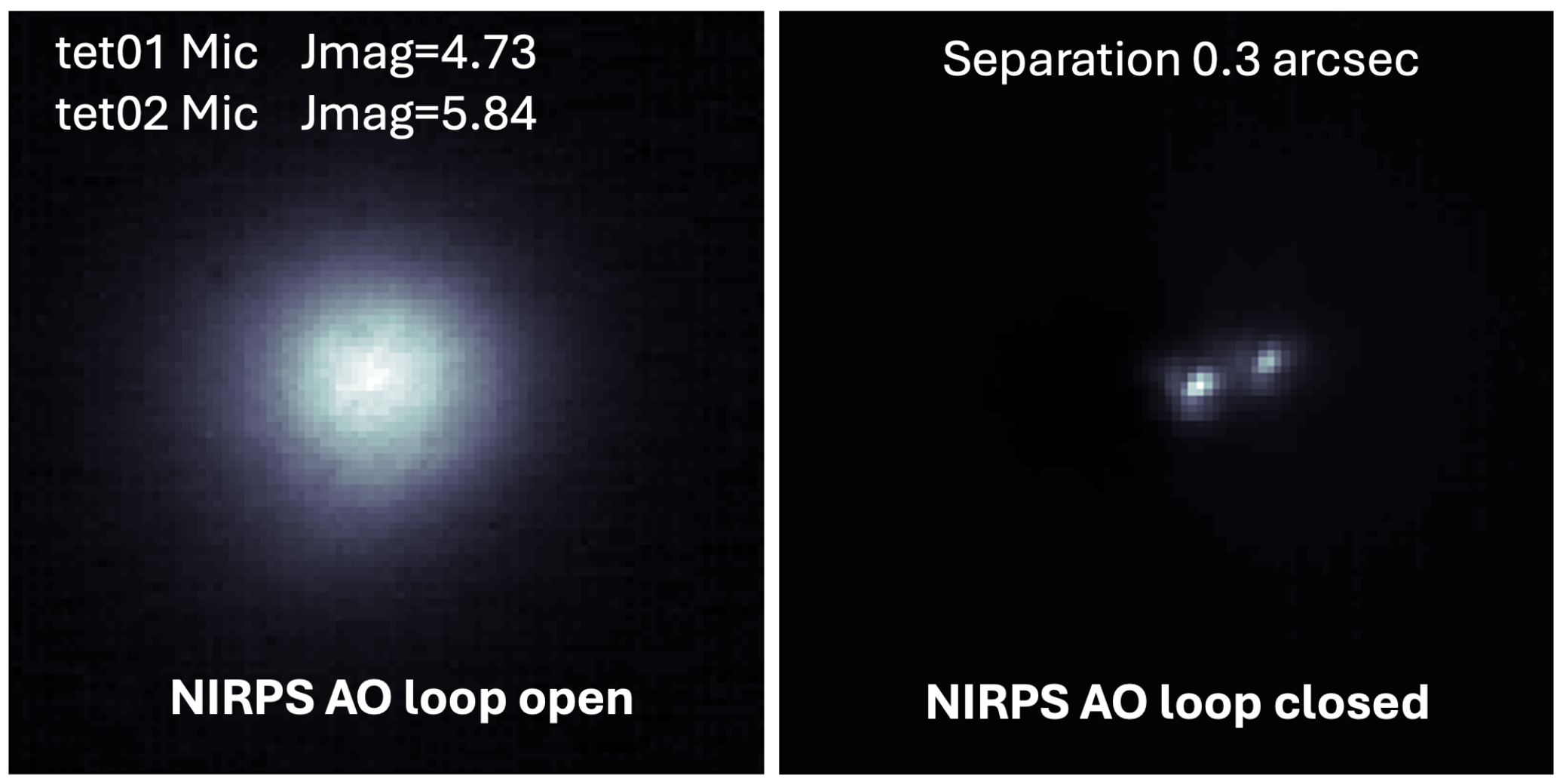}
  \caption{Illustration of the NIRPS AO performance on the $\theta Mic$ binaries with two components separated by 0.3\arcsec.}
\label{fig:AO_perf1}
\end{figure}

\begin{figure}[h]
  \centering
  \includegraphics[width=0.50\textwidth]{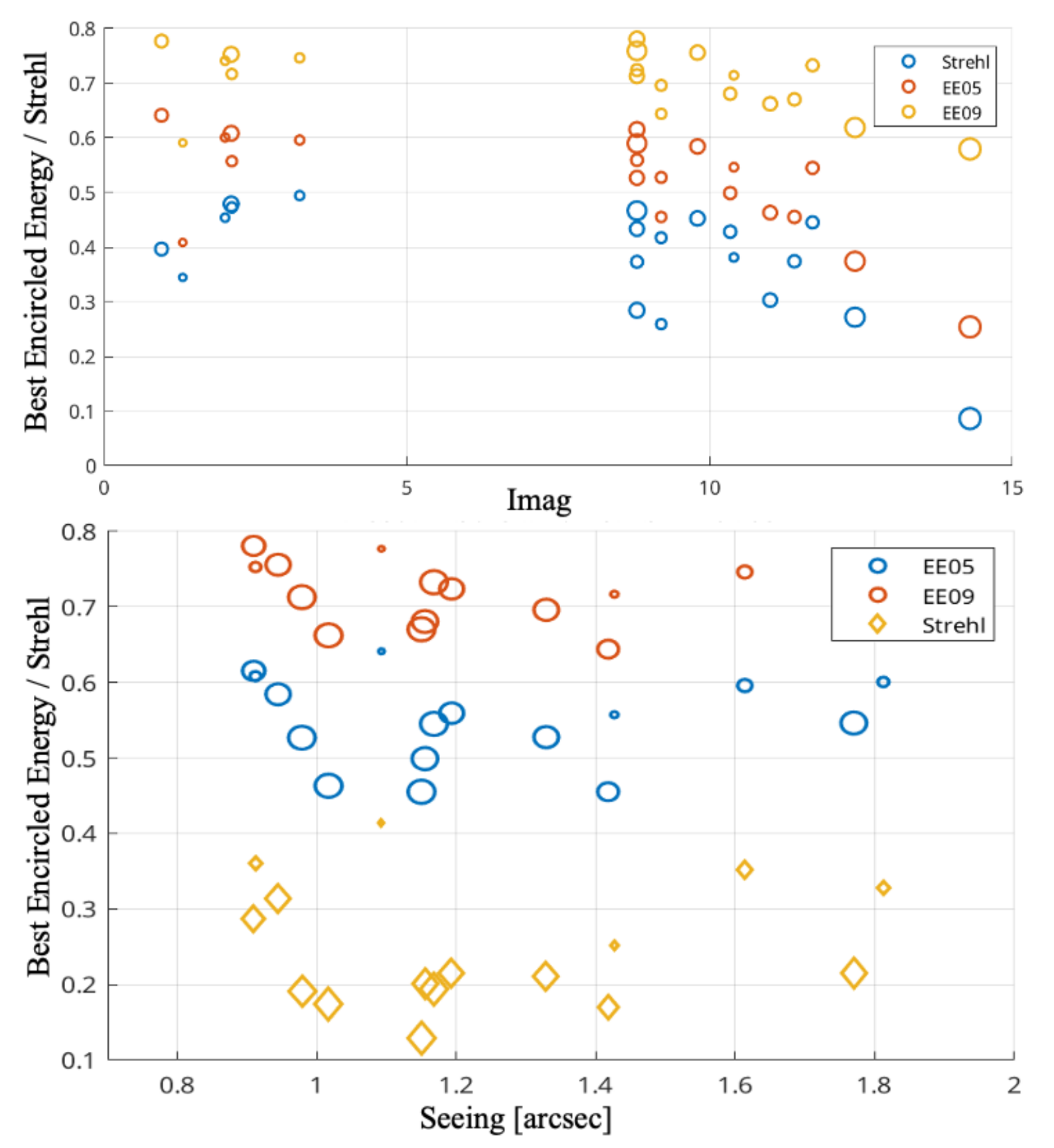}
  \caption{NIRPS adaptive optics performance measured during commissioning \#4. Top: Encircled energy for 0.5\arcsec\ and 0.9\arcsec\ and Strehl versus $I_{\rm mag}$. Dot size is proportional to Fried parameter, $r_0$, corresponding to seeing ranging from 2.2\arcsec\ to 0.7\arcsec\. (Bottom) Encircled energy for 0.5\arcsec\ and 0.9\arcsec\ and Strehl versus seeing from the AOP telemetry. Dot symbol size is proportional to the I mag.}
\label{fig:AO_perf2}
\end{figure}

The NIRPS AO performance was described by \cite{Blind2022}, who reported tests performed during commissioning phase \#4. Figure \ref{fig:AO_perf1} illustrates the AO performance on the bright binary system $\theta$\,Mic featuring two components (HR8151 and HR8180) separated by 0.3\,arcsec. 
Strehl at 1350\,nm typically reaches 30-40\%, slightly lower than the expected value of 55\% from simulations. However, this does not significantly impact fibre coupling, with the encircled energy reaching 55\% and 70\% for the 0.4\arcsec\ and 0.9\arcsec\ fibres, respectively and  these values remaining quite constant up to $I=11$ (Fig.~\ref{fig:AO_perf2}). The encircled energy linearly decreases down to 20\% and 55\% at $I=14.5$ for the 0.4\arcsec\ and 0.9\arcsec\ fibres respectively. Here, $I=14.5$ corresponds to the faintest magnitude on which the AO loop can be closed and the measured coupling values are close to the seeing-limited coupling since only tip-tilt is corrected.

The AO performance also shows low dependency with seeing up to 1.8\arcsec. A deeper analysis of the impact of seeing conditions on the global throughput was reported by \cite{Artigau2024} by comparing the S/N in the $H$ band of all Proxima measurements done in HE mode with seeing values reported by the AO. The S/N upper envelope stays constant for seeing up to 1.25\arcsec\ and slightly decreases by $\sim$10-12\% for seeing up to 2.25\arcsec,\ corresponding to a throughput loss of 20-25\% to be compared with a throughput loss of 90\% for a seeing-limited coupling.  

The NIRPS AO system has also been tested and can be used to observe Solar System objects as long as they have an angular size smaller than the WFS sub-apertures FoV ($\sim$2\arcsec), such as the moons of Saturn and Jupiter. 

Two linear stages are used to insert the light from the spectral calibration sources into the science fibre and/or reference fibre. A linear stage allows for swapping between both HA and HE fibres. The telescope guiding is always performed by the HARPS Technical CCD (TCCD) covering a FoV of $77\times77$ arcsec$^2$ and the HARPS tip-tilt device. The NIRPS guiding system then only manages the offset between HARPS and NIRPS optical paths. The guiding camera uses a NIR InGaAs camera (FLI C-Red2\footnote{\href{https://andor.oxinst.com/products/first-light-imaging-cameras}{andor.oxinst.com/products/first-light-imaging-cameras}}) and a high magnification lens, providing a fine plate scale of 35\,mas/pixel (i.e. 2.5\,pixel per diffraction element at 1550\,nm) and a FoV of $18\times22$ arcsec$^2$. This high sampling allows for an accurate guiding onto the fibres, which optimises 
coupling and limits modal noise on NIRPS few-mode fibres (see Sect.~\ref{sec:modalnoise}). It was also used during commissioning to perform diffraction-limited imaging and characterisation of the AO PSF and performance. The guiding on the NIRPS fibre entrance is done by sending tip-tilt offsets to the WFS reference slopes. The DM is offloaded at low frequency to the tip-tilt mount, which provides a total excursion of $\pm$40\arcsec and does not need to be offloaded to the telescope.
In case the HARPS tip-tilt does an offload to the telescope, the NIRPS AO system quickly recovers from it. In the NIRPS guiding camera, the angle of north is $148.4\pm3.4^\circ$, with the angle measured in the TCCD from the horizontal X axis to the vertical Y axis counterclockwise.

\subsection{The fibre link and the fibre stretcher}

The fibre-link design was based on the concepts developed for SOPHIE \citep{Bouchy2013}, HARPS \citep{LoCurto2015}, and ESPRESSO \citep{Megevand2014}. The NIRPS fibre link subsystem carries light from the telescope's front end of the spectrograph. Functionally, the fibre link performs the following tasks:
\begin{itemize}
\item Converts the F/8.09 front-end beam into an F/4.2 beam for optimal coupling in optical fibres;\item Mixes the phase between the fibre modes in an in-line, low-loss fibre stretcher;
\item Scrambles the image and pupil to improve the illumination stability;
\item Feeds the spectrograph located inside of a cryostat.
\end{itemize}

As specified in the previous section \ref{sec:front-end}, the fibre link supports two observing modes. Each mode requires a specific pair of fibres, each pair being mounted in a specific head:
\begin{itemize}
\item HA mode with an octagonal fibre of 29\,$\mu$m diameter with a conjugate size of $0.4\arcsec$ on the sky;
\item The HE mode using a 66\,$\mu$m diameter octagonal fibre whose conjugate size is $0.9\arcsec$ on the sky. In the double scrambler, the pupil image of this fibre is sliced in two halves, feeding a rectangular fibre of $33\times132$\,$\mu$m. This allows us to maintain high resolution and stable illumination. 
\end{itemize}
Given their relatively small size and the NIR wavelengths, those fibres work in regime of just a few modes. As few as 30 modes can propagate through the HA fibre at H-band wavelengths (compared to 1000\,s in HARPS fibres for instance), making modal noise management one of the challenges of NIRPS.

For both the HA and HE modes, two fibres illuminate the spectrograph: one fibre (A) carries the light from the science target, while the second one (B) of 29\,$\mu$m diameter carries either the light from the sky background (37\arcsec\ away) or the light from a reference source for simultaneous drift measurements. At the back of each of the two pierced mirrors of the fibre heads in the front end
 (see Fig.~\ref{fig:NIRPS_FE}), two relay optics re-image the light onto the target (A)  and reference (B) fibres. Each fibre individually carries the light to the double scrambler, which is mounted on the vacuum vessel. The main functionality of the double scrambler is to exchange the far field and the near field of the fibres. In combination with the use of octagonal fibres, this ensures a homogeneous and stable illumination of the spectrograph, which is fundamental for repeatable RV measurements \citep{Bouchy2013}. The double scrambler also integrates a vacuum window that provides the optical feed-through of the vacuum vessel. All fibres are then routed from the scrambler to the focal plane of the spectrograph to form a single entrance `slit' 
 bolted inside the vacuum vessel.
A and B fibres of each mode are aligned vertically along the cross-dispersion direction, while HA and HE pairs of fibres are arranged side-by-side horizontally along the main-dispersion direction. The output end of the fibre link converts the F/4.2 fibre beam to the F/8 spectrograph beam using an air-spaced triplet. 

Fibre stretchers were implemented on the four fibres. These units stretch a 20\,m spooled section of optical fibres with a total amplitude of 6 to 8\,mm at about $f$=0.3\,Hz. The modulation of the optical path between the fibre modes reduces the contrast of the output interference pattern (the so-called speckles in larger core fibres), ultimately minimising the modal noise \citep{Blind2022b,Frensch2022}. The modulation happens fully in the fibres, without a free-space interface, leading to a very high transmission >95\%. The benefits of this device on the measured modal noise are presented in Sec.\ref{sec:modalnoise}.\\

\subsection{The calibration and RV reference unit}

The purpose of the calibration unit is to characterise the spectrograph response to secure the highest possible RV stability, both for short-term (one night) and long-term (several years) activity. The calibration unit provides various sources to perform the following calibrations: 1) localisation, geometry and profile of spectral orders; 2) determination of the spectral flat-fielding and blaze profile response; 3) determination of the wavelength solution; and 4) determination of drift measurement in simultaneous-reference mode. The light from various calibration sources can be injected through the front end into any of the spectrograph fibres independently. A Tungsten lamp is used as a white lamp for order definition and spectral flat-fielding. Wavelength calibration is obtained by combining uranium-argon hollow-cathode lamp spectra with a Fabry-Pérot etalon illuminated with a bright fibre-fed Tungsten white lamp. The former provides absolute accuracy, and the latter delivers a high number (about 17\,800) of uniform spectral emission lines across the whole spectrum that ensure local wavelength precision \citep{Cersullo2019,Hobson2021}. 
The Fabry-Pérot etalon has a 12-mm cavity width housed in a temperature-controlled vacuum enclosure based on the same design used for the SPIRou one \citep{Cersullo2017}. A symmetrical set-up of parabolic mirrors couples the input fibre to the exit fibre, the etalon being located in the collimated beam between the two parabolas. The output fibre is connected to the calibration module. The RV reference unit is considered a light source of the calibration module. The calibration units also have the functionality of attenuating the light of the calibration source using two pairs of circular tunable neutral density filters. The calibration unit also includes two laser diodes for the AO calibration. There is no cold source in the calibration unit since the thermal background at ambient temperature in $YJH$ bands is negligible.

\subsection{The cryogenic spectrograph}

NIRPS is a cross-dispersed echelle spectrograph of the white-pupil type operating in quasi-Littrow conditions, located in the Coudé east room of the ESO 3.6\,m telescope. The optical design is described by \cite{Thibault2022}. The fibre beam (converted from 29\,$\mu$m$/$F4.2 to 55\,$\mu$m$/$F8.09) is collimated by the parabola and relayed to the echelle grating. The grating diffracts the collimated beam which is relayed back to the parabola. The parabola focuses the diffracted collimated beam to the flat mirror which folds it back to the parabola. The parabola collimates the diffracted beam to the cross disperser made of five refractive prisms which rotate the beam by 180 degrees. The refractive camera focuses the diffracted and cross-dispersed beam on the detector.
Each optical element (e.g. fold, parabola, grating, prisms) is bonded and assembled in its mechanical mount. The bonding is done using Armstrong A12 epoxy with a custom technique as described by \cite{Vallee2020}. The camera lenses are assembled in their cell using a retaining ring and beryllium-copper springs. 

The spectrograph is enclosed in a vacuum vessel (1.1\,m $\times$ 3.3\,m) maintained at a pressure $<$ 10$^{-6}$\,mbar. The cryogenic optical housing and optical elements are maintained at 75\,K with a stability better than 1\,mK RMS over 72 hours. To reach this stability, two CryoMech cryocoolers are coupled with 20 proportional-integral-derivative (PID) controllers. The thermal control system design responsible for setting control loop set points, safety interlocks, and merging temperatures is the Beckhoff PLC device, which monitors temperature from Lakeshore224, microK and opto22 devices. The high-level custom-made Twincat solution is described by \cite{Reshetov2020} and \cite{Malo2024}.

\begin{figure}[h]
  \centering
  \includegraphics[width=0.50\textwidth]{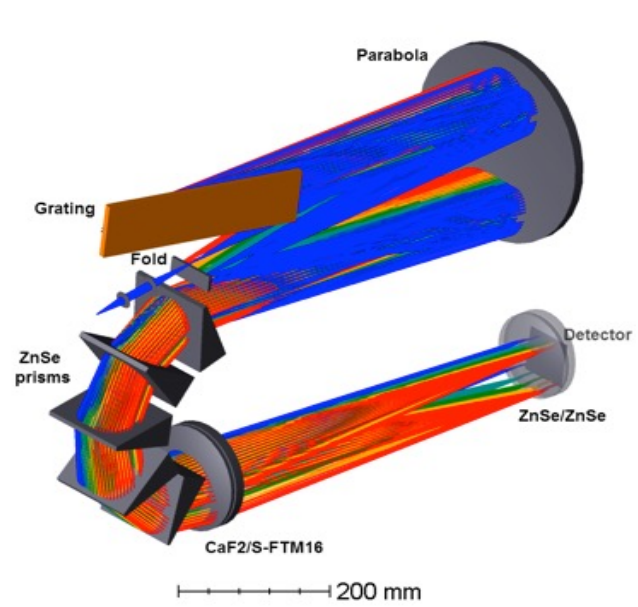}
  \caption{Schematic view of the optical design and layout of the NIRPS spectrograph and its components.}
\label{fig:NIRPS_FBE}
\end{figure}

Spectra from fibres A and B are formed by the spectrograph side by side on the detector. No moving parts are located inside the cryogenic enclosure. Figure \ref{fig:NIRPS_FBE} shows the optical design of the spectrograph. 

The initial echelle grating manufactured by Bach (76$^\circ$ blaze angle, with 13.333 lines/mm and efficiency $>$55\%.) used up to Commissioning \#5 was a traditional grating ruled in a thick layer of aluminum deposited on a thick slab of Zerodur blank of $80\times300$\,mm$^2$. Because the grating was chirped, its performance was not good enough, with efficiency in the 45-50\% range and numerous features and ghosts degrading imaging quality and spectral resolution. As such, a new grating made by wet-chemical etching applied to a crystalline silicon substrate was developed by Fraunhofer Institute for Applied Optics and Precision Engineering \citep{Sadlowski2023}. This technology workflow enables the creation of highly determined micro-facets and surfaces over macroscopic dimensions.  A gold coating was applied to increase the diffraction efficiency to about 70\%. The overall grating clear-aperture size is 78\,mm $\times$ 284\,mm providing a wave-front error of less than 70\,nm (RMS) measured throughout the full aperture. This new grating was tested during the commissioning phase \#6 and beyond. It allowed us to reach a much better image quality and a spectral resolution close to expectation.

During the commissioning phases, thermal coupling issues between the Coudé room and the optical bench were identified, motivating the installation of a thermally controlled insulation box around the vacuum vessel in April 2023. This allowed us to reach temperature stability well within 1\,mK, with daily changes below 0.1\,mK as shown by \cite{Artigau2024} and also described in Section \ref{sec:stability}.

\subsection{The NIR detector}

The broad simultaneous wavelength coverage of NIRPS, combined with its high spectral resolution, sampling of at least three pixels per resolution element, and sufficient inter-order spacing, necessitates the use of a large format detector. With its $4096\times4096$ pixel array and relatively low readout noise (RoN), the Hawaii-4RG (H4RG) detector, developed by Teledyne Scientific \& Imaging (TIS), is the optimal choice for this application. A similar H4RG detector is employed in the SPIRou spectrograph, with its performance described in \cite{Artigau2018}. The H4RG detector in NIRPS operates in a single-ended configuration, powered by TIS's SIDECAR ASIC. Unlike CCDs, hybrid CMOS IR detectors have a fixed sampling time. In the case of NIRPS, the minimum sampling time is 5.573\,s. Fortunately, the non-destructive readout nature of these detectors allows for multiple samples to be taken during a single integration. This sampling method, known as up-the-ramp, significantly reduces the RoN throughout an integration. The final data product is generated by performing an on-the-fly linear fit, non-linearity correction, and bias subtraction for each pixel. The FITS file extensions include the per-pixel slope, bias, and the number of samples used for each pixel. 

For correlated double sampling (CDS) images, the readout noise is $18.6 \pm 0.8$\,e$^-$. The RoN decreases as the sampling increases ($N_{d\_samples}$), potentially reaching a plateau close to 10\,e$^-$, as described by the following empirical equation:
\begin{equation}
    RoN=\sqrt{(245 / N_{d\_samples} + 100)} 
.\end{equation}

Inter-pixel capacitive coupling was measured in La Silla using subsections of hot pixels in a dark sequence. We found that 97\% of the flux received in a given pixel is detected within that pixel, while the remaining 3\% is distributed to adjacent pixels. 

The typical number of hot pixels determined from the DARK frames is 18\,000$\pm$200 corresponding to 0.108\% of the pixels. The typical number of bad pixels derived from the LED frames is 43\,700$\pm$400 corresponding to 0.26\% of the pixels. The Mean Dark level measured from a long dark exposure (830\,s) was estimated to be 1.7\,e$^-$/min. The typical cosmic rate was measured in the NIRPS Coudé room to be 0.0058\,hits/pixel/hour = 2575\,hits/cm$^2$/hour.  

While NIRPS integration times can be arbitrarily long, there is no reason to go beyond the timescale over which the dark current dominates over readout noise. The limit on integration times is fixed at 999\,s. Longer total integrations can be obtained by taking multiple exposures interleaved by resets. 
It is important to note that each integration includes an overhead of 11.15 seconds, resulting from the detector reset (5.573\,s) and the first readout (5.573\,s). Since at least two samples (CDS) are required to produce usable data, the minimum setup for an integration consists of one reset followed by at least two readouts.

Non-linearity was measured on illuminated frames using a 1550\,nm LED and a Lakeshore 121 current source. At an expected flux of 30\,000\,ADU, the deviation from linearity is around 3\%, increasing to 10\% at approximately 48\,000\,ADU. Linearity correction is applied on the fly by the low-level detector control software, ensuring accurate posemeter values and real-time feedback images for the user. In the data reduction pipeline, we fixed the quality control of saturation level at 45\,000 ADU per pixel.

We quantified the persistence of the NIRPS detector by performing a sequence of $5\times300$\,s DARK frames right after a Fabry-Pérot (FP) sequence. The data analysis was performed by stacking all FP peaks (about 17\,800) into a median FP peak, leading to a very-high-S/N version of the median FP, and tracing the evolution of persistence when it is well below the readout noise for any given persisting FP peak. The fractional flux falls off with an exponential decay from about 10$^{-4}$ of the initial flux within the first minute to about 2×10$^{-5}$ after 15\,min. A persisting flux at the level of 5×10$^{-5}$ is expected after the typical target-to-target slew time (2-3\,min). However, we recommend avoiding taking UrNe calibrations less than two hours before any scientific observations and avoiding observing a very faint target right after a very bright one.

\section{Observation and operation with NIRPS}
\label{sec:OPO}

\subsection{Observing configurations and instrument modes}

NIRPS was optimised for thermomechanical stability. By construction, the spectral format is fixed and the instrument configuration pre-defined. The only two aspects in which a user can configure NIRPS for a science objective are 1) the instrument mode (HA or HE), defining simultaneously spectral resolution, angle sub-tended in the sky and numerical sampling; and 2) the selection of the source to illuminate the reference fibre. 
The reference fibre can receive either light from the sky background or light from a calibration source for simultaneous instrumental drift measurement (by default the Fabry-Pérot etalon). In the first case, the reference fibre collects sky background from a second pinhole that points to a distance of 37\arcsec\ away from the main scientific target and is not visible in the nIR guiding camera. Considering the spectrograph intrinsic stability (see section \ref{sec:stability}), the usage of simultaneous drift reference is not helpful, even for high RV precision. Using the reference fibre to measure the sky spectra might be advisable to accurately characterise the sky background and detector noise contributions to the error budget. 

The acquisition templates must include the specific target parameters:
\begin{itemize}
\item Magnitude in $I$ and in $J$ band for WFS and guiding camera automatic preset respectively;
\item Target spectral type for an automatic selection of the cross-correlation-function (CCF) mask;
\item Target systemic RV for CCF computation; 
\item Selection of the source to illuminate the reference fibre.  
\end{itemize}

As in the case of other ESO instruments, the preparation of NIRPS observations is done using the usual P2 tool\footnote{\href{https://www.eso.org/sci/observing/phase2/p2intro.html}{eso.org/sci/observing/phase2/p2intro.html}} and assisted by the Exposure Time Calculator (ETC\footnote{\href{https://etc.eso.org/observing/etc/}{etc.eso.org/observing/etc/}}). The preset of the telescope and target acquisition is first done with the HARPS guiding camera. The target is selected by the operator and an offset is computed and sent to the telescope to centre the target in the HARPS fibre. The HARPS guiding system (via a dedicated tip-tilt plate) is activated. Thanks to the careful alignment, once the target is in the HARPS fibre, its nIR image falls very close (within 2\arcsec) to the NIRPS fibre. The operator then selects and clicks on the target identified on the NIRPS guiding camera. It automatically centres the target in the WFS FoV, closes the AO using the nominal parameters according to the I magnitude defined in the OB, and activates the guiding on the NIRPS fibre hole. For red objects too faint to be seen on the HARPS guiding camera, the guiding can be done off-axis on another target selected in the FoV of the HARPS guiding camera. There is also the possibility to do off-axis guiding on NIRPS for binaries or resolved objects with the constraint to be within the WFS FoV of $\pm$2\arcsec. The typical overhead\footnote{\href{https://www.eso.org/sci/facilities/lasilla/cfp/cfp116/overheads.html}{eso.org/sci/facilities/lasilla/cfp/cfp116/overheads}} for a NIRPS+HARPS acquisition ranges from 3\,min 30\,s to 5\,min 30\,s including telescope pointing and centring, target selection, HARPS guiding, instrument configuration, AO loop activation, and NIRPS guiding. Overhead between exposures on the same target but using a different observing mode is 3\,min. At the time of writing and as is the case with all instruments at La Silla, NIRPS is offered in visitor mode (VM) and delegated visitor mode (DVM) but not in service mode (SM).

\subsection{Daily and nightly calibrations}

Calibrations for HARPS and NIRPS are performed independently. The calibration exposure sequence has to be conducted according to the operation plan to provide calibration data for the automatic data reduction software (DRS) and consistency purposes over the lifetime of the instrument. 

The NIRPS back-end calibrations were performed during the day to determine the dark and bad pixels map, location and geometry of spectral orders,  blaze profile,  spectral flat‐field response, and the wavelength calibration. 
The detailed calibration plan is described in the NIRPS User Manual\footnote{\href{https://www.eso.org/sci/facilities/lasilla/instruments/NIRPS/doc/manuals/NIRPS-2000-GEN-UM-148-2.3\_User\_Manual.pdf}{eso.org/sci/facilities/lasilla/instruments/NIRPS/doc/manuals/NIRPS-2000-GEN-UM-148-2.3\_User\_Manual.pdf}}.
The calibration sequence must be completed at least two hours before the start of the night to avoid any persistence on the NIR detector from strong UrNe lines. A standard calibration OB containing the entire sequence is predefined for each instrumental configuration (HE and HA) to facilitate the daily calibration activities and it is available to the telescope Operators. The total duration of the calibration sequence for both HA and HE is about 2\,h. A set of predefined spectrophotometric standards and telluric standards (fast-rotating early-type stars) are regularly observed during twilight. A specific calibration sequence for the NIRPS front end is also performed every day. It includes Bias, Dark, Bad-pixel map, and Flat-Field of the NIRPS guiding camera, Dark calibration of the AO WFS, Calibration of the Interaction Matrix of the AO system and Bias of the AO deformable mirror. The total duration of the front-end
 calibration sequence is about 45\,min.

\subsection{Combining NIRPS and HARPS}

The HARPS and NIRPS spectrographs can be operated individually or jointly. The front end
 components that split the light between the two instruments are fixed and both instruments always see the relevant part of the spectrum of the science target. The default operation should be with both instruments operating simultaneously. There is no significant benefit in observing a target with only one of the spectrographs, even if all the science will be done with just one. There are two exceptions: 1) the polarimetric mode of HARPS (HARPS-pol), which requires the removal of the VIS-NIR dichroic and does not preserve polarisation properties; and 2) observations that require the highest S/N values between 435 and 455\,nm must consider using the HARPS-only mode.

NIRPS and HARPS are fed by different fibre links mounted on the front end; both instruments have a high-resolution and high-efficiency fibre pair. For HARPS, the high-resolution configuration known specifically as the high accuracy mode (HAM) and the high-efficiency configuration is EGGS. For NIRPS, the high-resolution configuration is NIRPS\_HA and the high-efficiency configuration is NIRPS\_HE. Users can switch from the high-resolution to the high-efficiency configuration at any time during the night, and the configuration used in one spectrograph does not constrain the configuration by the other spectrograph. This means that there are in total four acquisition modes for NIRPS+HARPS observations, allowing for the combination of all instrument modes. We note that the calibration sets of the different configurations must be obtained during the daytime.

To optimise the telescope time, the total integration times of both instruments should be roughly matched to minimise overheads. The user must take care to adapt exposure times and the number of exposures on both instruments to have the same total integration time. Furthermore, users ought to consider that exposure times longer than 1000\,s for NIRPS and longer than 3600\,s for HARPS are not recommended.

\subsection{Data flow and data reduction software\label{section_data_flow}} 

The raw data, composed of data cubes and  raw images, are archived in the ESO archive\footnote{\href{http://archive.eso.org/}{archive.eso.org/}} \citep{Delmotte_2006} and retrieved directly from there. The reduced data are also stored in the ESO archive. NIR guiding camera pre-reduced images corresponding to the average of the guiding images taken during the NIRPS exposure are also archived as an HDU extension of the NIRPS raw image.

NIRPS comes with data-reduction software (DRS) to deliver high-quality science-grade reduced spectra. The NIRPS-DRS is based on and adapted from the publicly available ESPRESSO pipeline \citep{Pepe2021}. Various updates to the ESPRESSO pipeline have been made to allow for the reduction of observations at IR wavelengths. {\modif Results presented on this paper are based on NIRPS-DRS version 3.2.0.} Among these changes is the proper handling of amplified cross-talk and low-spatial-frequency noise patterns \citep{Artigau2018}, the telluric absorption correction \citep{Allart2022}, and OH sky emission lines correction. The NIRPS-DRS\footnote{\href{https://www.eso.org/sci/facilities/lasilla/instruments/NIRPS/doc/manuals/espdr-pipeline-manual-2.4.16.pdf}{eso.org/sci/facilities/lasilla/instruments/NIRPS/doc/manuals/espdr-pipeline-manual-2.4.16.pdf}} is the nominal pipeline for NIRPS data reduction for the ESO science archive through the VLT data flow system (DFS). The NIRPS-DRS processes the data immediately after the observation and potentially issues a warning if values are out-of-bounds. The calibration products that have passed the quality control are copied to the local calibration database to be used in further data reduction. 

Data processing is organised as a sequential reduction cascade in which each raw data type is associated with a dedicated DRS recipe (i.e. a processing algorithm). The pipeline generates intermediate data products at each stage, many of which are then used as inputs in the subsequent steps of the cascade. The final task is the reduction of science spectra. The main steps of the science processing are: 

\begin{itemize}
    \item dark subtraction;
    \item subtraction of inter-order background;
    \item optimal extraction of spectral orders using master flat field as the order profile, with masking of dead, hot, and saturated pixels and rejection of cosmic rays;
    \item creation of extracted spectra in (order, pixel) space with associated error and quality maps;
    \item spectral flat-fielding and de-blazing of extracted spectra (S2D);
    \item correction of the OH sky lines for the observations with the simultaneous sky on fibre B;
    \item computation of barycentric correction and shift of wavelength solutions to the Solar System barycentre;
    \item instrumental drift computation for the observations with simultaneous reference on fibre B;
    \item creation of telluric-corrected S2D spectra;
    \item creation of resampled and merged 1D spectra (S1D) with associated error and quality maps;
    \item creation of flux-calibrated S1D spectra using estimated absolute efficiency and extinction curves;
    \item computation of the cross-correlation function (CCF) of the S2D spectrum with a stellar binary mask;
    \item fit of the CCF with a Gaussian model to derive RV and associated CCF parameters (FWHM, contrast, bisector span, asymmetry).
\end{itemize}

\subsection{APERO implementation for NIRPS}
\label{sec:apero}

Utilising two pipelines for data analysis is invaluable in cross-validating reductions and assessing the robustness of scientific results. This approach has become standard in JWST data analysis (e.g. \citealt{Lim2023}), as results confirmed by two pipelines reduce the risk of misinterpretation and significantly accelerate debugging and testing of new ideas. APERO\footnote{\href{http://apero.exoplanets.ca}{apero.exoplanets.ca}} is a publicly available Python package \citep{Cook2022} which was first developed for SPIRou data reduction \citep{Donati2020} and then designed to handle multiple instruments.  

APERO differs in several key ways from the NIRPS-DRS pipeline. Unlike NIRPS-DRS, which operates under a one-frame-in, one-frame-out requirement, APERO leverages multi-night data to optimise calibration and telluric correction. Its telluric correction uses a two-step process: the initial step follows the approach of \cite{Allart2022} using a \texttt{TAPAS} atmospheric model, applied to both science data and a set of hot star calibrators. The second step of the correction accounts for the spectrum of the science target by dividing it by an approximate stellar template, effectively removing the cross-term between telluric absorption and stellar features. From the hot stars, residuals to the telluric correction are derived and correlated with airmass and water absorption. First-order residuals are then further subtracted from the science spectrum using a PCA-based approach \citep{Artigau2014}.

Another significant difference lies in APERO’s accounting for the orientation and curvature of the fibre image along orders, which is particularly important for the HE fibre, characterised by a $1\times4$ aspect ratio and a tilt of up to 3$^\circ$ in parts of the orders. APERO uses a constant weight extraction method along the spatial profile, independent of the S/N. This contrasts with the \cite{horne1986} method used in NIRPS-DRS, which optimises for S/N propagation but introduces S/N-varying weights in the spatial dimension. APERO’s photon-limited regime approach, optimised for bright targets, leads to small differences in reported fluxes while maintaining similar reported S/N values. Radial velocity measurements in APERO are typically performed using the line-by-line (LBL\footnote{\href{http://lbl.exoplanets.ca}{lbl.exoplanets.ca}}) algorithm \citep{Artigau2022}. Although a CCF is computed for each observation and remains a relevant data product for stellar properties (e.g. v$\sin$i, detection of SB2 binaries), it is suboptimal for RV measurement in the NIR. The LBL algorithm can also handle NIRPS-DRS S2D spectra for cross-validation. Having a single RV measurement code that processes data from both pipelines enables the identification of discrepancies that arise from the reduction process (e.g. issues with telluric correction or extraction) or from the RV measurement itself (e.g. differences between template fitting, CCF fitting, or LBL).

\section{On-sky performance}
\label{sec:on-sky}

In this section, {\modif all} analyses were performed using NIRPS-DRS version 3.2.0. Some results were additionally obtained through post-processing with the APERO pipeline.

\subsection{Spectroscopic performance}

\subsubsection{Spectral format, image quality, and resolving power}\label{sec:resolution}

\begin{figure}
    \centering
    \includegraphics[width=0.5\textwidth]{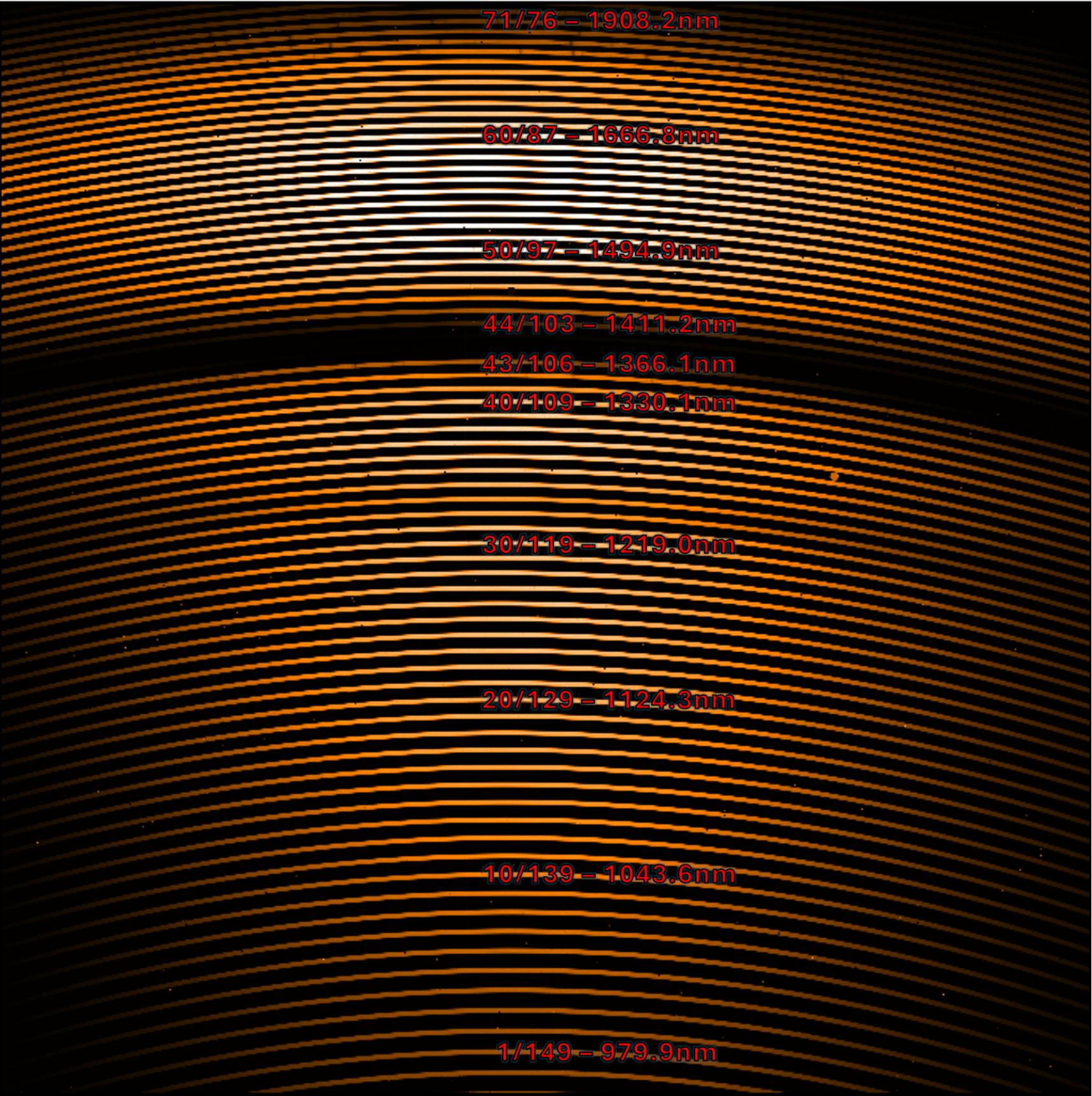}
    \caption{Raw frame of NIRPS with HE\_A fibre illuminated by the Tungsten lamp showing the spectral format with the location of the 71 detected spectral orders. Spectral orders are numbered with their localisation order (from 1 to 71), physical diffraction order (from 148 to 76) and central wavelength. 
    }
    \label{fig:raw_frame}
\end{figure}

\begin{figure}
    \centering
    \includegraphics[width=1.0\linewidth]{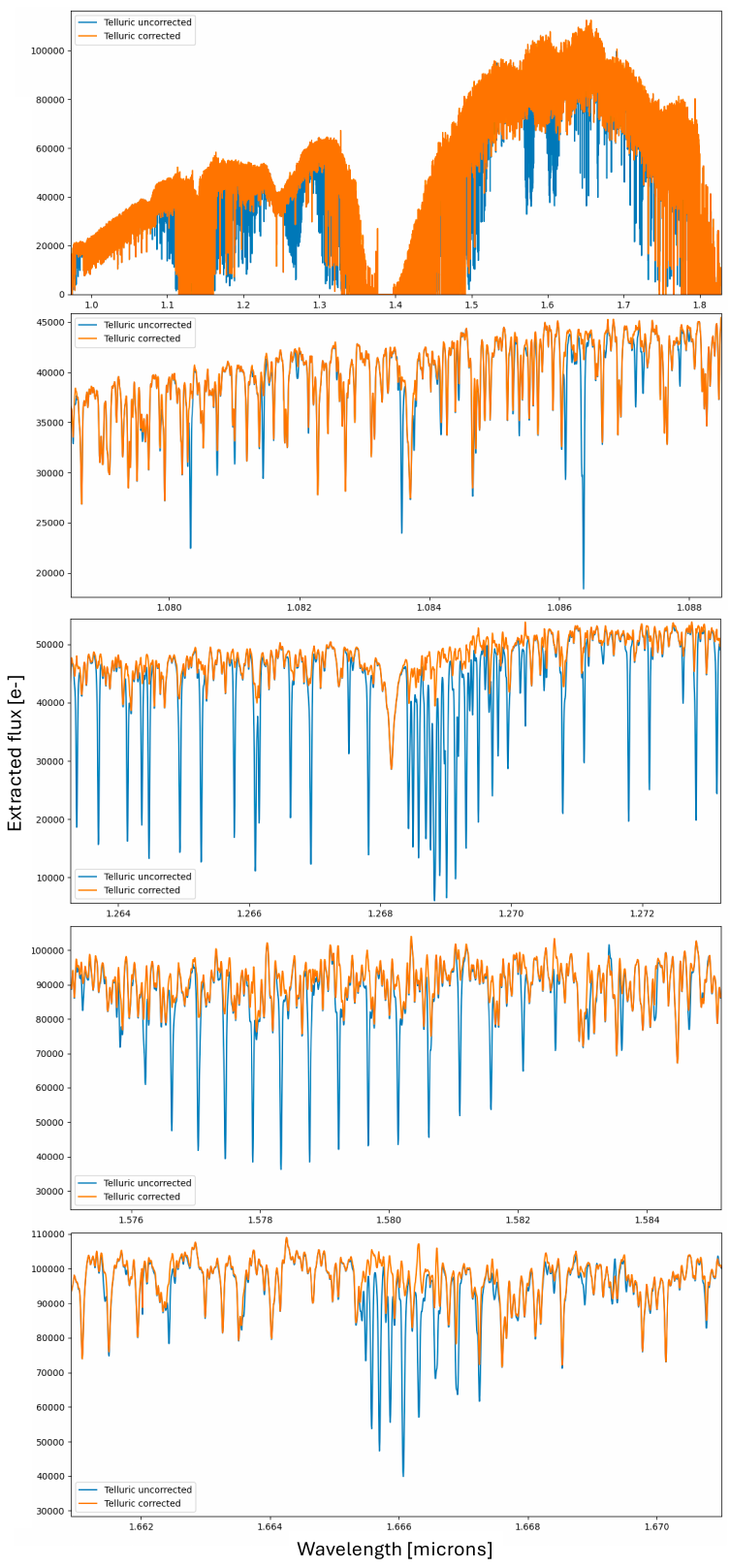}
    \caption{Top panel: Full extracted and wavelength calibrated spectrum with NIRPS-DRS of Proxima observed in HE mode. (4 Bottom panels) Zoom in on four different spectral domains. Blue curves and orange curves correspond to telluric uncorrected and telluric corrected spectra, respectively.}
    \label{fig:proxima_s1d}
\end{figure}

Figure~\ref{fig:raw_frame} shows the spectral format of NIRPS with 71 spectral orders. Two spectral orders (echelle orders 104 and 105) between 1376.5 and 1397.3\,nm are missing due to OH-doped absorption in the optical fibre train. These two orders also correspond to the unusable domain of the deep telluric water band between $J$ and $H$ photometric bands. A tabulated extract of the wavelengths of HE mode is given in Table~\ref{wave_format}. The spectral format of HA mode is shifted by about 36 pixels (36\,\kms) towards the blue relative to the HE mode. The NIRPS spectral ranges continuously cover the domain from 972.4\,nm to 1919.6\,nm. To illustrate the spectral range, richness, and quality, we show the extracted and wavelength-calibrated 1D-spectrum of Proxima as observed in the HA mode in Fig. \ref{fig:proxima_s1d}.

\begin{table}
    \centering                          
    \caption{NIRPS spectral format:\ HE mode}             
    \label{wave_format}      
    \begin{tabular}{lllll}  \hline\hline 
    Order & $\lambda_{c}$ & $\Delta\lambda$ & start-end & res. bin \\ 
          & [nm] & [nm] & [nm] & [nm/pix] \\ \hline
    71/76 & 1908.2 & 25.6 & 1894.0 - 1919.6 & 0.00626 \\
    70/77 & 1883.1 & 25.3 & 1869.4 - 1894.7 & 0.00619 \\
    & & & & \\
    60/87 & 1667.1 & 22.5 & 1654.4 - 1677.0 & 0.00553 \\
    59/88 & 1647.6 & 22.3 & 1635.6 - 1657.9 & 0.00546 \\
    & & & & \\
    50/97 & 1494.9 & 20.4 & 1483.8 - 1504.2 & 0.00499 \\
    49/98 & 1480.2 & 20.2 & 1468.6 - 1488.8 & 0.00494 \\
    & & & & \\
    44/103 & 1411.1 & 19.3 & 1397.3 - 1416.6 & 0.00472 \\
    43/106 & 1366.1 & 18.8 & 1357.7 - 1376.5 & 0.00460 \\
    & & & & \\
    40/109 & 1330.1 & 18.3 & 1320.3 - 1338.7 & 0.00450 \\
    39/110 & 1318.2 & 18.2 & 1308.3 - 1326.5 & 0.00445 \\
    & & & & \\
    30/119 & 1218.9 & 16.9 & 1209.3 - 1226.2 & 0.00413 \\
    29/120 & 1208.9 & 16.7 & 1199.2 - 1216.0 & 0.00411 \\
    & & & & \\
    20/129 & 1124.3 & 15.6 & 1115.5 - 1131.2 & 0.00384 \\ 
    19/130 & 1115.4 & 15.5 & 1107.0 - 1122.5 & 0.00379 \\
    & & & & \\
    10/139 & 1043.8 & 14.5 & 1035.3 - 1049.9 & 0.00355 \\
    9/140  & 1036.1 & 14.4 & 1027.9 - 1042.3 & 0.00352 \\
    & & & & \\
    2/147 & 986.7 & 13.6 & 979.0 - 992.6 & 0.00333 \\
    1/148 & 980.0 & 13.5 & 972.4 - 985.9 & 0.00330 \\ \hline
     \end{tabular}
     \tablefoot{For some representative extracted/diffraction orders, we provide the central wavelength ($\lambda_{c}$) in the middle of the order, the total spectral range ($\Delta\lambda$), the start and end wavelength of the total extracted spectral range, and the spectral coverage of a detector pixel located in the centre of the order.}
\end{table}

\begin{figure}
    \centering
    \includegraphics[width=1.0\linewidth]{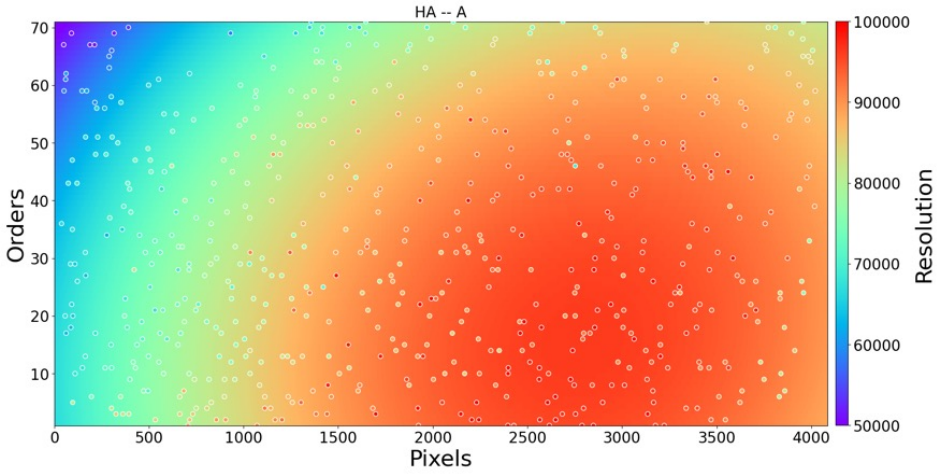}
    \includegraphics[width=1.0\linewidth]{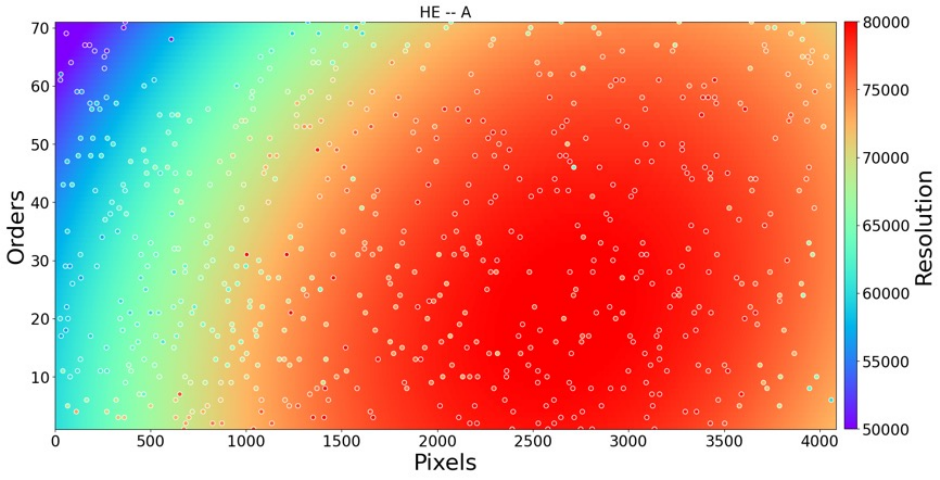}
    \caption{2D maps of NIRPS's resolving power for (top) fibre A of HA mode and (bottom) fibre A of HE mode.}
    \label{fig:resolution}
\end{figure}

The spectral resolution ($\lambda / \Delta\lambda$) of NIRPS was measured on both modes and each fibre using the uranium-neon lamp calibration frames assuming that most of the uranium emission lines are not or are only marginally resolved by the spectrograph. For all non-saturated uranium lines with sufficient signal, we measured the line width and converted it into resolving power. Table \ref{tab:res} presents for each fibre the minimum, median and maximum resolutions. Using a 2D interpolation (Equation 16 of \citealt{Allart2022}), we produced a 2D map of the resolving power as a function of the pixel and echelle order. Figure \ref{fig:resolution} represents the spectral resolution of HA\_A and HE\_A fibres. We can note how the spectral resolution varies across the detectors with a decrease towards the left of each order and towards the reddest orders. This drop on the left part of spectral orders was not expected, it cannot be reproduced by any optical models and is not explained so far.  

\begin{table}
    \centering                          
    \caption{NIRPS spectral resolution}             
    \label{tab:res}      
    \begin{tabular}{llll}  \hline\hline 
    & \multicolumn{3}{c}{$\lambda$ / $\Delta\lambda$} \\
     & Minimum & Median & Maximum \\ \hline
     HA\_A & 48\,500 & 88\,000 & 96\,300 \\
     HA\_B & 51\,100 & 88\,500 & 96\,300 \\
     HE\_A & 46\,700 & 75\,200 & 80\,300 \\
     HE\_B & 50\,100 & 88\,600 & 95\,400 \\ \hline
     \end{tabular}
\end{table}

The cross-talk from calibration channel B to science channel A was estimated with the Fabry-Pérot lamp injected into fibre B only. The contamination level was estimated to be below 10$^{-3}$ and $2\times10^{-3}$ for HA and HE mode, respectively. To assess the leaking of science channel A into calibration channel B, we examined OBJ-SKY observations of Barnard’s star (GL699). We combined about 200 frames in HE mode and 45 frames in HA mode to increase the S/N and we computed the CCF of both the science channel and the sky channel using an M4V mask. We took care to mask OH lines on channel B and to use the A-fibre wavelength solution for channel B as any leakage from fibre A onto fibre B will happen at the position of fibre A. The CCF of the leakage spectrum onto fibre B has an amplitude that is $3\times10^{-4}$ and $1.4\times10^{-4}$ times that of the science channel for HE and HA  illustrating the extremely low level of cross-talk between channel A and B, respectively.

\subsubsection{Spectral intrinsic stability}
\label{sec:stability}

\begin{figure}
    \centering
    \includegraphics[width=1.0\linewidth]{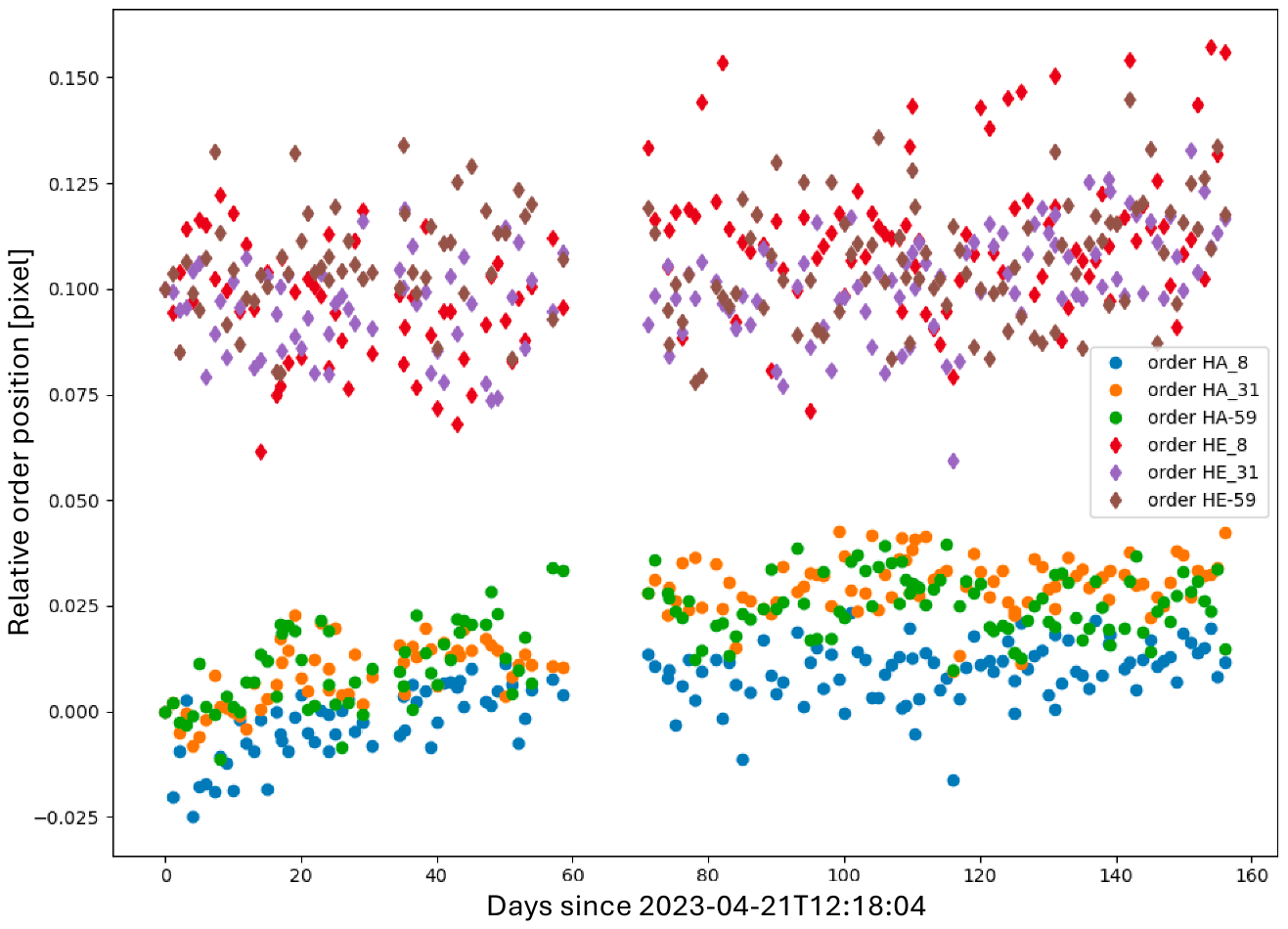}
    \caption{Relative position of the centre of orders 8, 31, and 59 of the HA and HE modes from April 2023 to Sept 2023.}
    \label{fig:loc_stab}
\end{figure}

Several indicators demonstrate the extremely high intrinsic stability of the NIRPS spectrograph over months. Figure \ref{fig:loc_stab} shows the relative locations of three spectral orders (8, 31, 59) located in the centre of $Y$, $J,$ and $H$ bands of HA and HE modes from late April 2023 (after the installation of the insulation box) to September 2023 (before a power cut). The central position of spectral orders does not move by more than 0.05\,pixels (0.75\,$\mu$m) over five months. The dispersion ranges from 10 to 13\,mpixels RMS in HA and from 12 to 18\,mpixels RMS in HE.  

\begin{figure}
    \centering
    \includegraphics[width=1.0\linewidth]{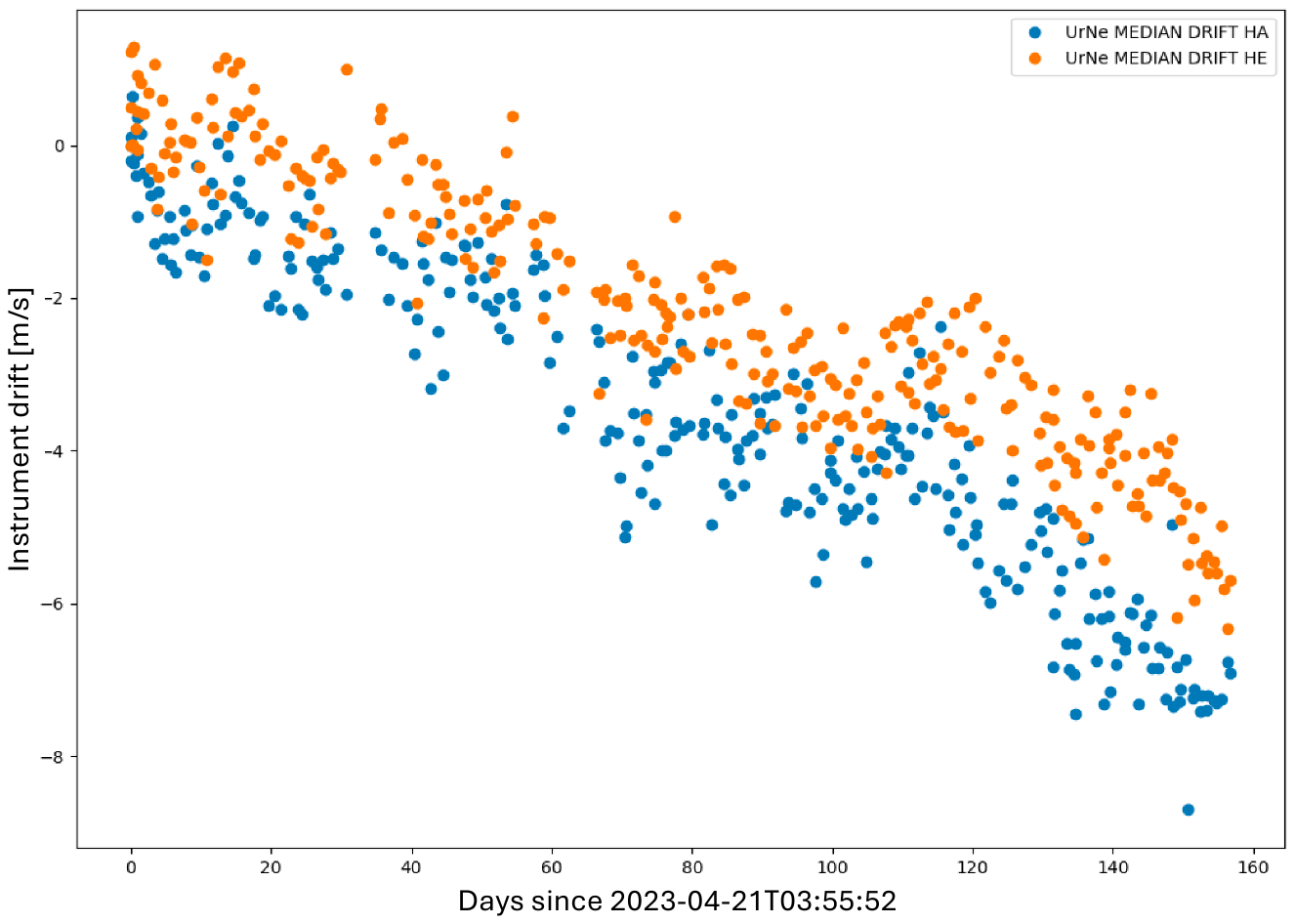}
    \caption{Intrinsic drift of the NIRPS spectrograph measured on uranium-neon lines from April 2023 to Sept 2023.}
    \label{fig:wave_stab}
\end{figure}

Figure~\ref{fig:wave_stab} shows the intrinsic drift of the instrument measured on the UrNe lines over five months. The typical drift is of 4.1\,{\cms}/day and 3.4\,{\cms}/day in HA and HE mode, respectively. The dispersion after removing the long-term linear drift is 72\,{\cms} and 64\,{\cms} for HA and HE mode, respectively. With such high intrinsic stability, the use of simultaneous Fabry-Pérot during science observation is no longer required nor justified. The recommended observing mode is OBJ-SKY which enables a monitoring and correction of OH emission lines (see Section \ref{sec:telluric}).

\begin{figure}
    \centering
    \includegraphics[width=1.0\linewidth]{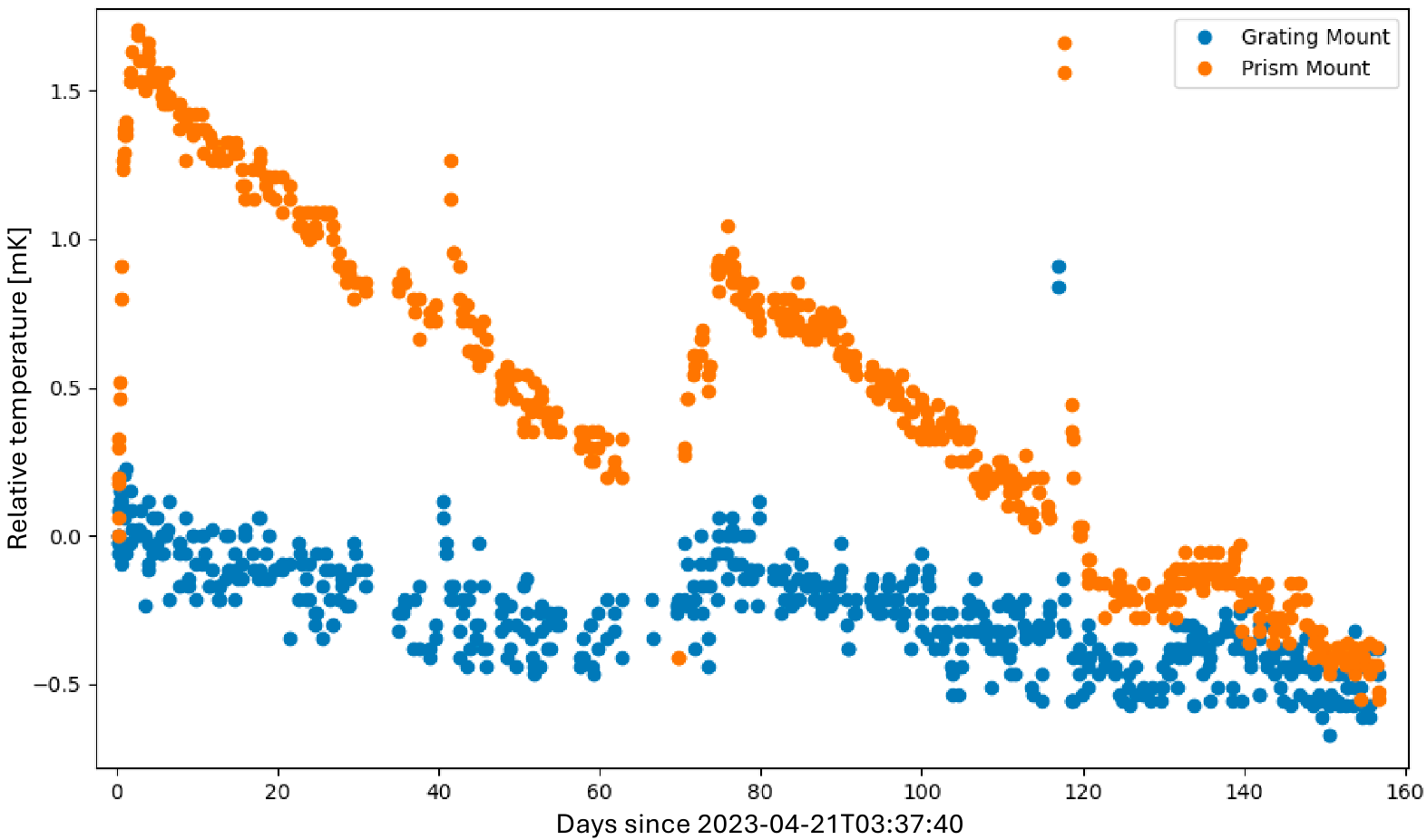}
    \caption{Relative temperature changes of the grating and prism mounts from April 2023 to Sept 2023. The mean temperature value is 75.16\,K and 75.37\,K for the grating and prism, respectively. The outliers and jumps are due to short power cuts.}
    \label{fig:temp_stab}
\end{figure}

Figure~\ref{fig:temp_stab} shows the temperature changes of the grating and prism mount inside the cryogenic spectrograph from April 2023 to Sept 2023. During this  five-month period, the NIRPS room temperature was fluctuating by 3.7$^\circ$C (11.9 - 15.6$^\circ$C). The dispersion of the Grating Mount temperature, after removing a few outliers, is 0.18\,mK RMS. The prism mount shows a slightly lower thermal stability and/or a longer timescale to reach the set point than the grating which may comes from a less efficient thermal coupling with the thermal-controlled optical bench.

\begin{figure}
    \centering
    \includegraphics[width=1.0\linewidth]{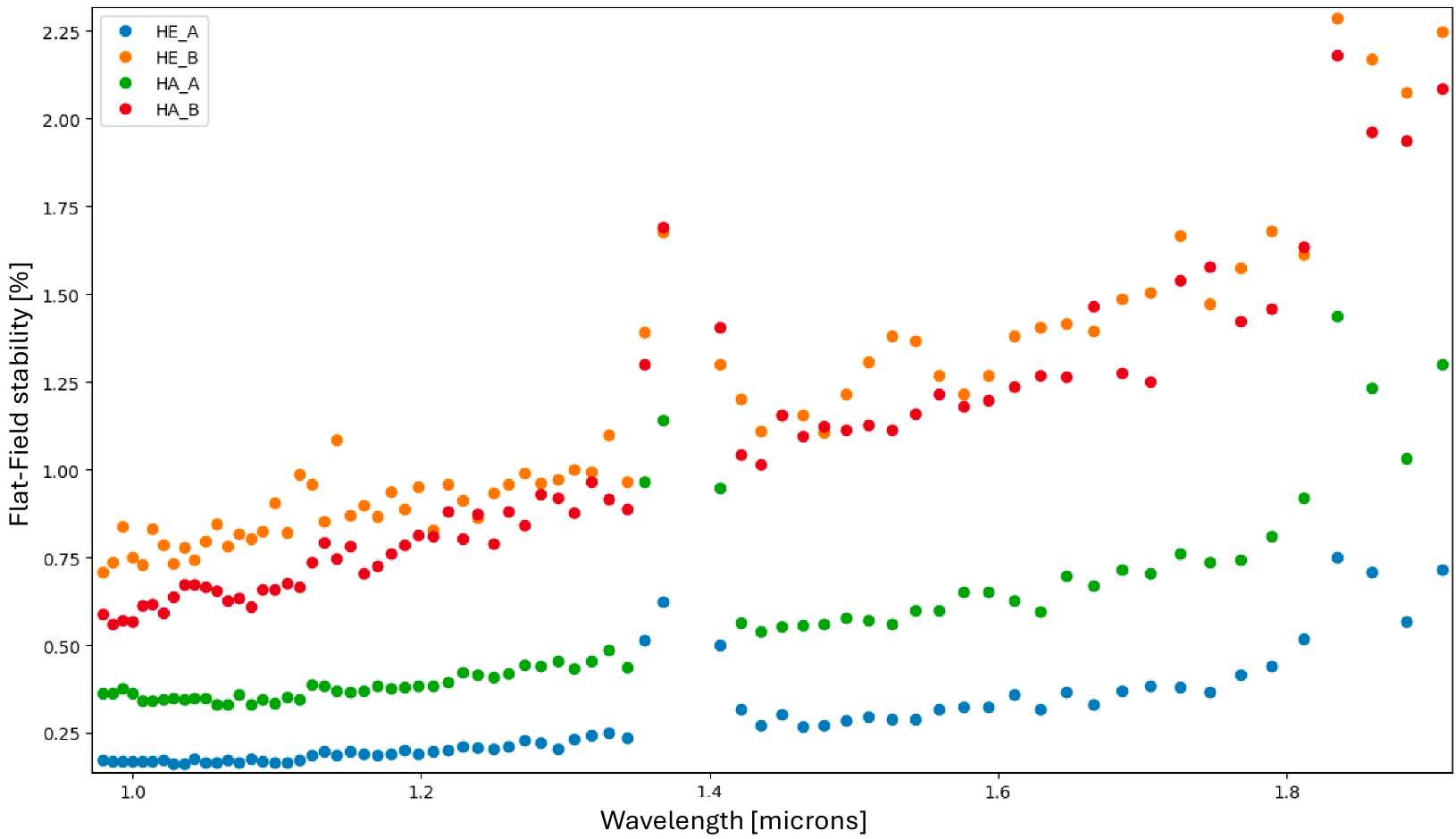}
    \caption{Relative flat-field stability over one month as a function of spectral orders for the four NIRPS fibres.}
    \label{fig:flat_stab}
\end{figure}

We compared two spectral flat fields separated by one month (from 2023 April 22 and 2023 May 21), to check the flat-field stability for the four fibres. Figure~\ref{fig:flat_stab} shows the dispersion of the relative difference between two spectral flat-fields for fibres A and B of modes HA and HE. We did not see a significant change of behaviour when comparing two flat fields separated by one day, one week and one month for fibres HE\_A and HA\_A. We see a clear increase of dispersion from the blue to red side by about a factor of 2 and it is proportional to the wavelength. This reveals modal noise, which is expected to scale with wavelength. The slight increase of dispersion around 1.38\,$\mu$m (orders 42-43-44) and above 1.81\,$\mu$m (above order 67) is due to water absorption lines visible in the spectral flat-fields coming from a few centimetres of air passing through the beam. The corresponding spectral domains on the science channel are completely saturated by telluric absorption. The best channel in terms of flat-field stability is HE\_A with dispersion from 0.17\% to 0.45\%, which is about a factor of 2 better than HA\_A. In the $H$ band, the flat-field stability is 0.33\% and 0.62\% for HE and HA mode, respectively. Both SKY channels (HA\_B and HE\_B) present significant instabilities. According to this result, we may expect the flat-field instabilities introduced by modal noise may affect our spectra for an S/N higher than 300 and 150 in the $H$ band for HE\_A and HA\_A respectively. \\    

\begin{figure}
    \centering
    \includegraphics[width=1.0\linewidth]{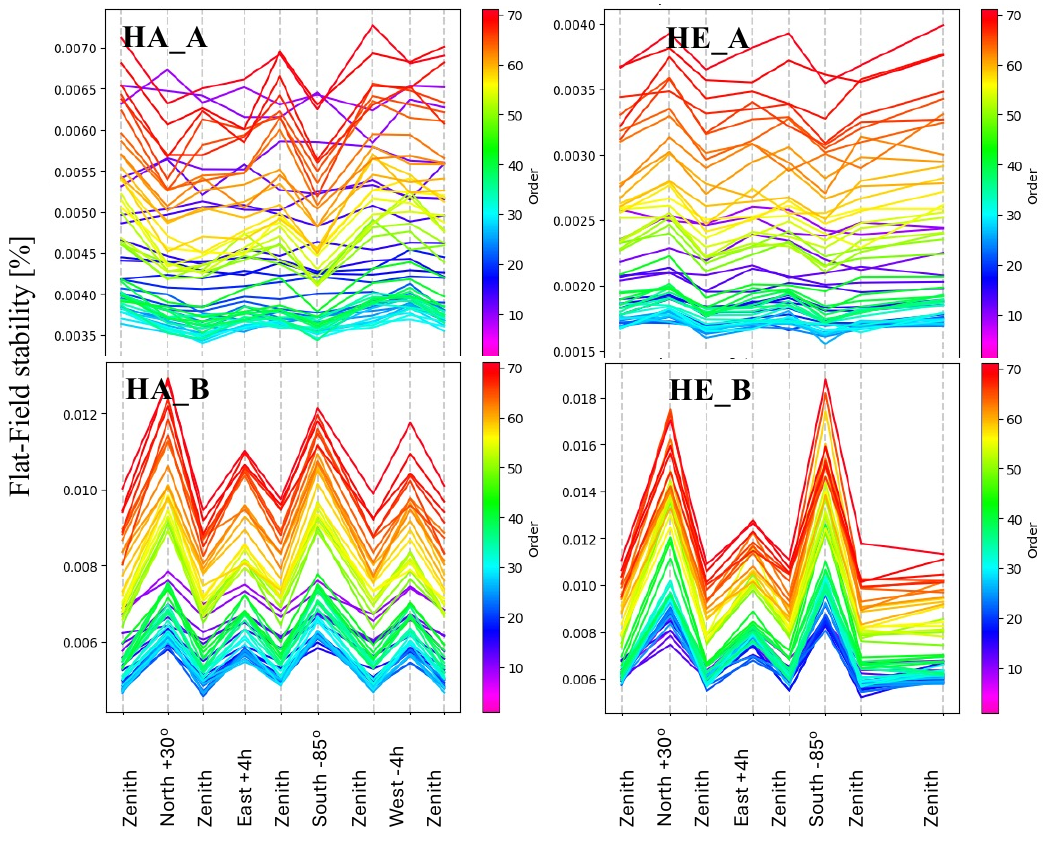}
    \caption{Relative flat-field stability as a function of the telescope position for the four NIRPS fibres.}
    \label{fig:flat_stab_tel}
\end{figure}

We checked the Flat-Field stability as a function of the telescope position by repeating Flat-Flat calibration sequences for different telescope positions (Zenith, North dec=+30, East HA=+4\,h, South Dec=-85 and West HA=-4\,h). We see in Fig.~\ref{fig:flat_stab_tel}, the impact of telescope position for both fibre HA\_B and HE\_B which means a strong sensitivity to fibres bending. We checked the impact of such instabilities of fibres B for the measurement of instrumental drift using simultaneous calibrations. We measured the apparent instrument drift from a NIRPS FP-FP sequence of several hours during HARPS standard operation, which means with telescope displacement every 15-30\,min. We see on fibre B, RV jumps of up to 50\,{\cms} correlated with the telescope position. 
This instability of fibre B due to modal noise does not permit the measurement of RV drifts to better than 50\,{\cms}. The use of simultaneous Fabry-Pérot during science observation is not justified nor recommended considering the spectrograph is more stable than 10\,cm/s over a night.   
The relatively strong modal noise observed on fibres HE\_B and HA\_B, twice worse than fibre HA\_A is not well understood. It may come from mechanical constraints of the fibre connectors, and/or slight misalignment in the fibre head. 

\begin{figure}
    \centering
    \includegraphics[width=1.0\linewidth]{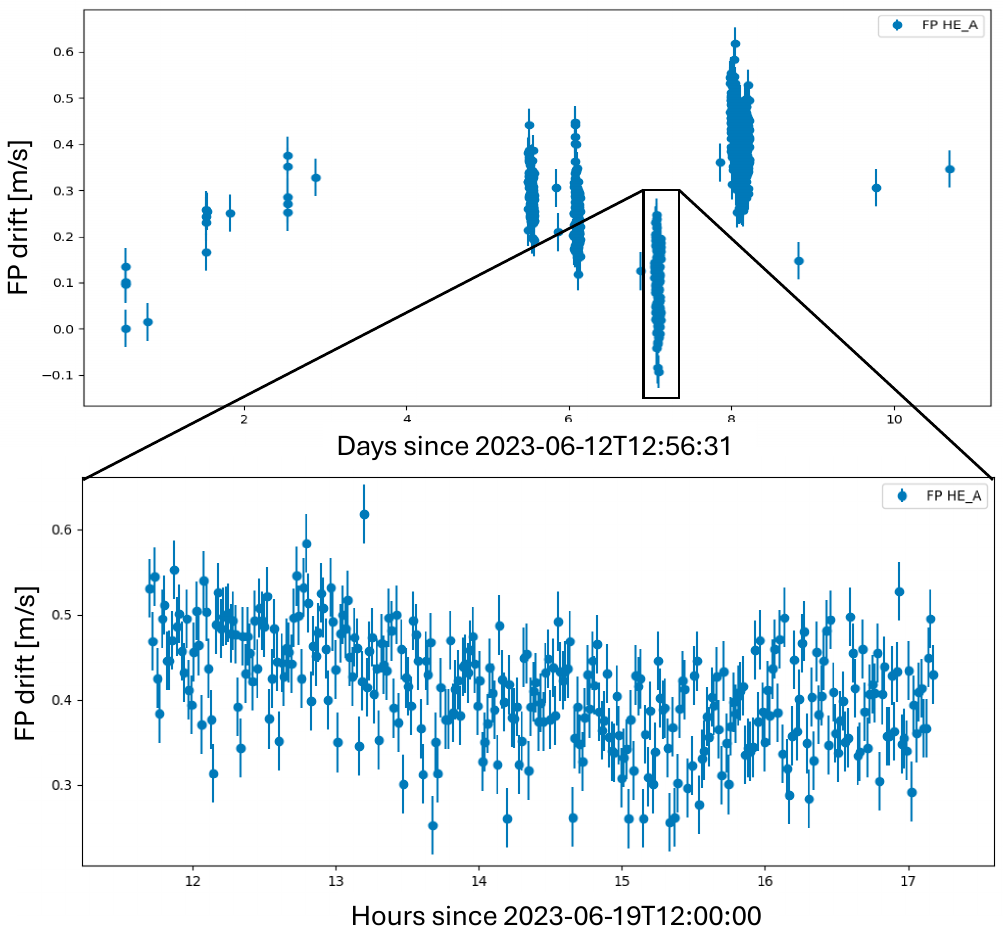}
    \caption{Top: Fabry-Pérot drift measurement overtendays on fibre HE\_A. Bottom: Zoom on the sequence done on 2023-06-19 with a dispersion of 6.5 {\cms}.}
    \label{fig:fp_stab}
\end{figure}

Figure \ref{fig:fp_stab} shows the RV drift measured on the FP-FP sequence overtendays using HE mode. The first exposure on 2023 June 12 is taken as reference and RV drifts are measured relative to this reference. On fibre HE\_A the overall measured drift presents modulation at the level of $\pm$30\,{\cms} and a global dispersion of 13.3\,{\cms}. The FP intrinsic stability cannot be guaranteed over several days. However, this sequence demonstrates that both the instrument and the FP cavity are stable well below 1\,{\ms} over a timescale of 10\,days.  During the night of 2023 June 19, a continuous sequence of 5.5 hours was done with a dispersion of only 6.5\,cm/s. The photon-noise uncertainty for this sequence is estimated to 3.5\,cm/s. Fibre HE\_B shows a global dispersion of 60\,{\cms} overtendays and the difference between fibre A and B can reach up to 1\,{\ms} due to the modal noise affecting fibre HE\_B five times more than it does fibre HE\_A. \\

The NIRPS wavelength solution is derived by first calibrating the FP cavity width with the UrNe lines (about 600) and then by using all the FP lines (about 17\,800) to calibrate each spectral order. We estimated with the NIRPS-DRS the uncertainty of the FP cavity width calibration to be 48\,{\cms} and 51\,{\cms} in HE and HA mode, respectively. We measured the dispersion of the FP cavity width measurements based on a sequence over several days with no significant drift of the instrument to be 55\,{\cms} and 69\,{\cms} in HE and HA mode, respectively. These numbers represent the current limitation of the NIRPS wavelength solution, or RV zero point, due to the relatively limited number of UrNe lines (about 600) available for the FP cavity calibration.

\subsubsection{Throughput efficiency and S/N}

\begin{figure}
    \centering
    \includegraphics[width=1.0\linewidth]{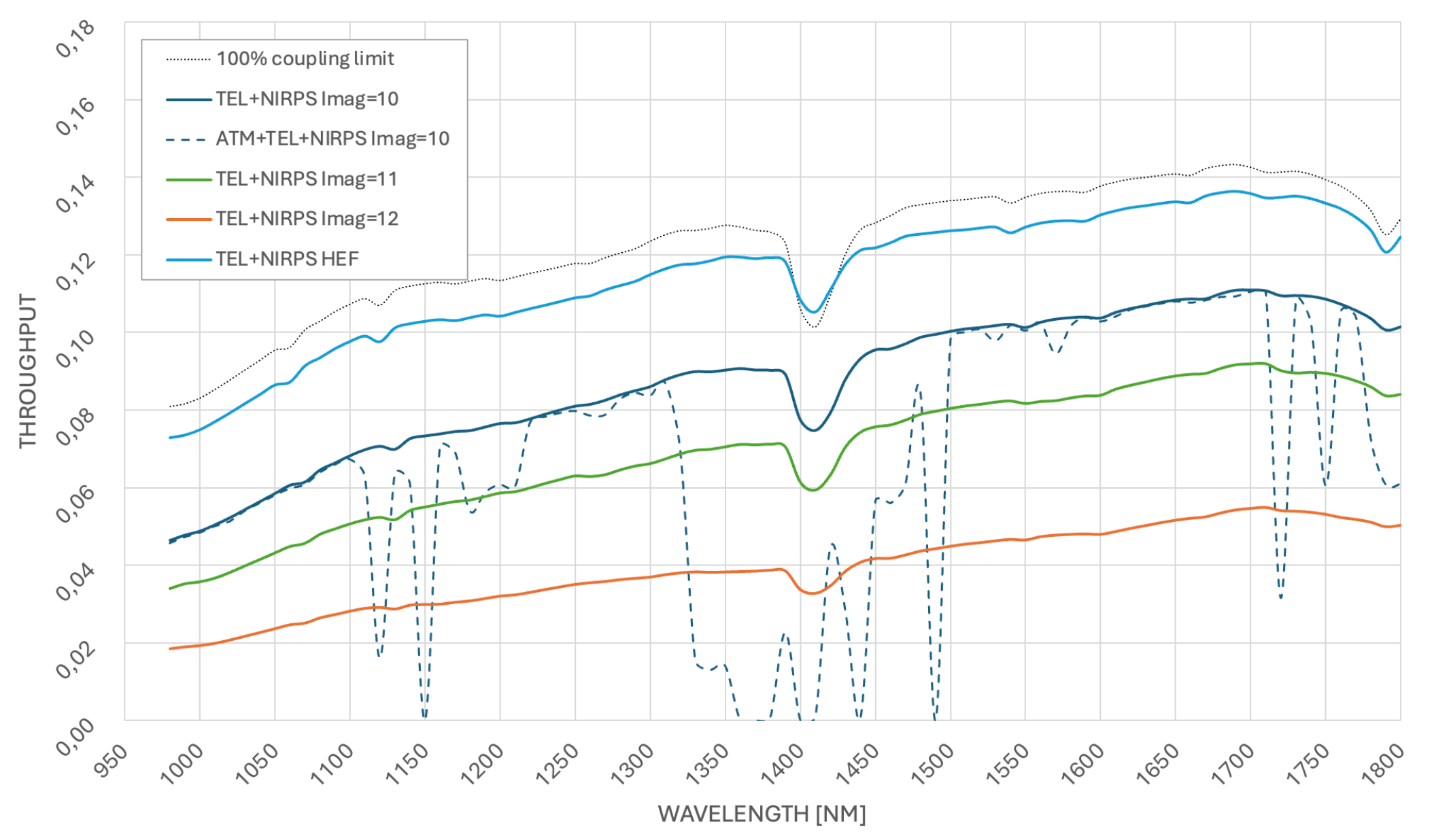}
     \caption{Expected overall throughput of NIRPS instruments for HA mode and different $I$ magnitude for a 0.9\arcsec\ seeing. The dark blue, green, and orange curves correspond to the overall throughput for $I$ of 10, 11 and 12, respectively. The dashed curve represents the atmospheric absorption bands on top of the overall throughput computed for $I=10$. The dotted curve shows the 100\% coupling into the HA fibre. The blue curve corresponds to the overall predicted throughput for the HE mode for targets brighter than $I$=9.}
    \label{fig:Throughput}
\end{figure}

Figure~\ref{fig:Throughput} shows the calculated and predicted overall throughput of NIRPS over its spectral domain at zenith and for a seeing of 0.9\arcsec. These curves were made assuming the atmospheric transmittance from \texttt{TAPAS} model \citep{Bertaux2014}, the transmission of the 3.6-m telescope after re-coating, the throughput of the front end
, fibre train, and spectrograph, the sensitivity of the H4RG detector and the coupling efficiency of the AO system for different $I$ magnitudes for HA and HE modes.

\begin{figure}
    \centering
    \includegraphics[width=1.0\linewidth]{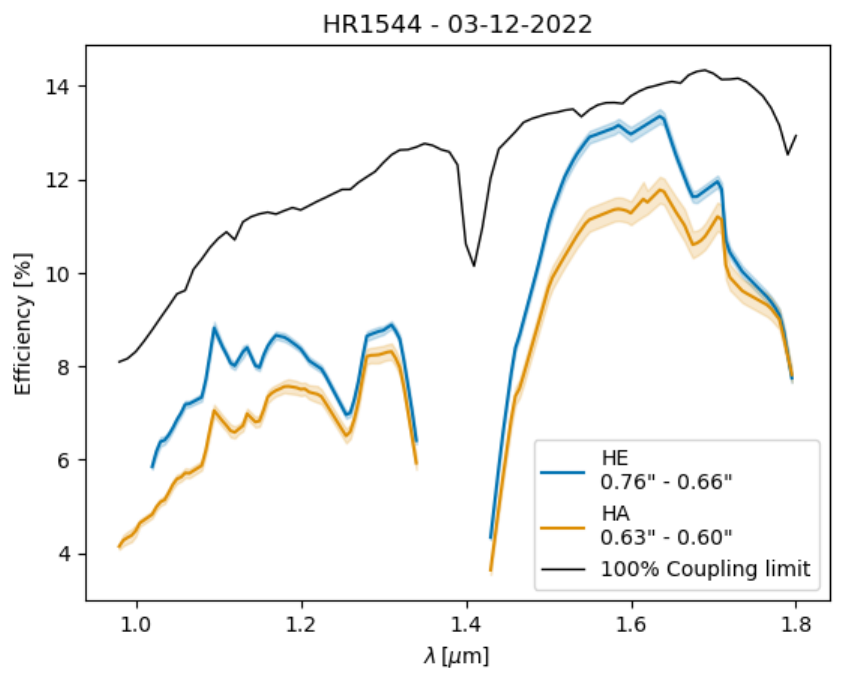}
     \caption{Overall efficiency curves of NIRPS measured on the spectrophotometric standard star HR1544 observed on 3 December 2022 under extremely good seeing conditions. The solid line shows the upper envelope of the resulting efficiency curves, and the shaded area indicates the standard deviation. The values in the legend show the seeing at the beginning of the first observation and the end of the last observation. The black curve represents the ideal case of 100\% coupling limit for seeing of 0.9\arcsec\ in the $YJH$-band.}
    \label{fig:efficiency}
\end{figure}

Figure~\ref{fig:efficiency}
shows the overall efficiency for both the HE and HA modes as a function of wavelength as measured by the NIRPS-DRS on the standard spectrophotometric star HR1544 during commissioning \#7. For both curves, 12 observations of HR1544 are combined and then divided by a CRIRES optical model for the same star. CRIRES models, ranging from 950 to 5300\,nm, are based on M. Hamuy/CTIO secondary spectrophotometric standards \citep{Taylor1984}, and are part of the static calibration frames, accessible via the ESO website\footnote{\href{https://www.eso.org/sci/facilities/paranal/instruments/crires/tools.html}{eso.org/sci/facilities/paranal/instruments/crires/tools.html}}. The efficiency reaches in the $H$ band $\sim$13\% and $\sim$11.5\% in HE and HA, respectively.\\

\begin{figure}
    \centering
    \includegraphics[width=1.0\linewidth]{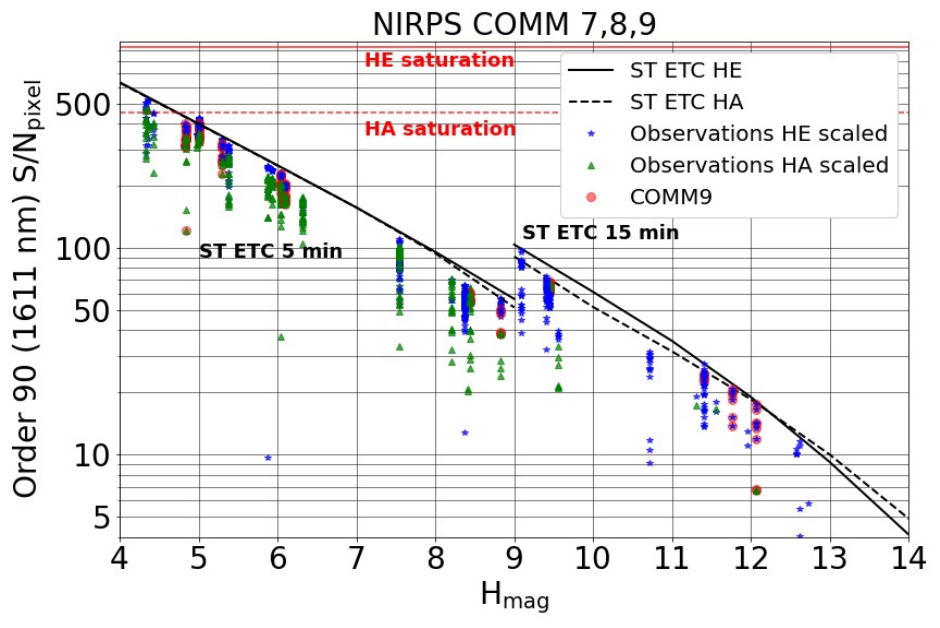}
     \caption{S/N measured in the $H$ band as a function of target magnitude observed during commissioning phases \#7, \#8 and \#9 for HE mode (blue) and HA mode (green). S/N values are scaled to 5\,min for bright ($H<9$) and 15\,min for fainter targets. The full line and dotted line represent the expected S/N from the Exposure Time Calculator for HE and HA mode, respectively. The upper envelope of the measured S/N, corresponding to optimum astroclimatic conditions, fits with the expected value.}
    \label{fig:SNR_ETC}
\end{figure}

Figure~\ref{fig:SNR_ETC} shows the S/N results obtained at 1611\,nm ($H$ band) in an extracted-pixel bin of 5.4\,nm (1.00\,{\kms}) as a function of target magnitude after renormalisation to an exposure time of five min for bright stars ($H < 9$) and 15\,min for fainter stars. S/N was measured during the commissioning phases \#7, \#8 and \#9 in both HA and HE modes, showing an excellent agreement with expectation. The upper envelope of our measurements, corresponding to optimum astroclimatic conditions, fit well with the expectation and the official ETC\footnote{\href{https://etc.eso.org/observing/etc/}{etc.eso.org/observing/etc/}}. In the best cases, we measured S/N=130 for $H=6$ in 5\,min, S/N=100 for $H=9$ in 15 min and S/N=20 for $H=12$ in 15\,min. Starting at $H\sim12$, we see the transition between photon-noise and detector readout noise limit. While the S/N obtained in HA and HE modes are relatively similar for bright targets ($H < 9$), HE mode shows better performance for fainter targets, illustrating the decrease of AO performance for faint targets. We note that the median distribution of S/N corresponds to $\sim$80-85\% of the upper envelope due to various factors such as airmass, seeing, absorption, AO instabilities, and so on. \\

\subsubsection{Modal noise limitation}
\label{sec:modalnoise}

\begin{figure}
    \centering
    \includegraphics[width=1.0\linewidth]{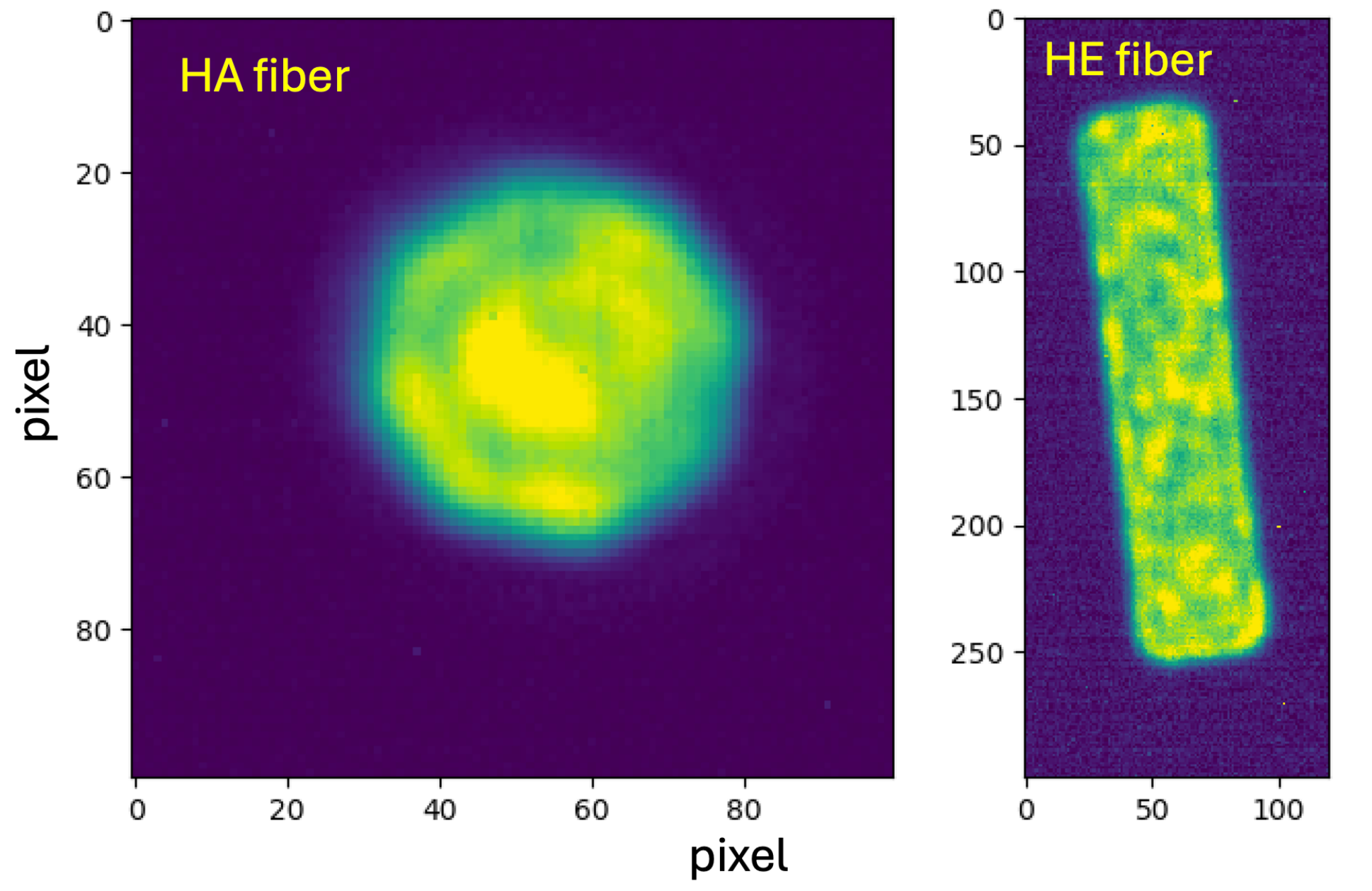}
     \caption{Near-field images of the 29-$\mu$m octagonal HA fibre (left) and the $33\times132$\,$\mu$m rectangular HE fibre (right) illuminated with a 1.55\,$\mu$m laser and the diffraction limited PSF of the front end. No fibre stretching nor AO scanning are applied. The pixel size on these images corresponds to 0.67\,$\mu$m.}
    \label{fig:fibre}
\end{figure}

During the commissioning \#4, we had the unique opportunity to inspect the output of the fibre link coupled to the front end
 before integrating it into the spectrograph. As described in \cite{Frensch2022}, we re-imaged the fibre head on a Xenics-Xeva camera. Figure~\ref{fig:fibre} shows the images of the near-field of the output section of the 29-$\mu$m octagonal HA fibre and the $33\times132$\,$\mu$m rectangular HE fibre illuminated with a 1.55\,$\mu$m laser at the diffraction-limited PSF of the front end
.  We see a speckle-like pattern due to the limited number of propagating modes into the fibres as discussed by \cite{Blind2022b}. Any displacements of the injection position of the spot at the fibre entrance and/or any change in the mechanical stresses on the fibre introduce changes in the centroid of the near-field fibre output corresponding to several 10\,{\ms}.  

\begin{figure}
    \centering
    \includegraphics[width=1.0\linewidth]{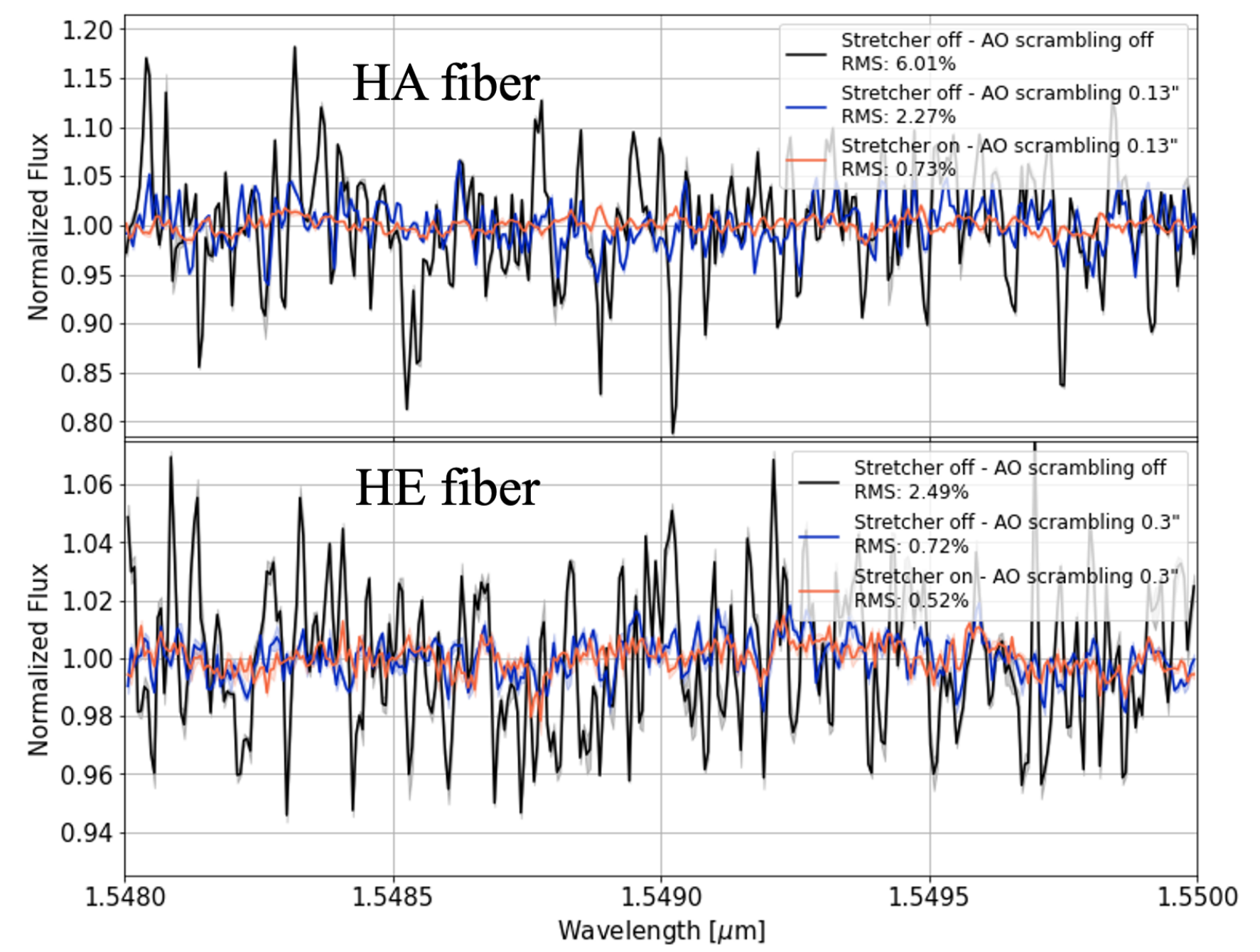}
     \caption{Modal noise structure seen on the continuum of featureless stellar spectra of HD18979 on HA mode (top) and HE mode (bottom) with different configurations of the fibre stretcher and the AO tip-tilt scanning mode.}
    \label{fig:modal_noise}
\end{figure}

To minimise the impact of the modal noise and to have the most uniform and stable illumination at the fibre output, we developed two approaches. First, a 20\,m long section of the fibre is physically stretched with piezoelectric actuators with a total amplitude of 6 to 8\,mm at a frequency of 0.3\,Hz: the modulation of the path length between modes strongly reduces the contrast of the exit interference pattern. Second, we use the AO tip-tilt to scan the fibre entrance and fill up uniformly as many fibre modes as possible with limited losses. The amplitude of the AO scanning mode was set to 0.1\arcsec\ for HA mode and 0.2\arcsec\ for HE mode (corresponding to about one-quarter of the fibre diameter) to find a compromise between the gain of scrambling and the loss of flux. Different scanning patterns were tested without showing significant differences. To quantify the impact of both the stretcher and the tip-tilt scanning mode, we measured during commissioning \#7 the continuum of the featureless stellar spectra of HD\,18979, a fast-rotating A3 star. Figure~\ref{fig:modal_noise} shows a small spectral region around 1.55 $\mu$m in both HA and HE modes, with and without the fibre stretcher and with and without the AO tip-tilt scanning. Without stretcher and tip-tilt scanning, the modal noise measured as the relative dispersion in the continuum can reach up to 6\% in HA and 2.5\% in HE. Both the fibre stretchers and the AO scanning introduce a clear and significant improvement. The measured dispersion in the $H$ band with both the fibre stretcher and the AO scanning mode is 0.52\% and 0.73\% for HE mode and HA mode, respectively. The photon noise is at the level of 0.29\% and 0.33\% of the continuum for HE and HA mode respectively. This means that the additional noise from the fibre mode propagation is at the level of 0.43\% and 0.65\% for HE and HA mode, respectively. These numbers are comparable to or slightly higher than the flat-field stability discussed in Sect.~\ref{sec:stability} (0.33\% for HE mode and 0.62\% for HA mode). 

As shown in Sect.~\ref{sec:RV}, the NIRPS RV precision is slightly better in HE mode. We suspect this comes from the modal noise residuals of the small HA fibre not being completely averaged by the fibre stretcher and the AO scanning mode. The modal noise residuals in the HA mode are at the level of 0.65\%, which means an apparent S/N is limited to 150. 
The modal noise residuals in the HE mode are at the level of 0.43\%, which means an apparent S/N is limited to 230. 
S/N of 150 in HA and 230 in HE can be reached under nominal conditions with an exposure time of 30 min on an M dwarf of $H=8.8$ and 8.0, respectively. According to RV photon-noise measured in Sect.~\ref{sec:RV}, these S/N limitations due to modal noise should introduce RV limitation at the level of 1.4\,{\ms} and 0.9\,{\ms} for HA and HE mode, respectively. 

The high sensitivity of the observed spectral structures with the fibre injection condition (AO scanning) and the fibre stretcher both indicate that we are in the presence of modal noise. Furthermore, the difference between the two modes and between the four fibres (as shown in Section~\ref{sec:stability}) clearly indicates that the structured noise pattern is coming from the fibres and not the spectrograph. The clear dependence on wavelength  also provides evidence that scrambling in fibres of only a few modes is not complete, thereby  limiting the performance in the $H$ band. 

We expect that the modal noise pattern is evolving slowly in time and is quite stable when the telescope does not move or is just in tracking mode or back in the same position. Therefore, the impact on differential measurements such as spectroscopic transits and dayside observations is expected to be negligible. 

\subsection{Telluric and OH line corrections}
\label{sec:telluric}

The telluric correction was performed with the method described in \cite{Allart2022}, available on GitHub\footnote{\href{https://github.com/RomainAllart/Telluric\_correction}{github.com/RomainAllart/Telluric\_correction}} and now part of the official NIRPS-DRS pipeline. The molecules included here are H$_2$O, O$_2$, CO$_2$, and CH$_4$ and the resolution kernels used are the ones described in Section \ref{sec:resolution} and shown in Fig.~\ref{fig:resolution} for both HA and HE. The correction reaches similar precision to the first telluric correction step done by the APERO pipeline (see Sect.~\ref{sec:apero}) with residuals below 1\% peak to valley across the $Y$, $J$ and $H$ bands for the four main molecules. An example is shown for Proxima observed during the commissioning phase in Fig.~\ref{fig:proxima_s1d}.

For the sky emission lines correction, mainly due to strong OH lines, we first built a master template combining all our night SKY-SKY observations to identify each OH emission line at high S/N. For each science observation made in OBJ-SKY mode, the emission lines flux on fibre B is compared to the flux of the fibre B SKY-SKY master template. The fibre A of the SKY-SKY master template is then rescaled to the previously obtained flux ratio and subtracted locally from the science channel. This approach avoids being affected by the difference in spectral resolution between fibre A and B in the HE mode, as well as any differences in the efficiency of the two fibres. It assumes that the flux ratio between fibres does not evolve with time.

\subsection{Empirical RV content of M dwarfs}
\label{sec:RV}

The fundamental RV precision achievable under photon-noise-limited conditions can be estimated using the relations detailed in \cite{Bouchy2001}. This RV precision not only depends on photon count and measurement S/N but also on the depth and profiles of spectral lines used in RV measurements. Therefore, RV precision is influenced by the spectrograph’s resolving power, the stellar properties (temperature, surface gravity, and metallicity; and defining line number and contrast), and its rotational velocity (impacting line broadening).

\begin{figure*}
    \centering
    \includegraphics[width=1.0\linewidth]{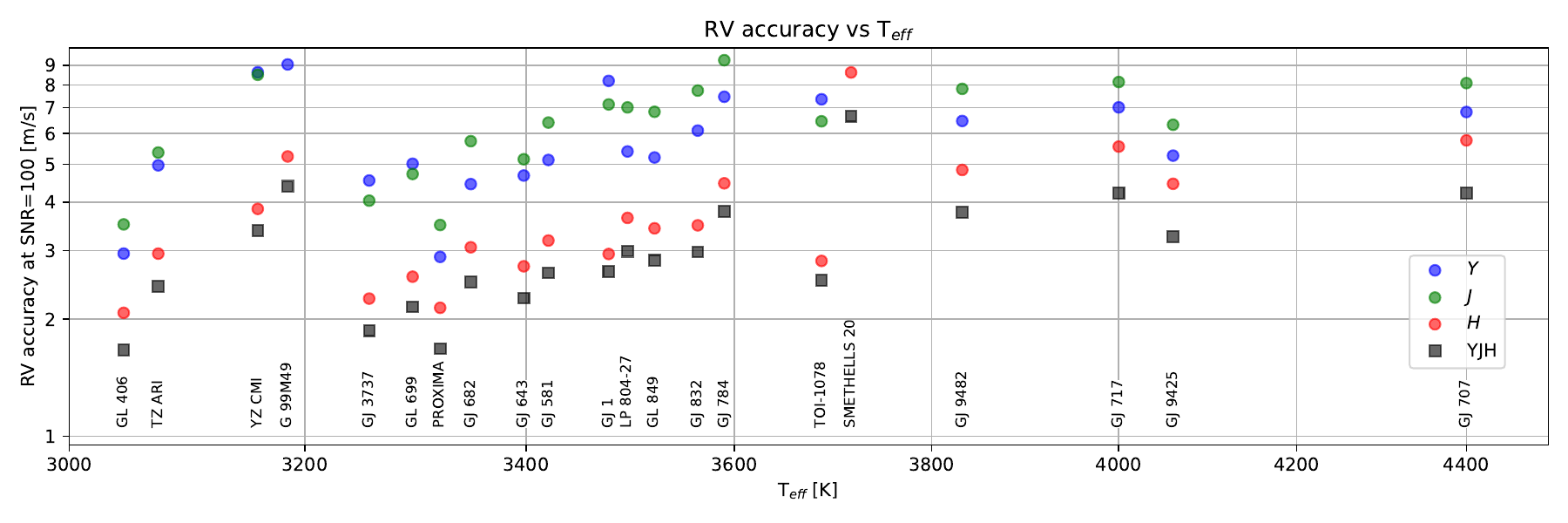}
     \caption{RV accuracy as function of effective temperature achievable with an S/N of 100 in $Y$, $J$ and $H$ bands for M dwarfs having high-S/N empirical spectral templates from NIRPS HE observations. The RV accuracy obtained by combining all three bands is expressed for an S/N=100 in $H$, the S/N in the other bands being typically slightly lower for M dwarfs.}
    \label{fig:rv_accuracy}
\end{figure*}

The RV photon limits achievable at a given S/N have been debated, with model-based estimates \citep{Figueira2016, Reiners2020} and hybrid methods combining limited domain observations with models \citep{Artigau2018_rvcontent}. Figure~\ref{fig:rv_accuracy} illustrates the attainable RV photon-noise uncertainties in $Y$, $J$, and $H$ bands (each at an S/N of 100) and when combining all three bands (S/N = 100 in $H$). The RV content determination here is empirical, using templates built from on-sky data. We selected all M and late K dwarfs observed during commissioning phases as well as during the first year of operation with over 50 visits, observed through a line-of-sight velocity excursion $\Delta_{BERV}$>50 {\kms}. The resulting photon-noise RV accuracy aligns closely with LBL-determined RVs. One concern in predicting RV accuracy in the  NIR was defining the usable domain relative to telluric absorption. Masking all regions where the barycentric motion intersects with a telluric line (up to $\pm$30 km/s based on the target’s ecliptic latitude) could lead to prohibitive domain loss if thresholds are set conservatively (e.g. excluding lines deeper than 2\%). Within the LBL framework, we include all available signals within photometric bandpasses and moderately from the domain between nominal bandpasses (see Fig.~3 in \citealt{Artigau2022}), corresponding to `condition 1' (non-polluted spectra) in \cite{Figueira2016}. High-S/N templates reveal RV photon-noise limit difference in Fig.~\ref{fig:rv_accuracy} due to different stellar properties. For instance, fast rotators, such as YZ CMi ({\vsini}$\sim$7\,{\kms}), show worse photon noise floors ($\sim$3.5 m/s) compared to slower rotators such as GL 406 ($\sim$1.7\,{\ms}) at similar temperatures. Stellar metallicities may also contribute to difference. Gl\,699 (Barnard's star, metal-poor) shows a slightly worse RV photon-noise limit than Proxima, reflecting the higher RV content at elevated metallicity as noted in \cite{Artigau2018_rvcontent}. 

Overall, at S/N=100 in $H$, the floor stands at $\sim$1.7\,{\ms} for solar metallicity, a significant improvement over predictions. This measured RV photon-noise floor can be compared with previous studies. The NIRPS HE mode resolution of $\sim$75\,000 corresponds in \cite{Artigau2018_rvcontent} to the \textrm{YJH80} with absorption $<10\%$ case (a conservative scenario, given usable deeper tellurics), with a predicted RV photon noise of 3.1\,{\ms} when scaling to an S/N=100 in $H$. The predictions made by \cite{Figueira2016} align closer, estimating $YJH$ RV photon limit at 2.8\,{\ms} when scaling to an S/N=100 in $H$. For an M4 dwarf at 3200\,K, \cite{Reiners2020} (Table~3) suggests photon-noise limits of 4.2\,{\ms} combining $Y$, $J$, and $H$ bands when scaling to a S/N=100 in $H$. All these estimates   
under-predict NIRPS performance by a factor between 1.7 and 2.5. 

It is remarkable that model-based RV content predictions for M dwarfs significantly underestimated on-sky performance. The RV content of M dwarfs, as empirically determined from NIRPS templates, enables RV photon-noise limit of 1\,{\ms} for slow rotating M4 dwarfs for an S/N $\sim$170 in $H$ band. This limit increases to $\sim$1.\,{\ms} and $\sim$2.4\,{\ms} for early M and late K, respectively.   

\begin{figure}[!htbp]
    \centering
    \includegraphics[width=1.0\linewidth]{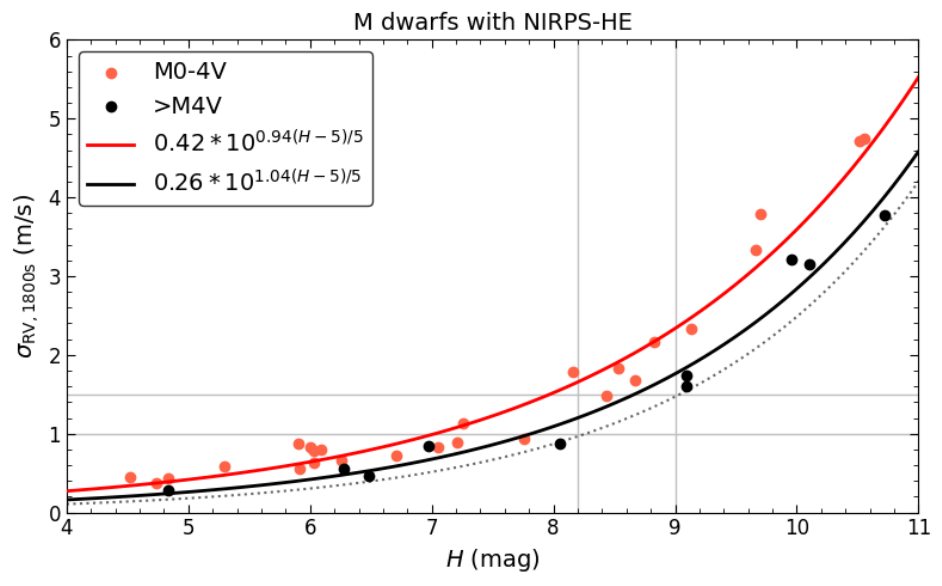}
     \caption{Estimated RV photon-noise rescaled to a 30\,min exposure for M dwarfs observed during the commissioning and the first year of operation as a function of $H$ magnitude. Black and red lines correspond to empirical relationships for the RV uncertainty obtained with NIRPS-HE for late (>M4) and early M dwarfs, respectively.}
    \label{fig:photon_noise}
\end{figure}

Figure~\ref{fig:photon_noise} shows the RV photon-noise uncertainties as a function of $H$ magnitude measured on M dwarfs and rescaled to a 30\,min exposure. The lower envelope of the RV photon-noise shows that in the best cases (non-rotating late M dwarfs), we can reach a photon noise of 1\,{\ms} in 30\,min for a $H=8.2$ target, 1.5\,{\ms} in 30\,min on a $H=9.0$ target. This empirical calibration was determined from stars observed under median observing conditions during commissioning but also during the first year of operation. 

\subsection{RV performance}
\label{radvelpef}

During commissioning \#8 and \#9, RV standards with known planetary systems such as Proxima (M5.5V) and GJ 581 (M3V) were observed to demonstrate the NIRPS RV performance under nominal operation over {\modif several} weeks. 
{\modif Proxima is known to display a RV scatter of about 2 {\ms} over more thanfiveyears \citep{Anglada2012}.}

\begin{figure}[!htbp]
    \centering
    \includegraphics[width=1.0\linewidth]{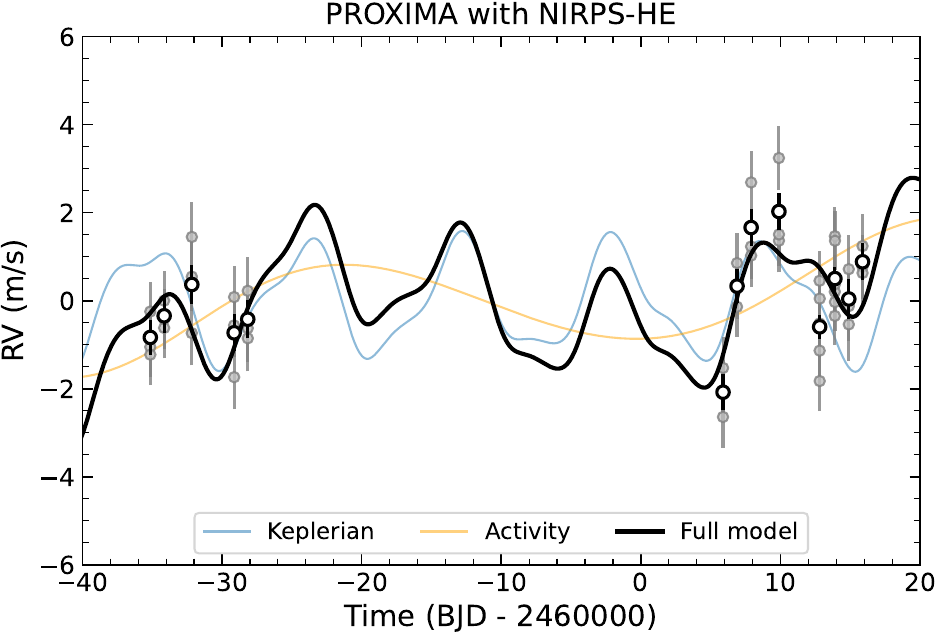}
     \caption{Proxima sequence with NIRPS (HE) over two commissioning phases (\#8 and \#9) separated by about 40\,days. The Keplerian components (Proxima b and d) and activity GP model are taken from \cite{Faria2022} derived from ESPRESSO/VLT measurements. White dots correspond to the night-binning RVs.}
    \label{fig:proxima_rv_full}
\end{figure}

\begin{figure}[!htbp]
    \centering
    \includegraphics[width=1.0\linewidth]{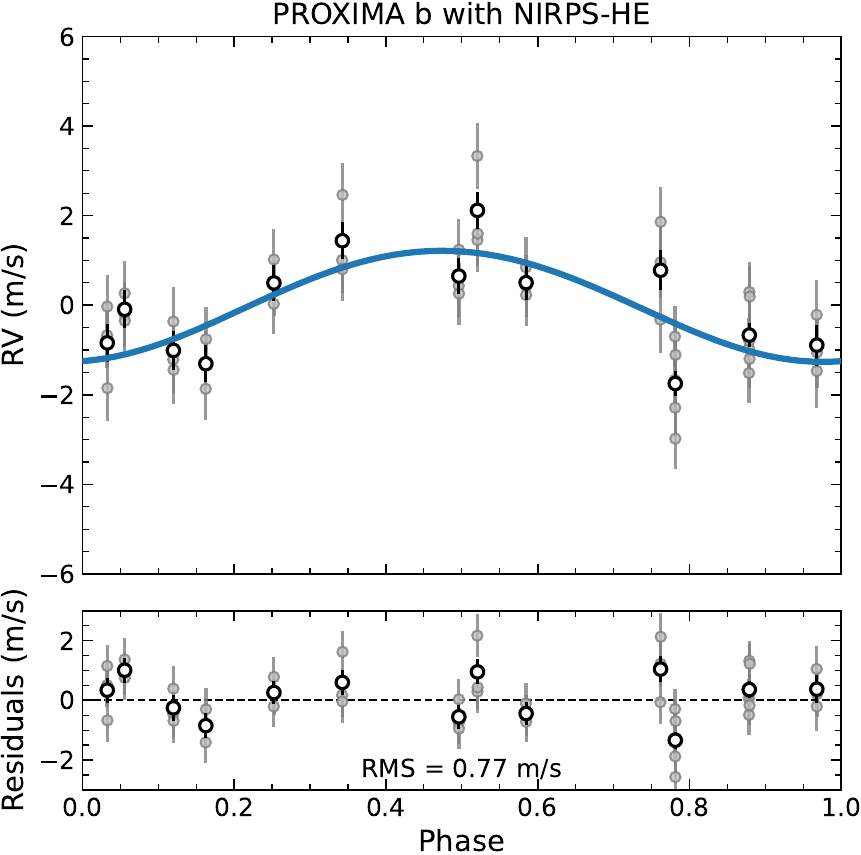}
     \caption{Proxima b Keplerian motion observed with NIRPS during commissioning with stellar activity and Proxima d signals removed. The residuals of the ESPRESSO model (not a fit) are below 1 {\ms}. The exposure time is $3\times200$\,s. The estimated photon-noise is 0.8\,{\ms}. White dots correspond to the night-binning RVs.}
    \label{fig:proxima_rv}
\end{figure}

A high-cadence sequence of Proxima of three to six exposures per night is presented in Fig.~\ref{fig:proxima_rv_full}. The RVs from NIRPS follow closely the Keplerian solutions and activity GP models of \cite{Faria2022} derived from ESPRESSO measurements. As shown in Fig.~\ref{fig:proxima_rv}, the Keplerian signature of Proxima\,b ($K = 1.24$\,{\ms}) is well detected with residuals of the ESPRESSO model (not a fit except the RV offset) below the 1\,{\ms} consistent with the estimated photon-noise of 0.8\,{\ms}. The full RV time series of Proxima, {\modif illustrating NIRPS RV precision over two years,} is presented by \citet{Suarez2025}. 

\begin{figure}[!htbp]
    \centering
    \includegraphics[width=1.0\linewidth]{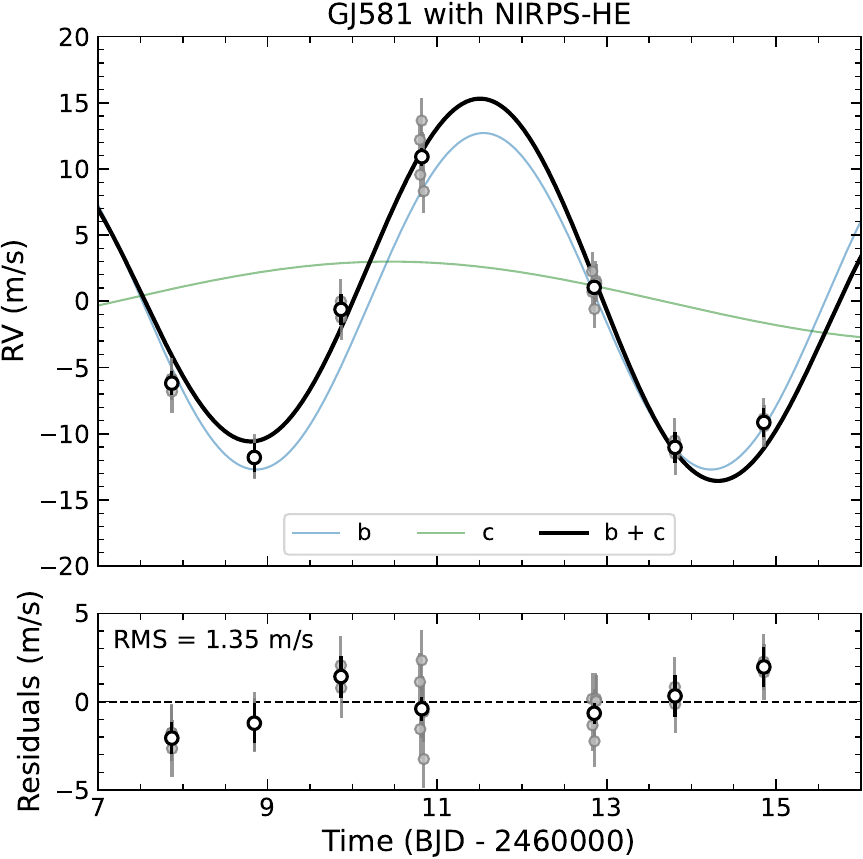}
     \caption{RV  sequence of GJ\,581 with the orbits of GJ\,581\,b ($P = 5.37$\,days) and c ($P = 12.91$\,days) is shown in the same plot using the solutions of \cite{Rosenthal2021}. Our measurements are consistent with the known planetary solutions with a residual RMS of 1.35 {\ms}. The exposure time is $2\times300$\,s. The estimated photon-noise is 1.5\,{\ms}. White dots correspond the night-binning RVs.}
    \label{fig:gj581_rv}
\end{figure}

GJ\,581 is an M3V dwarf known to harbor a 3-planet system \citep{Bonfils2005b,Mayor2009}. The raw RMS of 10.12\,{\ms} is significantly larger than the RV errors of 1.50\,{\ms}. This variation is in phase with the known planet GJ 581\,b ($P = 5.37$\,days) that we detect, even with onlysevenindependent epochs covering one orbit (Fig.~\ref{fig:gj581_rv}).

\begin{figure}[!htbp]
    \centering
    \includegraphics[width=1.0\linewidth]{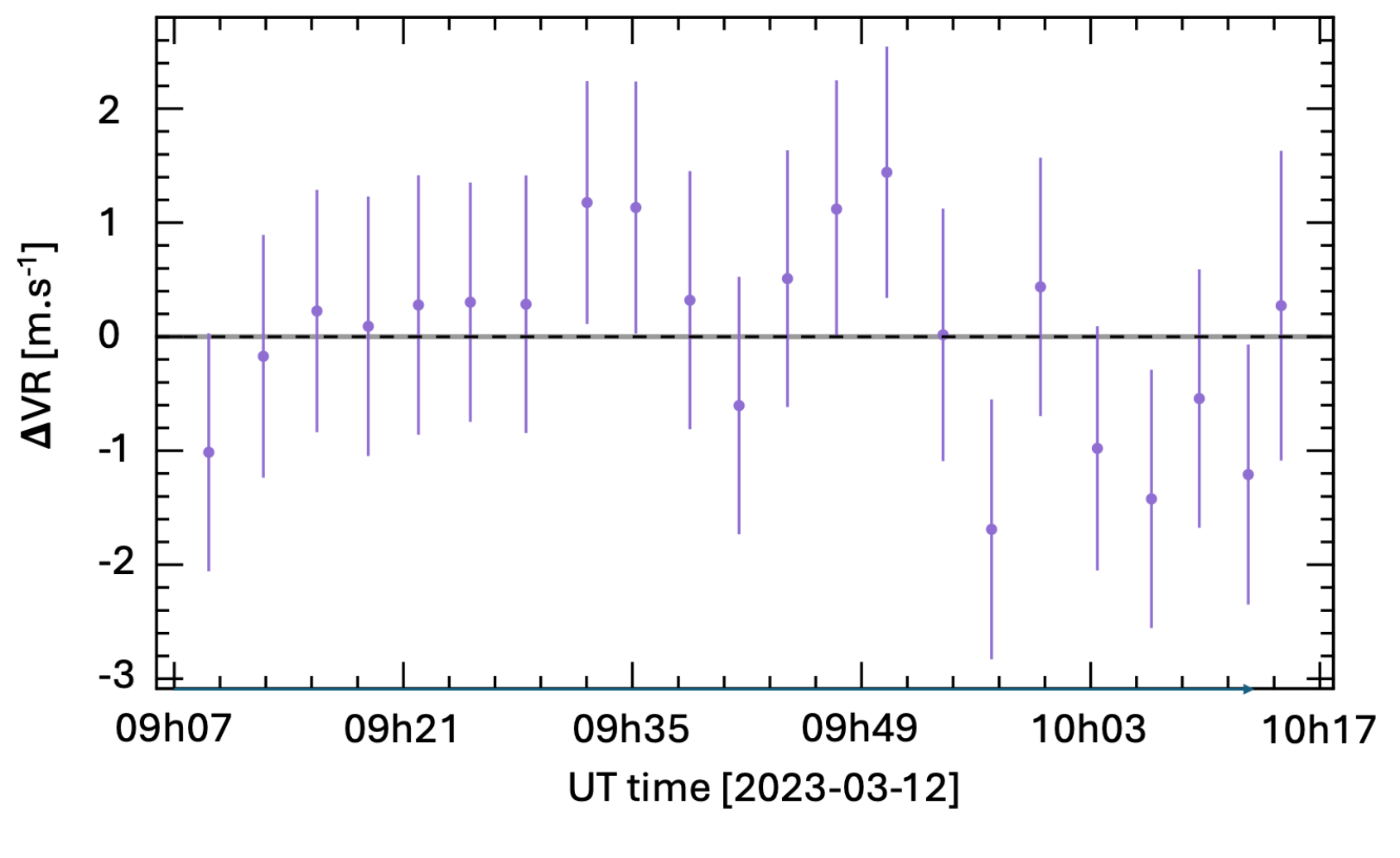}
     \caption{RV sequence of $\alpha$ Cen B obtained on 12 March 2023 on lasting 68 min. The dispersion is 0.85 {\ms} with a binning of three minutes.}
    \label{fig:acenb_rv}
\end{figure}

A short sequence of about one hour was done on the K1V star $\alpha$ Cen B with the minimum exposure time (5.57\,sec). A binning of 3-min was applied to reduce the impact of p-modes oscillation \citep{Carrier2003}. The dispersion of this short sequence shown in Fig.\ref{fig:acenb_rv} is 0.85\,{\ms}.

\begin{figure}
    \centering
    \includegraphics[width=1.0\linewidth]{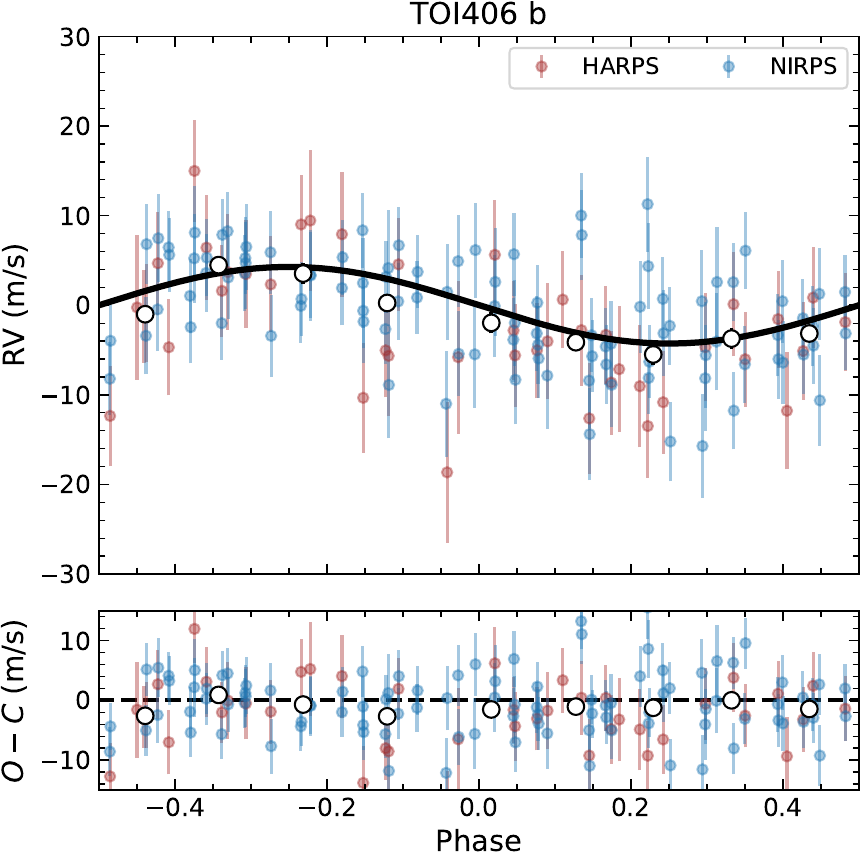}
     \caption{RVs obtained for TOI-406 phase-folded to the 13.176\,days of the sub-Neptunes TOI-406\,b. The dispersion of the residuals is 3.4\,{\ms} and 5.1\,{\ms} for NIRPS and HARPS, respectively. }
    \label{fig:toi406}
\end{figure}

The {\modif relatively quiet} M3V star TOI-406 ($V=13.8$ and $J=9.7$) was observed as part of our {\modif SP2} programme (see Sect.\ref{sec:wp2}) to measure the mass of the transiting sub-Neptune revealed by TESS with a period of 13.176 days. An additional transit signal of a super-Earth with a period of 3.307\,days was detected later on as a Community-TOI (CTOI). This target was also observed during the same season (P111) with ESPRESSO (PI: E. Pallé). We combined our data sets and published the masses of TOI-406\,b and TOI-406\,c \citep{Lacedelli2024}. Figure~\ref{fig:toi406} shows the NIRPS and HARPS RVs phase-folded to the 13.176 days of the 2.1\,{\Rearth} sub-Neptune TOI-406\,b with a semi-amplitude $K=4.27\pm0.8$\,{\ms}, slightly higher than but compatible at (1.7-$\sigma$) with the semi-amplitude of $K=2.83\pm0.3$\,{\ms} found by ESPRESSO only. Scaled to the same exposure time of 1800\,s, the median RV uncertainty is 3.1 and 3.7\,{\ms} for NIRPS and HARPS, respectively, {\modif illustrating the gain in photon-noise uncertainty between the two instruments}. The dispersion of the residuals is 3.4\,{\ms} and 5.1\,{\ms} for NIRPS and HARPS, respectively. The detection of the inner 1.3\,{\Rearth} super-Earth is not significant with NIRPS+HARPS ($K=1.25\pm0.6$ {\ms}) but very clearly with ESPRESSO only ($K=1.60\pm0.12$ {\ms}). 

{\modif Other planetary systems monitored with NIRPS in these two last years and illustrating the long-term RV precision are presented by Lamontagne et al. (in prep) and Weisserman et al. (in prep).}

\subsection{Sun observed as a star with HELIOS}

The HARPS Experiment of Light Integrated Over the Sun
(HELIOS) is a copy of the HARPS-N solar telescope \citep{Dumusque2015} installed outside the 3.6\,m telescope building and it has been operational since September 2018. This small robotic telescope focuses the disk-integrated Sun’s light onto an integrating sphere and feeds it to both HARPS and NIRPS spectrographs via optical fibres. This telescope therefore allows for the acquisition of high-precision solar spectra at a cadence of on minute for several hours per day, on any clear day. The Plexiglass dome over HELIOS introduces as expected some absorption bands in the NIR and mainly around 1.7\,$\mu$m (with $\sim$80\% of absorption). By considering the solar coordinates at the photocentre of each exposure, the data reduction pipelines provides the RV correction required to transform from the Earth’s reference frame to the Solar System barycentre. To isolate the Sun’s intrinsic variations, we need to additionally transform from the barycentre to the heliocentre to remove the influence of the Solar System planets, primarily the annual synodic signal of Jupiter with a semi-amplitude of $\sim$12 {\ms}. This is done by querying ephemeris from the JPL Horizons interface\footnote{\href{https://ssd.jpl.nasa.gov/horizons}{ssd.jpl.nasa.gov/horizons}} following the work presented in \citet{Collier-Cameron:2019aa}. As described in that paper, differential extinction is also corrected for.

\begin{figure}
    \centering
    \includegraphics[width=1.0\linewidth]
    {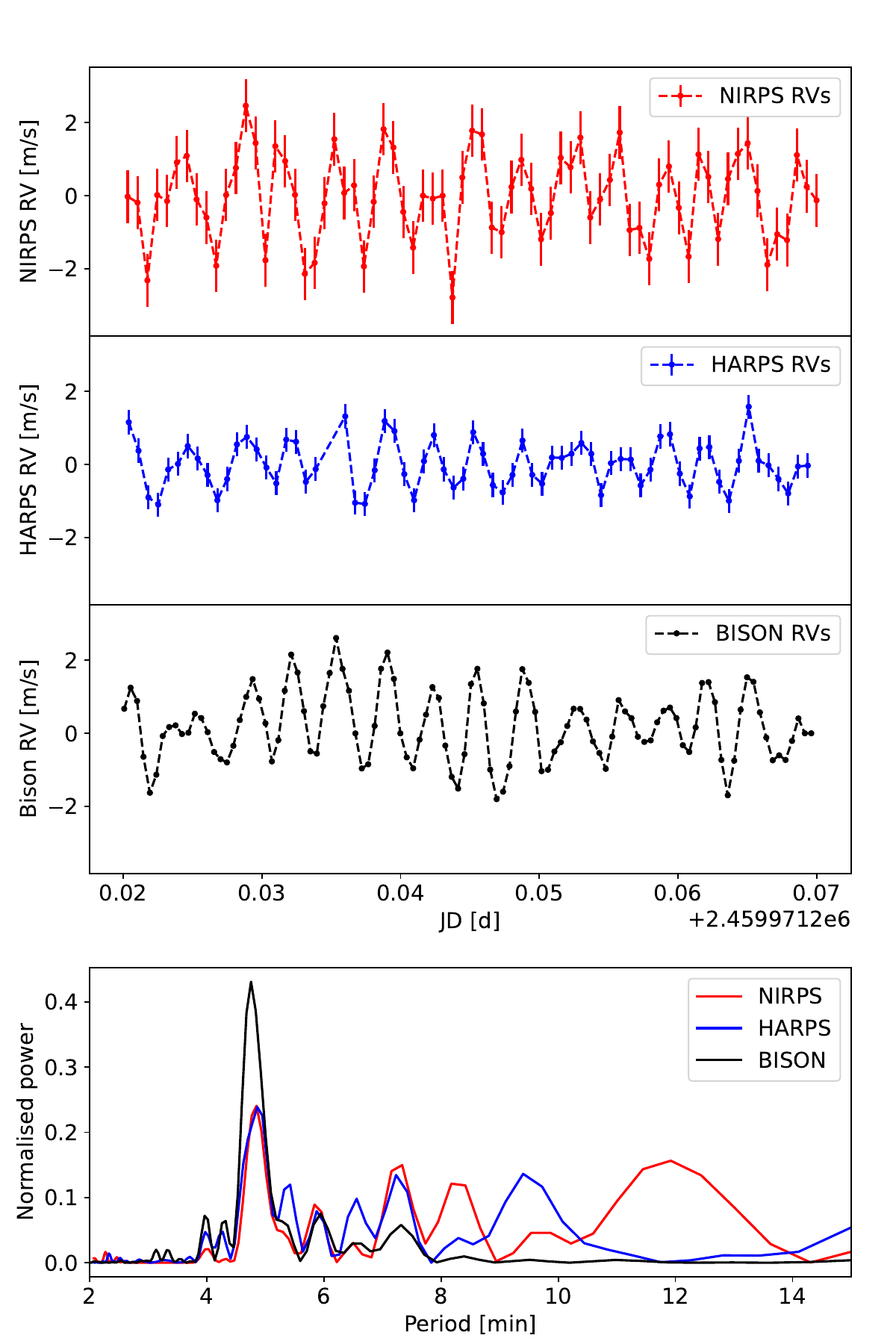}
     \caption{Top:  one-hour sequence of simultaneous NIRPS, HARPS (both taken with HELIOS) and BISON RVs of the disk-integrated Sun’s light showing the 5-mn acoustic oscillations, shown from  top to bottom. The data were taken on January the 26th 2023. Bottom:\ Power spectral density of the NIRPS, HARPS and BiSON RVs all showing the same clear peak aroundfiveminutes (3.45, 3.40 and 3.50 mHz, respectively).}
    \label{fig:sun_1day}
\end{figure}

\begin{figure}
    \centering
    \includegraphics[width=1.0\linewidth]{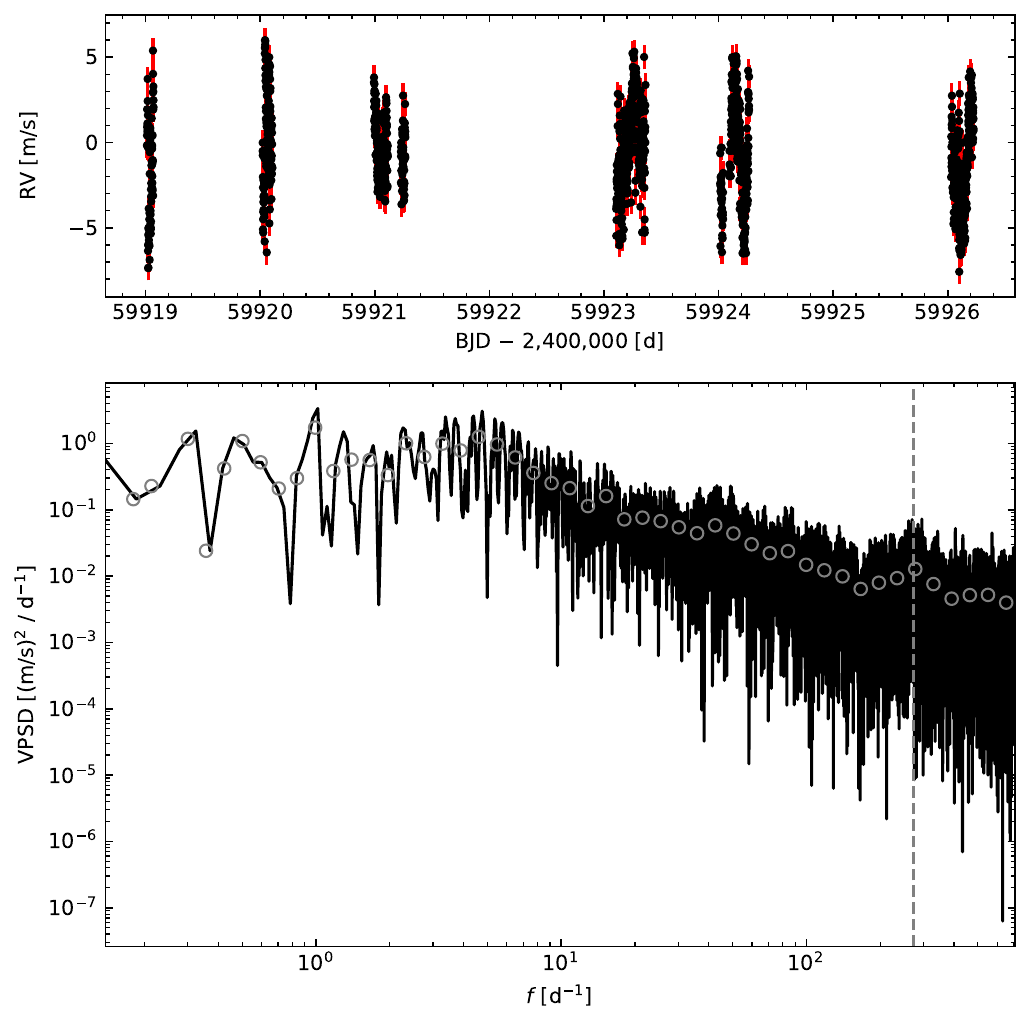}
     \caption{Top: NIRPS RV sequence of the disk-integrated Sun’s light with HELIOS over one week. Bottom: Power spectral density of the HELIOS RV sequence showing a bump around 3.5 mHz.}
    \label{fig:sun_1week}
\end{figure}

Figure \ref{fig:sun_1day} shows the NIRPS, HARPS and Birmingham Solar Oscillations Network \citep[BiSON -][]{Chaplin1996} RV time series obtained during one hour on 2023 January 26, during commissioning \#8. The p-mode oscillations around 5\,min are visible in this sequence illustrating the high precision of NIRPS on short timescales. We note that this is the first time that p-mode oscillations are so clearly detected with a NIR spectrograph. The p-mode oscillation signal as observed with NIRPS is completely in phase with the signal observed with HARPS and BiSON. We note that the p-modes amplitude seems slightly larger in the  NIR. The amplitude of solar p-mode oscillations is governed by stochastic excitation and mode damping, both of which take place in the surface convection zone. \cite{Ronan1991} showed that the relative intensity fluctuations in the p-mode band increase with height in the solar photosphere. The difference of p-modes amplitude may reflect that NIRPS is probing different heights than HARPS in the solar photosphere. Figure~\ref{fig:sun_1week} shows the NIRPS-HELIOS RV over one week (5-12 Dec 2022) after subtracting a daily mean to remove some day-to-day jumps present in the data and suspected to be due to residuals of telluric correction. The dispersion measured on individual days ranges from 1.6 to 3.1\,{\ms} but is dominated by p-mode oscillations. A smoothing of the data with a 15-minute window reduces the dispersion {\modif between 1.2 and} 2.4 {\ms}. The bottom panel of Fig.\,\ref{fig:sun_1week} shows the power spectral density of the HELIOS RV sequence. The dashed vertical line shows the central frequency of the oscillation bump at 3.16 mHz, as obtained from HARPS-HELIOS RVs analysed in \cite{AlMoulla2023}.

NIRPS-HELIOS commissioning data were also used to study the variability of the He triplet, a key tracer of evaporating exoplanet atmospheres. Multi-epoch observations of the Sun obtained with NIRPS were analysed to investigate temporal variability of the He triplet due to telluric contamination and solar activity and their impact on the retrieval of planetary atmospheric parameters \citep{Mercier2025}. In Summary, although the He triplet is strongly sensitive to stellar activity on the stellar rotational timescale (a few weeks), the variability over the timescale of a transit (a few hours) is small and does not seem to impact atmospheric parameter retrieval for a Sun-like star.

\subsection{Spectroscopic characterisation of exoplanet atmospheres}

\begin{figure}
    \centering
    \includegraphics[width=1.0\linewidth]{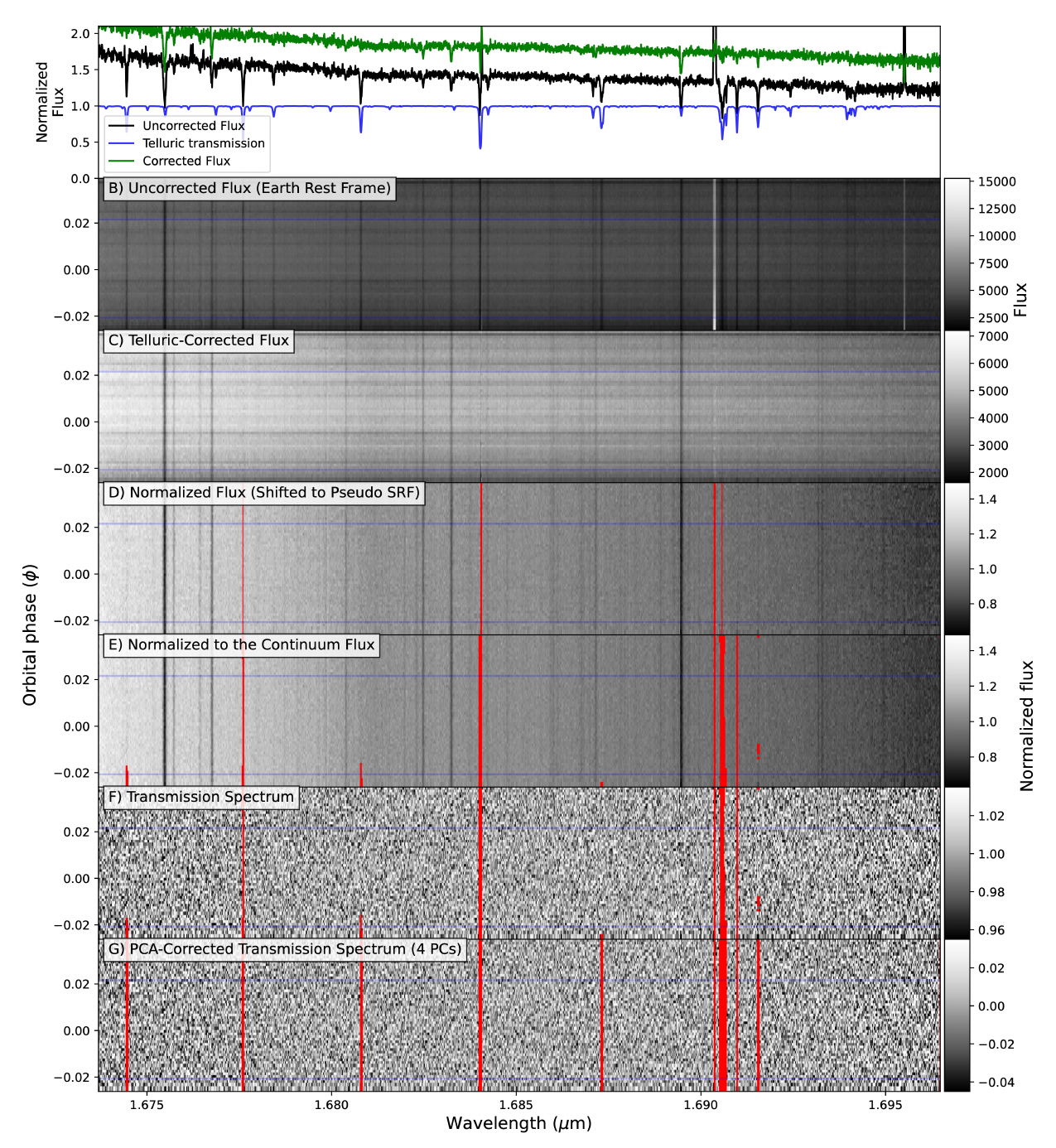}
    \caption{Full additional reduction steps and transmission spectra computation from \texttt{STARSHIPS} for telluric lines below 70\% transmission masked up to 90\% transmission in their wings and four PCs removed for the echelle spectral order 86 with the NIRPS-DRS data products. \textit{Top:} Single exposure of the uncorrected and telluric corrected flux along with the reconstructed telluric transmission spectrum, with offsets for visibility. \textit{B)} Uncorrected flux, normalised by the instrumental blaze function. \textit{C)} Telluric-corrected flux (correction from the NIRPS-DRS - \citealt{Allart2022}). \textit{D)} Spectra normalised by their median (over wavelengths) and shifted to the pseudo-SRF, with a first sigma-clipping of pixels showing abnormally high variations in time. \textit{E)} Spectra normalised to the continuum of the \textit{master-out}. \textit{F)} Spectra divided out by the normalised \textit{master-out} (transmission spectra). \textit{G)} Transmission spectra after final PCA-based correction, additional telluric masking, mean subtraction and a second sigma-clipping in the time dimension.}
    \label{fig:WASP-127b_reduction_steps}
\end{figure}

To assess the instrumental performance for spectroscopic characterisation of exoplanet atmospheres (either in transit or dayside), we analyse one transit of the hot Saturn WASP-127 b \citep{Lam2017} observed on the night of 2023-03-10 in the HE mode, during NIRPS commissioning \#9. We perform the analysis using data products from both the APERO (version 0.7.289 - \citealt{Cook2022}) and NIRPS-DRS (version 3.2.0). We use telluric- and blaze-corrected spectra. We apply the additional reduction and analysis procedure from the \texttt{STARSHIPS} code outlined in \cite{Boucher2023} to obtain the transmission spectra. The main steps are 1) Doppler-shifting the spectra to the stellar rest frame (SRF), 2) building a \textit{master-out} reference stellar spectrum by combining spectra out-of-transit and removing it to obtain transmission spectra, 3) sigma-clipping pixels with bad or outlying behaviour before and after stellar spectrum removal, and 4) using principal components analysis (PCA) to remove any remaining stationary signal in the SRF. The main reduction parameters we vary are the minimum level of transmission for telluric lines past which they are masked, the width of the masking (using the level of transmission in the wings from the reconstructed telluric absorption spectra), and the number of principal components (PCs) to remove in the final step. We use the reconstructed telluric transmission spectrum from each pipeline to inform the additional telluric masking. The full process is shown in Fig.~\ref{fig:WASP-127b_reduction_steps}.

We searched 
for a water vapour detection using the best-fit atmosphere described in \cite{Boucher2023} for the CCFs: a water volume mixing ratio of $10^{-5}$, a cloud top pressure of 0.1\,bar, and a Guillot temperature profile \citep{Guillot2010} with equilibrium temperature $T_{eq} = 1200$\,K, internal temperature $T_{int} = 500$K, IR opacity $\kappa_{IR} = 10^{-3}$ and optical to IR opacity ratio $\gamma = 10^{-1.5}$. We compute the corresponding transit spectrum using petitRADTRANS \citep{Molliere2019}. We cross-correlate it with the transmission spectra from -50\,\kms to 50\,\kms for a range of orbital $K_p$ to produce a 2D CCF map. Before performing the cross-correlation, we apply the PC removal step to the model as well, since that reduction step may affect some of the planetary signal in the data. We find a clear correlation peak near the known planet $K_p=129.8$ \kms and with a blueshift of about 8\,\kms, consistent with the results of \cite{Boucher2023}, obtained from three SPIRou transits. When computing the CCF S/N, the region near the peak is excluded.  

\begin{figure*}
    \centering
    \includegraphics[width=1.0\linewidth]{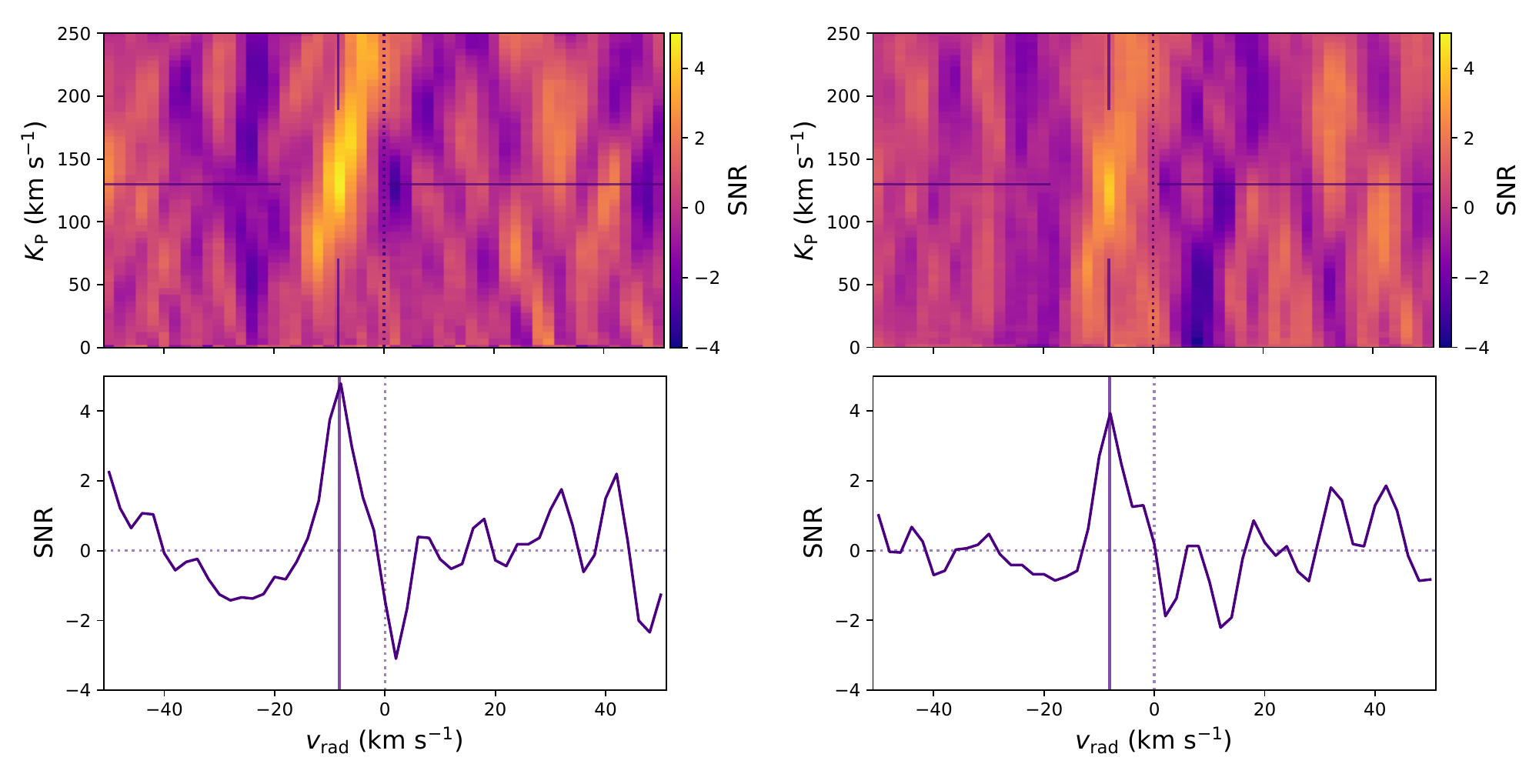}
    \caption{2D CCF detection maps (\textit{top}) for water absorption in the atmosphere of WASP-127 b, along with a 1D CCF curve (\textit{bottom}) corresponding to a slice of the map at the known planetary $K_p$ (horizontal line in the top panels). \textit{Left}: APERO DRS reduction. \textit{Right}: NIRPS DRS reduction.}
    \label{fig:WASP-127b_ccf_maps_both_DRS}
\end{figure*}

We find comparable results with both data reduction pipelines with a CCF peak S/N of around 4 $\sigma$ with the NIRPS DRS products (and 5$\sigma$ with APERO) in a single transit. We optimise the reduction parameters (extra telluric masking and number of PCs removed) to maximise the water detection: in both cases, we mask telluric lines reaching below 70\% in transmission up to where the wings reach 90\% transmission, and we remove four PCs. We note that while we show results for these best values, we do recover detections (within 1.5$\sigma$ and 0.8$\sigma$ for the APERO and NIRPS DRS, respectively) for a wide range of reduction parameters: masking telluric lines from 20 to 70\% transmission, wings masked up to either 90 or 97 \%, and 2 to 5 PCs removed. The CCFs for each reduction are shown in Fig.~\ref{fig:WASP-127b_ccf_maps_both_DRS}.\\

These results showcase the capabilities of NIRPS to produce high-fidelity spectroscopic characterisation of exoplanets' atmospheres under a minimal number of nights and be competitive with other NIR high-resolution spectrographs such as {\modif GIANO-B \citep{Claudi2017},}SPIRou \citep{Donati2020}, CARMENES \citep{Quirrenbach2014}, CRIRES+ \citep{Dorn2023} or IGRINS \citep{Park2014}.

\subsection{Stellar characterisation}
\label{sec:stellar}

NIRPS's high resolving power and wide wavelength coverage provide a powerful means to determine stellar effective temperature ($T_{\rm{eff}}$) and abundance measurements. While $T_{\rm{eff}}$ and abundance measurements are prone to systematic errors (see section~\ref{sec:OS1}), these can be mitigated by taking advantage of the wide wavelength coverage, allowing for realistic uncertainty estimates. \citet{Jahandar2024} developed such an analysis framework for SPIRou data to derive the $T_{\rm{eff}}$ of Barnard's star, achieving consistency within 20\,K of interferometric measurements. This framework also enabled abundance measurements of 15 elements. Applied to a sample of 31 nearby M dwarfs, including 10 within FGK binaries, \cite{Jahandar2025} confirmed that $T_{\rm{eff}}$ can be estimated with an accuracy of $\sim$30\,K using this approach.

Using the analysis framework of \citet{Jahandar2024}, commissioning NIRPS data on LHS~1140, the host of the temperate planet LHS~1140\,b, was used to determine its $T_{\rm{eff}}$ and the abundance of several refractory elements (Fe, Si, and Mg), which are essential for internal structure models. These measurements were used to constrain both the core mass fraction and the water mass fraction of LHS~1140\,b, suggesting that this planet likely has a significant water envelope, comprising approximately 15\% of its mass \citep{Cadieux2024}.

\section{GTOs with NIRPS}
\label{sec:GTO}

In exchange for funding, building, and operating the NIRPS instrument, our consortium was awarded 725 nights of GTO over five years with the 3.6m telescope, following a well-established and successful ESO scheme for funding second-generation instruments for the La Silla and Paranal observatories. 
Our GTO proposals, approved by ESO’s Observing Programme Committee (OPC), follow the science programme proposed by the Consortium to ESO at the time of the signature of the construction agreement. The NIRPS GTO, started in April 2023, is organised into three main scientific {\modif sub-programs (SPs)} of 225 nights each and `Other sciences' (OS) small programmes totalling 50 nights. The overall scientific objectives of these three {\modif SPs} and OS programmes are described in the following sub-sections. Our list of protected targets is accessible via the ESO website\footnote{\href{https://www.eso.org/sci/observing/teles-alloc/gto/115.html}{eso.org/sci/observing/teles-alloc/gto/115.html}}.

\subsection{{\modif SP1} - Blind RV search of exoplanets around low-mass stars}

The {\modif sub-programme} 1 of the NIRPS GTO is dedicated to a blind search of planets around very low-mass stars (< 0.6 {\Msun}). Its objectives are centred on two major challenges: (1) identifying Earth-like planets for future atmospheric characterisation and (2) understanding the process of planet formation and dynamical evolution, and in particular its sensitivity to initial conditions in the proto-planetary disk.

\subsubsection{Search for the best Earth-like planets for future atmospheric characterisation.}

Our first goal is to detect the best system for future atmospheric characterisation by the new generation of facilities. Earth-like planets are incredibly faint in comparison to their star, and so close to it, that they represent challenges even with a 40-m class telescope. Earth-like planets in the HZ of M dwarfs have a planet-to-star luminosity contrast of about $10^{-7}$, compared to a few $10^{-10}$ for the Earth-Sun contrast. Over the next decade, systems around nearby ($d<15$ pc) M dwarfs are the only opportunity to obtain reflected spectra of Earth-like or Neptune-type planets by combining high contrast imaging and high-dispersion spectroscopy on the ELT \citep{Snellen2015}. 
Today the majority of planets in the close solar neighbourhood are yet to be discovered. Of the closest 100 M dwarfs \citep{Reyle2021}, only 25 are known to host planetary systems (based on the NASA Exoplanet Archive\footnote{\href{https://exoplanetarchive.ipac.caltech.edu/index.html}{exoplanetarchive.ipac.caltech.edu/index.html}}), despite statistical studies from RV and transit surveys demonstrating that the large majority of M dwarfs must have planets \citep[e.g.][]{Mignon2025,Bonfils2013,Dressing2015,Sabotta2021}.

We have identified the nearest southern stars, at less than 6 pc, for which RV measurements are to date still too few or too imprecise to detect small mass planets. In general, this corresponds to the latest spectra types (beyond M5) or most active stars. Near-infrared observations with NIRPS will make it possible to obtain more precise measurements for these very red stars and to better filter out the effect of stellar activity. We are measuring around twenty of these very close systems at high cadences. 
{\modif Proxima Centauri is one of the {\modif SP1} golden target and was intensively observed. The 149 nightly-binned RVs are presented by Suárez Mascareño et al. (this same edition of the journal) confirming Proxima b and finding evidence of the presence of the sub-Earth Proxima d. The standard deviation of the residuals of NIRPS RVs after the modeling is $\sim$0.80 {\ms}, showcasing the potential of NIRPS to measure precise RVs in the NIR. The active star GJ581 was monitored with NIRPS and HARPS as part of {\modif SP1} to revisit the properties of its planetary system and to explore the differences and similarities of the RV activity signal and activity indiced between the NIR and the optical \citep[][Larue et al. in prep]{Gomes2025}.}  

\subsubsection{Planetary system formation and dynamical evolution as a function of the central mass}

The statistical study of planets orbiting very low-mass stars provides key insights into planetary formation \citep{Alibert2017}, particularly regarding its sensitivity to initial conditions in the protoplanetary disk, which strongly depend on the central temperature and thus on stellar mass. Several RV studies have now offered a solid understanding of planetary occurrence statistics for early and mid-M dwarfs. Leveraging the NIR capabilities of NIRPS, we {\modif extend} precise RV measurements to the ultra-cool dwarf (UCD) regime (M7V and later). Since M dwarfs span a factor of 7 in terms of mass (0.07 to 0.5 M$_{\odot}$) and a factor of 200 in terms of luminosity, this will allow us to compare the initial conditions that are significantly different from those of early and mid-M dwarfs.
This will also allow for a direct comparison with synthetic planet populations around low-mass stars calculated by the Generation III Bern global model of planet formation and evolution \citep{Burn2021}. 

Another approach we have adopted is to search for planetary systems around very low-mass young stars. This allows us to distinguish the effects of planetary formation from those of dynamical evolution. Once again, the ability to filter out the effects of stellar activity is particularly useful for detecting planets around these very low-mass young stars. We are monitoring a sample of stars in young associations or those surrounded by circumstellar dust debris disks, as NIRPS can detect Neptune-like planets around such targets.

\begin{figure}
    \centering
    \includegraphics[width=1.0\linewidth]{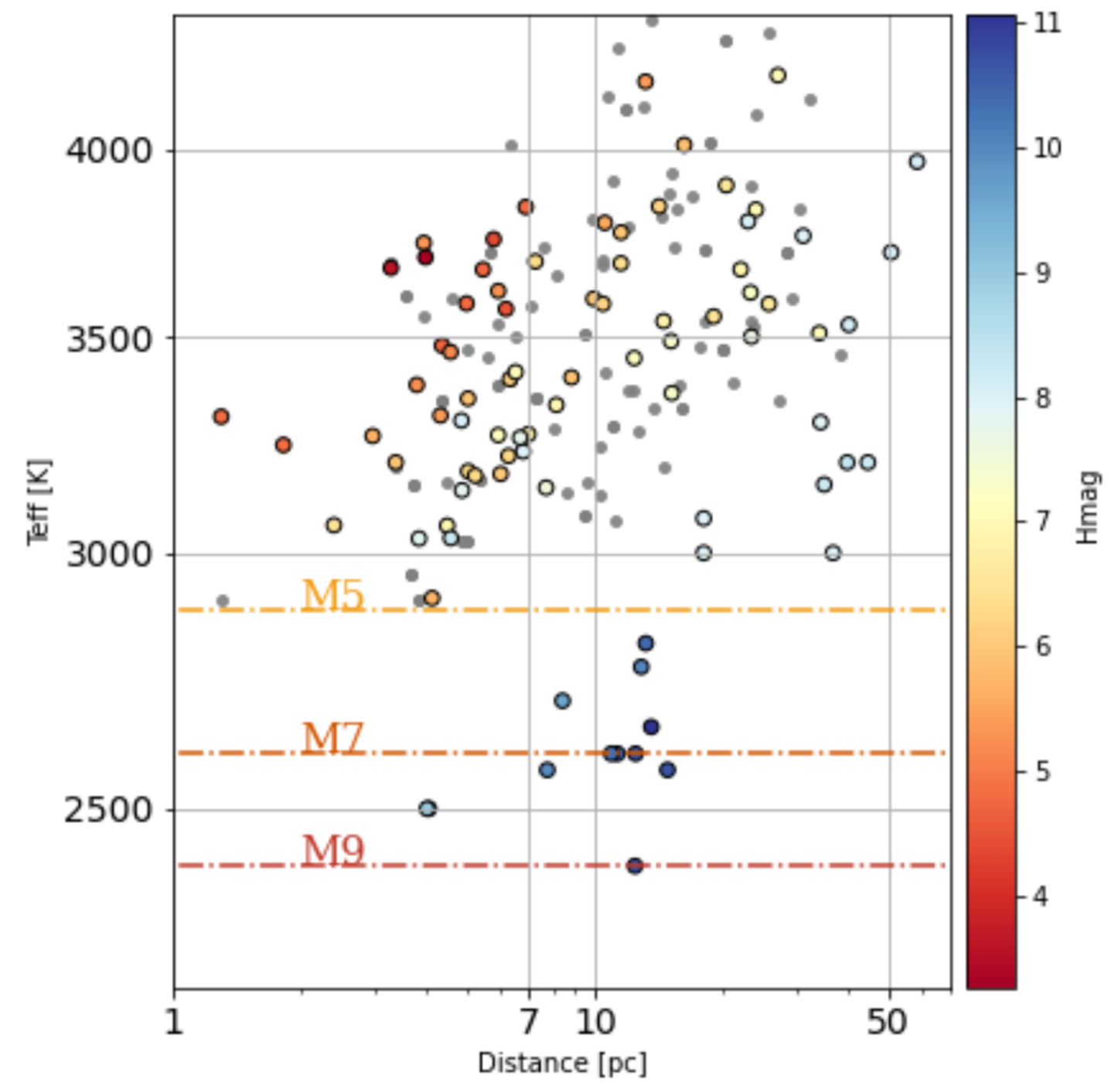}
    \caption{Effective temperature as a function of the distance of the {\modif SP1} selected sample colour-coded as a function of their H magnitude. Grey dots correspond to known exoplanets around cool stars (< 4500 K).}
    \label{fig:wp1}
\end{figure}

Finally, we  selected M dwarfs that host at least one massive planet with an intermediate orbital period to search for potential terrestrial planets in shorter orbits. The goal 
of characterising the overall architecture of planetary systems is crucial for understanding their formation history and dynamical evolution. While more than 5\,000 exoplanets have been detected, fully characterising both the inner and outer regions of planetary systems remains a significant challenge. As a result, the interaction between these regions during planetary formation and evolution is still poorly constrained by current observations. Figure~\ref{fig:wp1} shows our selected {\modif SP1} targets in the {\teff} - distance diagram. 

\subsection{{\modif SP2} - Mass characterisation of exoplanets transiting M dwarfs}
\label{sec:wp2}

The second main {\modif sub-programme} of NIRPS GTO is devoted to the mass and bulk density characterisation of exoplanets transiting M dwarfs. 
{\modif SP2} aims to constrain the internal composition of exoplanets around M dwarfs including their iron/rock/water fraction and to explore how the properties of exoplanets around M dwarfs vary with stellar irradiation, stellar mass, planetary architecture, and even stellar composition (see Section \ref{sec:stellar}), shedding new light on the formation and evolution pathways of these systems. 

Exoplanet's density and surface gravity are key parameters to characterise its internal structure and composition as well as to interpret direct imaging and atmospheric observations \citep[e.g.][]{Dorn2015,Brugger2017,Otegi2020,Batalha2019,Haldemann2024}. 
Kepler and TESS surveys have shown that planets smaller than 4 {\Rearth}, that is, super-Earths and sub-Neptunes are the most common type of exoplanets in the solar neighborhood. These small planets are characterised by a bimodal radius distribution \citep{Fulton2017,Cloutier2020} whose origin is unclear but probably related to various thermally-driven atmospheric escape mechanisms \citep{Jin2018,Gupta2020,Wyatt2020} or as a footprint of formation, migration and evolution processes \citep{Venturini2020,Burn2024}. 

\begin{figure}
    \centering
    \includegraphics[width=1.0\linewidth]{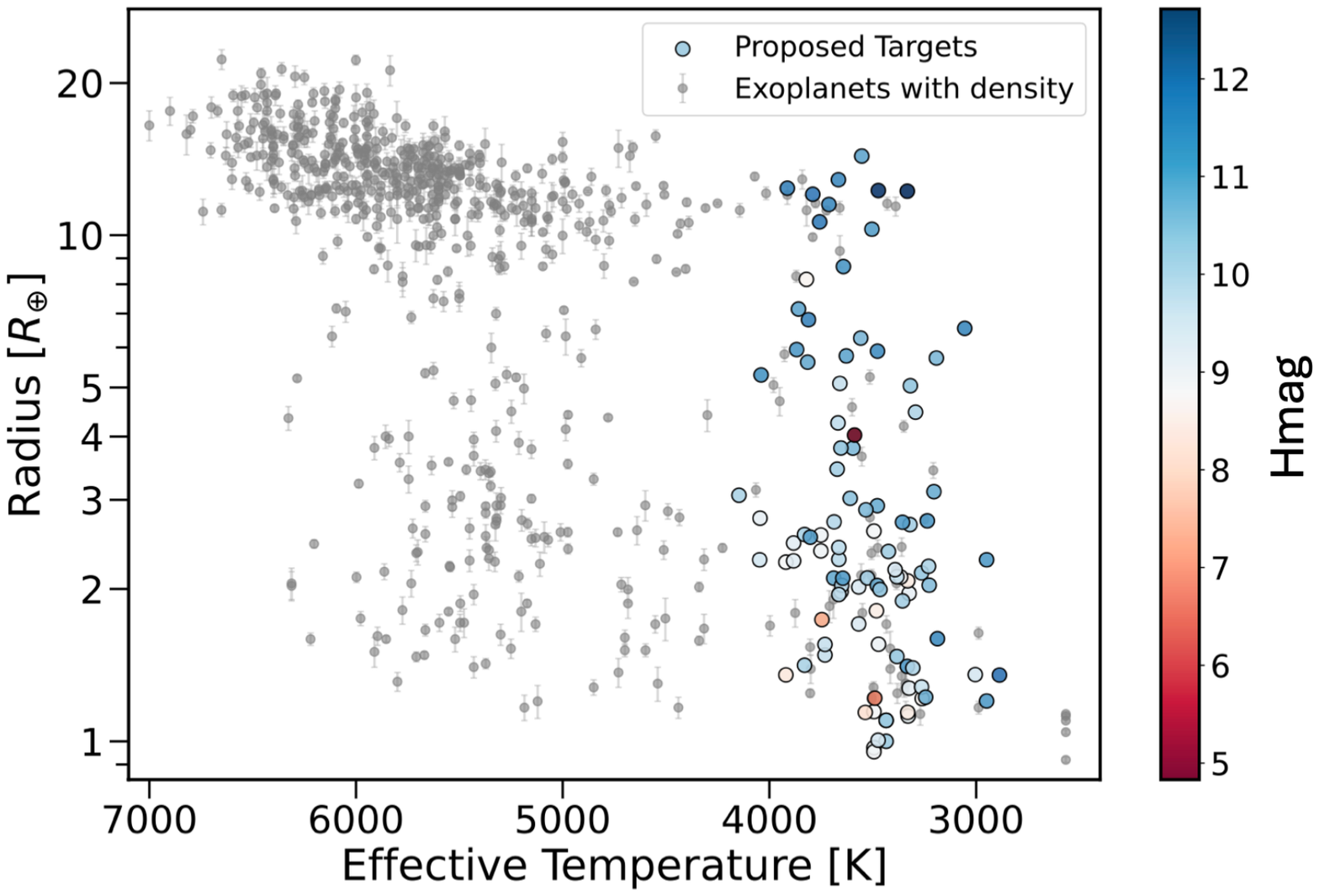}
    \caption{Radius versus host star effective temperature of exoplanets with mass precision < 25\% and radius precision < 8\% from $PlanetS$ catalog \citep{Parc2024}. SP2 targets are shown with a colour code linked to their H magnitude.}
    \label{fig:wp2}
\end{figure}

Our objective is to measure precise masses and bulk densities of the transiting exoplanet population orbiting M dwarfs discovered mainly by TESS. The current sample has been defined based on the following criteria: a) confirmed/validated planet candidate; b) host star $I < 14.5$; and c) no mass measurement available or very uncertain one (higher than 30\%). We plan to pave the mass-radius diagram with exoplanets transiting mid-to-late M dwarfs with a specific focus on the architecture of multi-planetary systems which can provide much stronger tests for planetary formation theories than single-planet systems. We also plan to determine the nature of temperate transiting small-size planets ($< 4 $R$_\oplus$) which represent a prime opportunity for follow-up atmospheric characterisation with JWST but require prior knowledge of their mass to correctly interpret their transmission spectra \citep{Batalha2019}. We also want to build up a sample of keystone planets with well-characterised bulk composition near the radius valley at short orbital periods which constitutes the transition between rocky exoplanets and sub-Neptunes with extended H/He envelope to constrain the origin of the M dwarf radius valley and discriminate the scenario of terrestrial planet formation in a gas-depleted environment \citep{Lopez2018} rather than by a thermally-driven mass loss process. A specific effort {\modif is} made on transiting rocky planets to measure ultra-precise planetary masses and consequently to recover those planets' iron core mass fraction at a ground-breaking level of precision (10\%). 
Although many rocky planets have bulk compositions consistent with that of the Earth (i.e. 33\% iron core + 67\% silicate mantle), residual scatter in the internal compositions remains: small planets appear to exhibit a broader distribution of Fe/Si ratios than their host stars \citep{Plotnykov2020,Adibekyan2021}, which themselves exhibit a remarkable homogeneity of Mg/Si and Fe/Si abundances among nearby stars \citep{Bedell2018}.
The lower mass and luminosity of M dwarfs compared to Sun-like stars opens the door to the mass characterisation of temperate Earth-size planets, inducing {\ms} RV signals rather than {\cms}. Moreover, their lower luminosity implies a closer-in habitable zone (HZ) corresponding to relatively short orbital periods (10–30 days). Systems comprising of an M-dwarf host with a transiting exoplanet inside or near the HZ represent golden targets and prime opportunities for follow-up atmospheric characterisation with JWST. We expect NIRPS to be able to characterise transiting rocky planets within the habitable zone of their host star in unprecedented detail. Figure~\ref{fig:wp2} shows our selected {\modif SP2} targets in the radius - {\teff} diagram. 

{\modif Several results from SP2 will soon be published: The discovery of a transiting sub-Neptune and a cold eccentric giant orbiting the M-dwarf TOI-756 \citep{Parc2025}; The updated characterisation of the multi-planetary system including a transiting sub-Neptune and orbiting the nearby M-dwarf GJ3090 (Lamontagne et al. in prep); The two giant planets transiting TOI-3832 and TOI-4666 (Frensch et al. in prep); The ultra-short period super-Earth TOI-4552b (Srivastava et al. in prep)}

\footnotetext{www.dace.unige.ch/exoplanets}

\subsection{{\modif SP3} - In-depth look into exoplanets through high-resolution time-series spectroscopy}
\label{sec:wp3}

The third main {\modif sub-programme} of the NIRPS GTO is devoted to high-resolution time-series spectroscopy of exoplanets \citep[e.g.][]{Birkby_2018}. Such observations allow for the detailed analysis of spectral line shapes and contrasts, offering an unparalleled window to understand the chemistry and dynamics of exoplanet atmospheres and their orbital architectures. 

Our objectives are to measure properties such as mass-loss rate, atmospheric composition, equilibrium versus disequilibrium chemistry, molecular dissociation, temperature structure, and day-to-night temperature contrasts to answer fundamental questions about these poorly understood processes in exoplanets atmospheres. Based on those constraints we want to build a better understanding of how high-resolution time-series spectroscopy can inform exoplanet formation and evolution models. Measurements of atmospheric compositions, such as the C/O ratio and metallicity can indicate where in the protoplanetary disk the planets formed and how they accreted material \citep[e.g.][]{Pelletier2025}. From formation theory, models such as core accretion, predict different outcomes for these properties. Therefore, comparing these models to observations helps distinguish between various planet formation scenarios \citep[e.g.][]{Chachan_2023}. {\modif Furthermore}, measurements of mass-loss rate \citep[e.g.][]{allart_2018}, atmospheric heating, atmospheric circulation \citep[e.g.][]{seidel_2023} and orbital architecture \citep[e.g.][]{Bourrier_2024} track the evolution of exoplanets. These processes can affect an exoplanet's climate, chemical balance, stability, mass, and radius over time. It is thus critical to study these evolutionary processes by observing a diversity of exoplanets across a range of masses, radii, and irradiation conditions. This helps refine models of atmospheric stability and escape to explain phenomena such as the absence of hot Neptunes \citep[e.g.][]{Mazeh_2016} and the differences between super-Earths and sub-Neptunes \citep[e.g.][]{Owen2017}.

\begin{figure}
    \centering
    \includegraphics[width=1.0\linewidth]{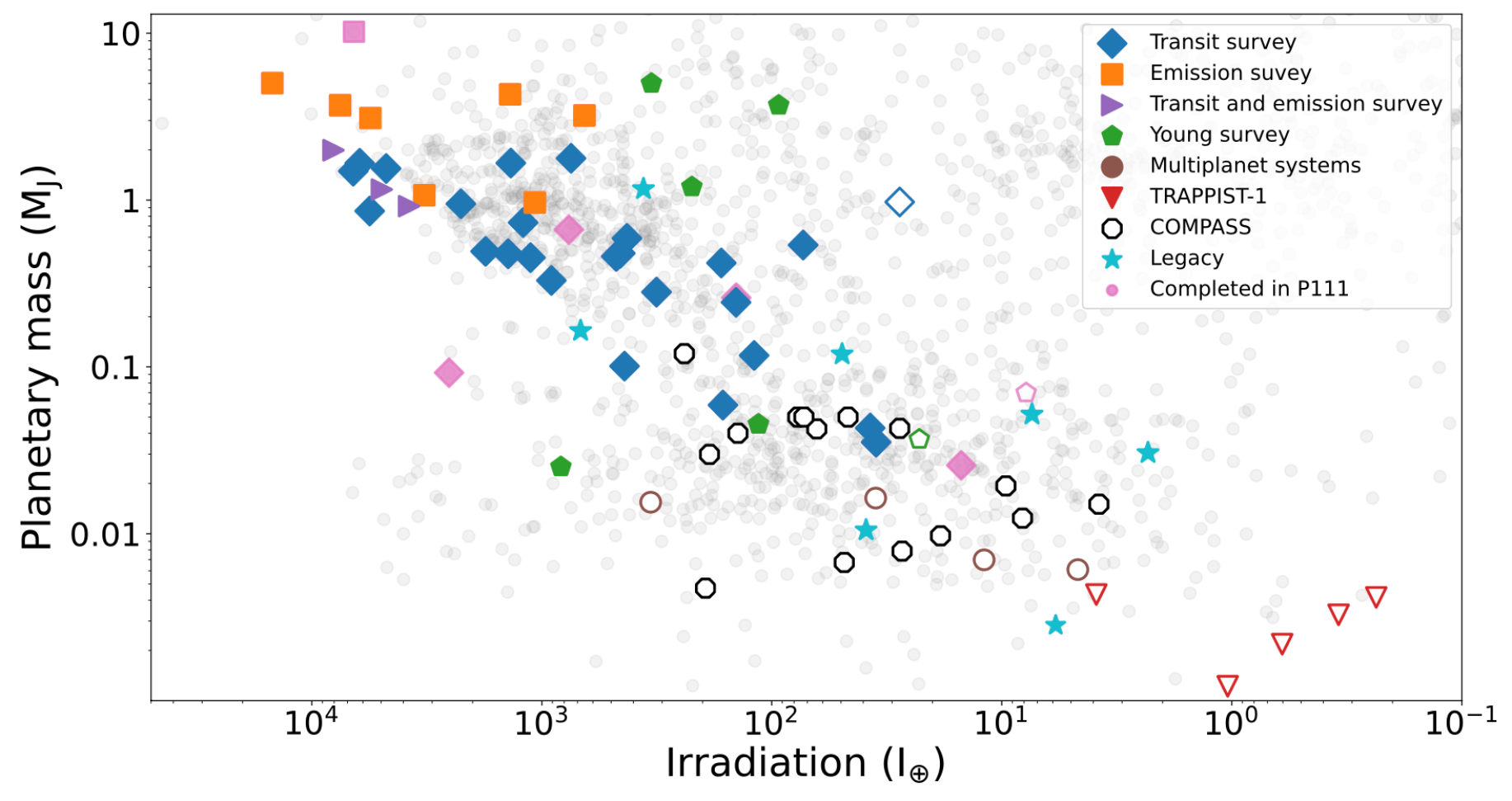}
    \caption{Exoplanet population in a mass-irradiation diagram. {\modif SP3} targets are highlighted with a colour code linked to the different sub-programmes. COMPASS
    investigates the frequency of mutually-misaligned small planets and has overlaps with other sub-programmes.}
    \label{fig:wp3}
\end{figure}

By dedicating 225 nights to time-series spectroscopy observations, the NIRPS consortium is conducting an in-depth look into exoplanets from their orbital architecture to their atmosphere both in transmission and emission. To do so, the consortium is building a large comprehensive population survey of exoplanets to study the aforementioned processes, bridging the gap between observations and models of the formation and evolution of exoplanets. This will be an important stone in the field of population-based studies built on a uniform data acquisition, reduction, and analysis strategy. In addition, to complement this comprehensive survey, the NIRPS consortium is allocating a significant fraction of time to build high-fidelity high-S/N spectra for a carefully selected sample of exoplanets which will significantly enhance our understanding of exoplanetary atmospheres and orbital architecture properties in preparation for the ELT era. Altogether, {\modif SP3} observations aim to shift the paradigm of exoplanet atmospheric chemistry and dynamics, formation, and evolution models for the next decades. Figure~\ref{fig:wp3} shows our {\modif SP3} selected targets in the mass-irradiation diagram.

{\modif As first results of SP3, \citet{Allart2025} 
analysed three transits of the warm Saturn WASP-69b. The study shows a clear detection of the helium triplet with variable properties along the transit chord, clearly indicating an important mass-loss rate and particles escaping into space in a cometary-like tail. In addition, we reach a similar precision on the measured projected obliquity between HARPS and NIRPS that led to the conclusion that WASP-69b has a slightly misaligned orbit with a true obliquity of $\psi$ = 28.7$^{+6.1}_{-5.3}$ $^{\circ}$.

The detailed analysis of the optical (with HARPS) and NIR (with NIRPS) transmission spectrum of the ultra-hot gas giant WASP-189b is presented by \citet{Vaulato2025}.
It reveals that hydride ion continuum hides absorption signatures in the NIR transmission spectrum of this exoplanet atmosphere. 

The atmospheric analysis of the daysides of the ultra-hot Jupiter WASP-121b from emission observations using NIRPS is presented by \citet{Bazinet2025}. 
It reveals evidence of thermal dissociation and redshifted water detection in this exoplanet atmosphere.}

\subsection{Other science programmes}

\subsubsection{OS1: Improving metallicity determination of M dwarfs with FGK binaries}
\label{sec:OS1}

The correct characterisation of exoplanets requires a complete and detailed knowledge of their host stars. In particular, the stellar parameters $T_{\rm{eff}}$, $\log$\,g and [Fe/H] can be derived with good precision by analysing high resolution and good S/N spectra. During the last decades, the methods used to derive such parameters in solar-type stars have significantly improved, reaching precision values below 50\,K for $T_{\rm{eff}}$ and below 0.03 for [Fe/H] \citep[e.g.][]{Sousa2008,Blanco2014,Bensby2014}. These parameters are not only essential to individually characterise  the planets around such stars but also to make statistical analysis of the populations of planet hosts \citep[e.g.][]{Sousa2019}. However, the characterisation of M dwarfs using optical spectra is very complicated due to the presence of molecular lines that blend with other atomic lines and depress the continuum \citep[e.g.][]{Neves2014}. Because of this, in the last years, great efforts have been made to take advantage of spectra in the NIR, which is significantly less blended than the optical region of the spectra \citep[e.g.][]{Lindgren2016,Veyette2017,Hobson2018,Passegger2019,Sarmento2021,Marfil2021,Cristofari2022,Jahandar2024}. Unfortunately, these parameters are still prone to large errors, typically above 100\,K for $T_{\rm{eff}}$ and above 0.1 dex for [Fe/H], which translates into larger uncertainties in their planet properties. More worrying is the fact that both [Fe/H] and T$_{\rm eff}$ derived by different methods provide different scales, especially for metallicity, making it also difficult to compare with those derived for G and K dwarfs \citep[e.g][]{Passegger2022}.
One way of tackling this issue is by analysing binary systems of an FGK primary and an M-dwarf companion, where the metallicity of the primary can be accurately determined. The results obtained for the M dwarf can then be calibrated based on the primary. This has been already done by using optical spectra \citep[e.g.][]{Bonfils2005a,Neves2012,Montes2018} and NIR spectra \citep[e.g.][]{Onehag2012,Lindgren2016}, where the latter used only small portions of CRIRES J-band spectra. Therefore, the large simultaneous wavelength coverage of NIRPS+HARPS provides an excellent opportunity to improve previous calibrations and advance towards a more homogeneous set of metallicities. We used the samples presented in those four papers to construct our target list.

\subsubsection{OS2: Origin of RV variations in cool giants}

To understand the planetary formation and evolution mechanisms around early F or A stars one has to look at K-M giants, the evolved counterparts of those massive stars, with generally lower rotation rates and displaying a much larger number of spectral lines \citep[e.g.][]{Lovis2007,Niedzielski2015}.
It is not clear if the frequency of massive planets increases as the stars get more massive or if there is an upper limit stellar mass to form planets \citep[e.g.][]{Johnson2010,Reffert2015,Wolthoff2022}. Attempting to tackle this issue, several programmes to discover planets around evolved stars have started up in the last years, both in open clusters and in the field \citep{Ottoni2022}. The advantage of open clusters is that the ages and masses of their member stars can be much better constrained than for field stars, leading to a more precise mass estimation for the planet. Furthermore, finding planets around stars with well-determined evolutionary stages may help to understand planetary evolution and engulfment processes. Despite many efforts to detect planets in clusters, only a few discoveries have been reported around MS stars \citep{Fujii2019}. At the same time, only four massive planets and a brown dwarf have been reported around red giants (in the Hyades - \citealt{Sato2007}; in M67 - \citealt{Brucalassi2014}; in NGC2423 and NGC4349 - \citealt{Lovis2007}; in IC4651 - \citealt{leao2018}), all of them having long periods. Indeed, current planetary search surveys have failed to detect close-in planets around massive evolved stars \citep[e.g.][]{Reichert2019}.
Another important concern when interpreting RV variations in red giants is the presence of intrinsic stellar jitter which shows a typical level of 10-20 {\ms} for stars at the base of the red giant branch (RGB) and increases as the stars further evolve \citep{Hekker2008}. The modulation of active regions in red giants can produce large amplitude RV variations. We note that the rotational velocities of these stars are not very high and considering their large radii the estimated rotational period can have similar values (hundreds to thousands of days) to the orbital period of their candidates in some cases. Therefore, analysing the stability of the signal over several orbits (and for a time span larger than the stellar rotational period) is essential to understand and disentangle the possible role of stellar activity on the RV variations.
\cite{Delgado2018,Delgado2023} found indications that the RV variability previously attributed to substellar companions in NGC4349 and NGC2423 (with periods of $\sim$700 days) might be caused by stellar activity. Among the post-main sequence stars, there are also some problematic cases where it has been challenging to establish whether the RV signals are of planetary origin or not, despite the extensive observational campaigns carried out. If these signals are of stellar origin, we would expect them to have a different amplitude in the optical and the NIR, due to the different contrast between the spot and the stellar disk \citep{Figueira2010}. Therefore, a combination of HARPS and NIRPS RVs is crucial to disentangling the variations induced by real planets and by stellar activity.

\section{Discussion and conclusion}
\label{sec:conclu}

NIRPS offers the astronomical community high-resolution and high-stability NIR spectroscopy. The instrument's unique features open up a new parameter space in ground-based observations that can address a large diversity of science cases. We present the NIRPS instrument and provided information on its characteristics, performance, and limitations, as well as advice on how to carry out observations. Using on-sky tests carried out during the commissioning phases, we demonstrated that all the main performance requirements have been met. First, a high resolving power of $R\sim$90\,000 is provided in HA mode configuration. In HE mode configuration, a resolving power of $R\sim$75\,000 is achieved. The spectral range continuously covers the domain from 972.4 nm to 1919.6 nm. The overall throughput, from the top of the atmosphere to the detector, peaks at 13\%. The RV precision, demonstrated on several M dwarfs, is close to 1 {\ms}. NIRPS and HARPS can be used simultaneously, offering  unique advantages: 1) to  simultaneously cover from 378 nm to 1920 nm, notably with a gap in between 691 nm and 972 nm; 2) to  significantly reduce the RV photon-noise uncertainty on M dwarfs; and 3) to {\modif explore different approaches to } disentangling stellar activity signals from real exoplanets. The NIRPS AO system offers the unique possibility of high angular resolution with a fibre acceptance of 0.4 \arcsec\ and NIRPS is also continuously monitoring the disk-integrated Sun's light with HELIOS. 

 We have shown that modal noise introduced some limitations, especially with the HA mode,  justifying the implementation of fibre stretchers and AO scanning mode. The difference in modal noise between the four NIRPS fibres indicates that some of them are more sensitive to flexure and/or injection conditions. Fibre connectors and stress at the fibre inputs could play a role in the mode propagation although in the few-mode regime, we may expect that modes are more robust to disturbance and stress. A more detailed and oriented analysis in the lab of the modal noise sensitivity would have potentially allowed for the selection of better fibres.  The double scrambler, very efficient for geometric scrambling of multi-mode fibres, may potentially reduce the effectiveness of the mode scrambling introduced by the AO scanning mode and the stretcher due to imperfect mode coupling between near  and far fields. The echelle 
 grating could also be under-sized for smooth far-field of few-mode fibres. The HE mode, although providing a slightly lower spectral resolution, is recommended to reach the highest S/N and RV precision. The HA mode is recommended in cases high angular resolution is requested (e.g. visual binaries with angular separation less than 2\arcsec). The fact that the spectrograph is intrinsically ultra-stable (with a typical drift of 0.1\,m/s/day) and that  fibre B (especially with the HE mode) presents a higher level of modal noise indicate the use of the simultaneous Fabry-Pérot no longer required. Furthermore, the use of the OBJ-SKY mode is recommended for the sky OH line corrections. We suspect that the modal noise residuals are stable on short timescales and when the telescope does not move or  when is just in tracking mode -- or back in the same position. We plan to investigate the possibility of correcting the residual structures of modal noise using bright telluric standards observed during the night. Another approach could be to correct our spectra using PCA from a library of telluric standards observed with different conditions of injection and telescope position. 

 As expected for nIR spectrographs, the telluric contamination is one of the main limitations on the RV precision. The standard cross-correlation method shows some limitations and post-processing techniques, such as LBL, are needed to optimise the RV measurements. The RV computed online and in real-time at the telescope should be considered as indicative (not nominal). 

 A laser frequency comb (LFC) based on electro-optic modulation \citep{Obrzud2018} was installed and connected to the NIRPS Calibration Unit in December 2023, but it still requires optimisation to make it adequate for daily usage. Combined with the uranium-neon lamp and Fabry-Pérot, it is expected to significantly improve the NIRPS wavelength solution, which is presently at the level of 55 {\cms} in HE mode.   

We present the NIRPS consortium GTO programme that has been ongoing since April 2023. The five-year duration of our GTO programme will offer us the unique opportunity of long-term RV monitoring and accumulation of spectroscopic transits, which is particularly important for the {\modif SP1 and SP3} subprogrammes, respectively.

\begin{acknowledgements}

We thank the  Swiss  National  Science  Foundation  (SNSF) and the Geneva University for their continuous support of our planet search programmes. This work has been in particular carried out in the frame of the National Centre for Competence in Research {\it PlanetS} supported by the SNSF. This publication makes use of The Data \& Analysis Center for Exoplanets (DACE), which is a facility based at the University of Geneva dedicated to extrasolar planets data visualisation, exchange, and analysis.\\   

RD, \'EA, LMa, FBa, RA, CC, NJC, JS-A, PV, TV, LB, BB, AB, LD, AD-B, PLam, AL, OL, LMo \& JPW acknowledge the financial support of the FRQ-NT through the Centre de recherche en astrophysique du Qu\'ebec as well as the support from the Trottier Family Foundation and the Trottier Institute for Research on Exoplanets.\\
RD, \'EA, LMa, FBa, JS-A, PV, TA, J-SM \& MO acknowledges support from Canada Foundation for Innovation (CFI) programme, the Universit\'e de Montr\'eal and Universit\'e Laval, the Canada Economic Development (CED) programme and the Ministere of Economy, Innovation and Energy (MEIE).\\
XDe, XB, ACar, TF \& VY acknowledge funding from the French ANR under contract number ANR\-18\-CE31\-0019 (SPlaSH), and the French National Research Agency in the framework of the Investissements d'Avenir programme (ANR-15-IDEX-02), through the funding of the `Origin of Life' project of the Grenoble-Alpes University.\\
The Board of Observational and Instrumental Astronomy (NAOS) at the Federal University of Rio Grande do Norte's research activities are supported by continuous grants from the Brazilian funding agency CNPq. This study was partially funded by the Coordena\c{c}\~ao de Aperfei\c{c}oamento de Pessoal de N\'ivel Superior—Brasil (CAPES) — Finance Code 001 and the CAPES-Print programme.\\
JRM acknowledges CNPq research fellowships (Grant No. 308928/2019-9).\\
RR, JIGH, JLR, ASM, FGT, NN, VMP \& AKS acknowledge financial support from the Spanish Ministry of Science, Innovation and Universities (MICIU) project PID2020-117493GB-I00.\\
NCS, SCB, EC, JGd, ED-M \& ARCS acknowledge the support from FCT - Funda\c{c}\~ao para a Ci\^encia e a Tecnologia through national funds by these grants: UIDB/04434/2020, UIDP/04434/2020.\\
Co-funded by the European Union (ERC, FIERCE, 101052347). Views and opinions expressed are however those of the author(s) only and do not necessarily reflect those of the European Union or the European Research Council. Neither the European Union nor the granting authority can be held responsible for them.\\
RA acknowledges the Swiss National Science Foundation (SNSF) support under the Post-Doc Mobility grant P500PT\_222212 and the support of the Institut Trottier de Recherche sur les Exoplan\`etes (IREx).\\
This work has been carried out within the framework of the NCCR PlanetS supported by the Swiss National Science Foundation under grants 51NF40\_182901 and 51NF40\_205606.\\
BLCM \& AMM acknowledge CAPES postdoctoral fellowships.\\
BLCM acknowledges CNPq research fellowships (Grant No. 305804/2022-7).\\
XDu acknowledges the support from the European Research Council (ERC) under the European Union’s Horizon 2020 research and innovation programme (grant agreement SCORE No 851555) and from the Swiss National Science Foundation under the grant SPECTRE (No 200021\_215200).\\
FG acknowledges support from the Fonds de recherche du Qu\'ebec (FRQ) - Secteur Nature et technologies under file \#350366.\\
TV acknowledges support from the Fonds de recherche du Qu\'ebec (FRQ) - Secteur Nature et technologies under file \#320056.\\
JLA, RLG, DOF, YSM \& MAT acknowledge CAPES graduate fellowships.\\
SCB acknowledges the support from Funda\c{c}\~ao para a Ci\^encia e Tecnologia (FCT) in the form of a work contract through the Scientific Employment Incentive programme with reference 2023.06687.CEECIND.\\
This project has received funding from the European Research Council (ERC) under the European Union's Horizon 2020 research and innovation programme (project {\sc Spice Dune}, grant agreement No 947634). This material reflects only the authors' views and the Commission is not liable for any use that may be made of the information contained therein.\\
NBC acknowledges support from an NSERC Discovery Grant, a Canada Research Chair, and an Arthur B. McDonald Fellowship, and thanks the Trottier Space Institute for its financial support and dynamic intellectual environment.\\
LD acknowledges the support of the Natural Sciences and Engineering Research Council of Canada (NSERC) [funding reference number 521489] and from the Fonds de recherche du Qu\'ebec (FRQ) - Secteur Nature et technologies [funding file number 332355].\\
DBF acknowledges financial support from the Brazilian agency CNPq-PQ (Grant No. 305566/2021-0). Continuous grants from the Brazilian agency CNPq support the STELLAR TEAM of the Federal University of Ceara's research activities.\\
ED-M further acknowledges the support from FCT through Stimulus FCT contract 2021.01294.CEECIND. ED-M acknowledges the support by the Ram\'on y Cajal grant RyC2022-035854-I funded by MICIU/AEI/10.13039/50110001103 and by ESF+.\\
DE acknowledge support from the Swiss National Science Foundation for project 200021\_200726. The authors acknowledge the financial support of the SNSF.\\
AL acknowledges support from the Fonds de recherche du Qu\'ebec (FRQ) - Secteur Nature et technologies under file \#349961.\\
ICL acknowledges CNPq research fellowships (Grant No. 313103/2022-4).\\
LMo acknowledges the support of the Natural Sciences and Engineering Research Council of Canada (NSERC), [funding reference number 589653].\\
CMo acknowledges the funding from the Swiss National Science Foundation under grant 200021\_204847 “PlanetsInTime”.\\
NN acknowledges financial support by Light Bridges S.L, Las Palmas de Gran Canaria.\\
NN acknowledges funding from Light Bridges for the Doctoral Thesis "Habitable Earth-like planets with ESPRESSO and NIRPS", in cooperation with the Instituto de Astrof\'isica de Canarias, and the use of Indefeasible Computer Rights (ICR) being commissioned at the ASTRO POC project in the Island of Tenerife, Canary Islands (Spain). The ICR-ASTRONOMY used for his research was provided by Light Bridges in cooperation with Hewlett Packard Enterprise (HPE).\\
CPi acknowledges support from the NSERC Vanier scholarship, and the Trottier Family Foundation. CPi also acknowledges support from the E. Margaret Burbidge Prize Postdoctoral Fellowship from the Brinson Foundation.\\
ARCS acknowledges the support from Funda\c{c}ao para a Ci\^encia e a Tecnologia (FCT) through the fellowship 2021.07856.BD.\\
AKS acknowledges financial support from La Caixa Foundation (ID 100010434) under the grant LCF/BQ/DI23/11990071.
\end{acknowledgements}

\bibliographystyle{aa}
\bibliography{bib_file}
\end{document}